\documentclass[useAMS, usenatbib]{mn2e}
\usepackage{graphicx}
\usepackage{longtable}
\usepackage{multicol}
\usepackage{lscape}
\usepackage{subfig}
\usepackage{natbib}
\usepackage{threeparttable}
\usepackage{gensymb}

\title[Coronal-Line Forest AGN]{Coronal-Line Forest AGN: the best view of the inner edge of the AGN torus?}
\author[M. Rose, M. Elvis \& C. N. Tadhunter]{Marvin Rose$^{1}$\thanks{E-mail:mrose@cfa.harvard.edu (MR)}, Martin Elvis$^{1}$ and Clive N. Tadhunter$^{2}$\\
$^{1}$Harvard Smithsonian Center for Astrophysics, 60 Garden St., Cambridge, MA 02138, USA\\
$^{2}$Department of Physics and Astronomy, University of Sheffield, Sheffield S3 7RH, UK\\}
\begin{document}

\date{}

\pagerange{\pageref{firstpage}--\pageref{lastpage}} \pubyear{2011}

\maketitle

\label{firstpage}

\begin{abstract}

We introduce Coronal-Line Forest Active Galactic Nuclei (CLiF AGN), AGN which have a rich spectrum of forbidden high-ionization lines (FHILs, e.g. [FeVII], [FeX] and [NeV]), as well as relatively strong narrow ($\sim$300 km s$^{-1}$) H$\alpha$ emission when compared to the other Balmer transition lines. We find that the kinematics of the CLiF emitting region are similar to those of the forbidden low-ionization emission-line (FLIL) region. We compare emission line strengths of both FHILs and FLILs to CLOUDY photoionization results and find that the CLiF emitting region has higher densities (10$^{4.5}$ $<$ n$_H$ $<$ 10$^{7.5}$ cm$^{-3}$) when compared to the FLIL emitting region (10$^{3.0}$ $<$ n$_H$ $<$ 10$^{4.5}$ cm$^{-3}$). We use the photoionization results to calculate the CLiF regions radial distances (0.04 $<$ R$_{CLiF}$ $<$ 32.5 pc) and find that they are comparable to the dust grain sublimation distances (0.10 $<$ R$_{SUB}$ $<$ 4.3 pc). As a result we suggest that the inner torus wall is the most likely location of the CLiF region, and the unusual strength of the FHILs is due to a specific viewing angle giving a maximal view of the far wall of the torus without the continuum being revealed.

\end{abstract}


\section{Introduction}

In the standard active galactic nuclei (AGN) unification model \citep{antonucci85} a large-scale height torus surrounds the inner AGN, obscuring our view of the central continuum source and the broad emission line region (BLR) from most directions. The inner wall of this torus lies at the sublimation radius of the most refractive dust - no dust can exist stably at smaller radii. This inner wall is heavily irradiated. It has been proposed that ablation of the surface produces a thermal wind of highly ionized gas \citep{krolik95}. This gas may produce an unusual spectrum, which may include forbidden high ionization lines (FHILs), also known as `coronal lines' \citep{appenzeller91}. 

From most orientations, either pole-on or edge on, the torus inner wall will be barely visible. However, there should be a small range of angles of which the torus only just hides the inner nucleus. For these type 2 (dust obscured) AGNs the rear inner wall will be maximally visible. Here we describe a population of AGN that are candidates to be this special population - the Coronal Line Forest AGN (CLiF AGN).

In this paper, we introduce the CLiF AGN as a distinct class of AGN, and present data for a sample of the first seven identified CLiF AGN. The cosmological parameters used throughout this paper assume a $\Lambda$ cold dark matter ($\Lambda$CMD) cosmology with parameters $\Omega_M$ = 0.3, $\Omega_\Lambda$ = 0.7 and $H_\circ$ = 70 km s$^{-1}$ Mpc$^{-1}$.

\section{Properties of CLiF AGN}\label{sect:clif}

Most AGN show some weak (EW $<$ 3 \AA ) spectral lines from forbidden transitions of highly ionized ions (FHIL, ionization potential, I$_{P}$ $>$ 54.4 eV; the HeII edge) in their spectra, e.g. [NeV], [FeVII], [FeX], [FeXIV] \citep{penston84}\footnote{In this work we define I$_P$ as the ionization potential required to achieve the current ionic stage by removing an electron from a lower-ionization stage \citep{derobertis84}.}. The presence of FHILs in the spectra of AGN unambiguously indicate the presence of an AGN, due to their high ionization potentials, reaching up to soft X-ray energies $\sim$0.1-0.2 keV. FHILs are not generally associated with LINER nuclei or starbursts \citep{filippenko84}. FHILs are physically distinct because their transitions have higher critical densities (n$_C$ $\sim$ 10$^{7.5}$ cm$^{-3}$; \citealt{osterbrock06}) than lower ionization species (n$_C$ $<$ 10$^{6.0}$ cm$^{-3}$; \citealt{osterbrock06}). In most AGN the FHILs are blueshifted ($\Delta$v $\sim$-250 km s$^{-1}$) with respect to the rest frame of the AGN and have substantial velocity widths (FWHM $\sim$1000 km s$^{-1}$), That are intermediate between those of the standard low ionization forbidden lines (FLIL) of the narrow emission line region (NLR) (FWHM $\sim$500 km s$^{-1}$, e.g. [OIII]$\lambda$5007), and the BLR (FWHM $\sim$5000 km s$^{-1}$). These kinematic properties support an origin for the FHIL emitting gas in a wind outflowing from the BLR \citep{appenzeller91} or the torus \citep{krolik95}. 

A few rare AGN have instead a dozen or more FHILs with relatively high fluxes (F(FHIL)/F(H$\beta$) $>$ 0.25): we call these the CLiF AGN. So far, CLiF AGN have only been identified serendipitously. Indeed, there are just 5 CLiF AGN in the literature (Table \ref{tab:prop}): Q1131+16 \citep{rose11}, III Zw 77 \citep{osterbrock81}, Mrk 1388 \citep{osterbrock85}, ESO 138 G1 \citep{alloin92} and Tololo 0109-383 \citep{fosbury83}. 

The optical spectra of these objects are remarkable in several ways. As well as the the typical FLILs, i.e. [OIII], [NII], [OI] and [SII], and the Balmer emission lines including H$\alpha$, H$\beta$, their spectra show many FHILs, including [FeV], [FeVI], [FeVII], [FeX], [FeXI], [NeV] and [ArV] all with F(FHIL)/F(H$\beta$) $>$ 0.2. This is more than found in typical AGN which just show the [FeVII]$\lambda\lambda$5721,6087lines \citep{appenzeller91}, and even these are weak (EW $<$ 3 \AA ).

We find ten unusual properties of CLiF AGN:\begin{enumerate}
\item[1.] Detection of the [FeVII] lines 3759\AA , 5159\AA , 5720\AA\  and 6087\AA\  with the flux ratio F([FeVII]$\lambda$6087)/F(H$\beta$) $>$ 0.25. Where F(H$\beta$) is the flux from the {\it narrow} H$\beta$ emission line.

\item[2.] F([FeX]$\lambda$6374)/F(H$\beta$) $>$ 0.2.

\item[3.] F([NeV]$\lambda$3426)/F(H$\beta$) $>$ 1.

\item[4.] F([FeV])/F(H$\beta$) \& F([FeVI])/F(H$\beta$) $>$ 0.2.

\item[5.] H$\alpha$/H$\beta$ $>$ 2.9 [the Case B value seen in the NLR; \citep{osterbrock06}] {\it and} H$\gamma$/H$\beta$ $\approx$ 0.47 (Case B), implying that the apparent level of dust extinction calculated using H$\alpha$/H$\beta$ is greater than that calculated using H$\gamma$/H$\beta$.

\item[6.] Higher level HI Balmer transitions (with wavelengths blueward of H$\delta$ and up to H$_{16}$; \citealt{rose11}) are present in emission.

\item[7.] The FHILs are not blueshifted with respect to the FLILs ($\Delta$v $<$ 100 km s$^{-1}$).

\item[8.] The velocity widths of the FHILs are narrow and the same as the FLILs (FWHM $\sim$ 300 km s$^{-1}$) using single Gaussian fits.

\item[9.] The FHILs maintain their high EWs for longer than 3 years. This is long compared with the observed emission line fading times for Type IIn supernovae \citep{smith09} and tidal disruption events for stars falling on to a supermassive black hole  (SMBH; \citealt{gezari06}; \citealt{komossa09}; \citealt{wang12}), two astrophysical phenomena that also exhibit rich spectra of narrow emission lines including both FLILs and FHILs. 

\item[10.] High EW ($>$3\AA ) emission lines from unknown/unexpected species are present.

\end{enumerate}

This cluster of properties appear to be important for CLiF AGN. However, at this stage it is premature to formally define the CLiF AGN. We need a larger sample of CLiF AGN in order to produce a well-crafted definition.

\section{A search for CLiF AGN in SDSS}

\begin{center}
\begin{table*}
\centering
\caption{Basic properties of the CLiF AGN sample. Redshifts (z$_{[OIII]}$) are determined by single Gaussian fits to the prominent [OIII]$\lambda$5007 emission line. The column r' gives the SDSS r' model magnitudes (https://www.sdss3.org/dr10/algorithms/magnitudes.php). We choose model magnitudes because these objects are type 2 AGN and therefore the photometry will be dominated by the host galaxy. In the cases of ESO 138 G1 and Tololo 0109-383, there are no available r' magnitudes, instead we use Cousins R band magnitudes taken from \citet{lauberts89}. `Spectrum' gives the origin of the spectrum. The column `Ref' refers to the first publication that discusses the optical spectrum of the CLiF AGN candidate.}
\begin{tabular}{lcccccc}
\hline
Name	&	RA (J2000)	&	Dec (J2000)	&	z$_{[OIII]}$	&	r'	&	Spectrum  &	Ref. 	\\
\hline									
Q1131+16	&	11:31:11.05	&	16:27:39.50	&	0.1732	& 17.31 & WHT/ISIS &	Rose et al. (2011)	\\
III Zw 77	&	16:23:45.87	&	41:04:56.69	&	0.0341	& 15.25 & SDSS &	Osterbrock (1981)	\\
Mrk 1388	&	14:50:37.85	&	22:44:03.61	&	0.0216	& 14.55 & SDSS &	Osterbrock (1985)	\\
ESO 138 G1	&	16:51:20.13	&	-59:14:05.20	&	0.0091	& 13.02$^a$ & ESO &	Alloin et al. (1992)	\\
Tololo 0109-383	&	01:11:27.63	&	-38:05:00.48	&	0.0118	& 12.38$^a$  & IPCS/RGO	& Fosbury \& Sansom (1983) \\
J1241+44	&	12:41:34.25	&	44:26:39.25	&	0.0422	& 15.98 & SDSS &	This work	\\
J1641+43	&	16:41:26.91	&	43:21:21.59	&	0.2214	& 18.60 & SDSS &	This work	\\
\hline									
\end{tabular}
\begin{tablenotes}
       \item[a]$^a$ The R-band magnitude is from \citet{lauberts89}.
     \end{tablenotes}
\label{tab:prop}
\end{table*}
\end{center} 

Given that CLiF AGN have very different spectral properties compared to typical AGN, one might expect that CLiF AGN would be easy to select. We used each of the unusual properties listed above that could be selected from the SDSS DR10 database\footnote{http://skyserver.sdss3.org/dr10/en/tools/search/sql.aspx}. 

The properties of the emission lines in the SDSS spectra are determined using the BOSS pipeline \citep{bolton12}. Each line is modeled as a single Gaussian. The amplitudes, centroids, and widths of each emission line are optimized non-linearly to obtain a minimum-$\chi$$^{2}$ fit to each emission line. All fitted emission lines are assumed to have the same redshift. The widths of the emission lines are assumed to have the same intrinsic velocity depending on the `width group' they belong to. The relevant groups for the emission lines used in this work are `Balmer' (consisting of the Balmer emission/absorption features) and `emission' (consisting of non-Balmer emission lines). The velocity widths are a strength weighted average over the width group. The continuum level is estimated either from the best-fit velocity-dispersion model if the spectrum is classified as a galaxy\footnote{The best-fit velocity dispersion is determined by fitting locally for the position of the minimum-$\chi$$^{2}$ versus trial velocity dispersion in the neighborhood of the lowest gridded $\chi$$^{2}$ value. See https://www.sdss3.org/dr10/algorithms/redshifts.php for a full description.}, or from a linear fit either side of the emission line if the spectrum is classified as a quasar. The parameters such as the fluxes, FWHM, redshifts, EW and continua levels, as well as their associated uncertainties, are all reported by the BOSS pipeline \citep{bolton12}. We use EW instead of a flux ratio with H$\beta$ because our definition for a strong FHIL as the ratio of the FHIL flux strength to the flux from the {\it narrow} H$\beta$ emission line is not usable with the SDSS database. As the SDSS database does not differentiate between BLR and NLR Balmer emission, we cannot perform searches based on the flux ratios. 

Six of the ten CLiF properties could not be searched using the BOSS database. Properties 1, 4 and 10 could not be selected because these Fe emission lines are not defined in the SDSS spectral fitting model (see Table 5 in \citet{bolton12} for the full list of emission lines fitted by the SDSS). The kinematic properties 7 and 8 are not available in the BOSS database either. Also, there is no timing information to determine whether property 9 is observed. 

We attempted to select CLiF AGN candidates from the BOSS catalog, making use of the on-line query tool, for properties 2, 3, 5 and 6:

\begin{itemize}

\item For property 2, \citet{gelbord09} noted that the [FeX]$\lambda$6375 broad blue wing is often blended with the [OI]$\lambda$6364 emission line which is included in the BOSS catalog. As [OI]6300/6364 = 3 from atomic physics, if [FeX]$\lambda$6375 is present and blended with [OI]$\lambda$6364, it will reduce the [OI]6300/6364 ratio. Hence we required that objects had to have [OI](6300/6364) $<$ 3. A caveat is that [OI]$\lambda$6300 can also be blended with [SIII]$\lambda$6312. This blend could result in a ratio [OI](6300/6364) $>$ 3. However, the [SIII]$\lambda$6312 emission line is included in the BOSS spectral fits \citep{bolton12}. This selection criterion will then be conservative because any contribution from [SIII]$\lambda$6312 will result in CLiF AGN not being selected.   

\item For property 3, we selected objects with strong [NeV]$\lambda$3426 emission (EW\footnote{While EW has been used as a standard indicator of emission line strength, it depends on the intensity of both the emission line and underlying continuum. Therefore there is a caveat when defining the properties of CLiF AGN using the EW because the large observed EWs could be due to a weak stellar or AGN continuum emission.} $>$ 10 \AA ), although this only works for z $>$ 0.109. 

\item For property 5 we required that H$\alpha$/H$\beta$ $>$ 2.9 AND H$\gamma$/H$\beta$ $<$ 0.5, so as not to exclude reddened objects. 

\item For property 6 we note that H$\epsilon$, which is in the BOSS spectral fits, has at $\lambda$=3970 a wavelength close to that of [NeIII]$\lambda$3968 even closer than [OI]$\lambda$6364 and [FeX]$\lambda$6375. Hence we created the criterion EW([NeIII]$\lambda$3968) $>$ 5 \AA\footnote{ Tests of lower ($\sim$3\AA) EW limits returned larger samples of several thousand unrelated objects, and higher ($\sim$10\AA) EW limits produced no CLiF AGN at all.} as H$\epsilon$ will boost the [NeIII]$\lambda$3968 flux.  

\end{itemize}

Each of the initial searches based on only one of these properties resulted in large samples of several thousand objects. These were of unrelated phenomena, such as typical quasars whose broad emission lines altered both the Balmer decrements and [OI] ratios, resulting in false detections for properties 2 and 5. Property 3 also selected quasars which lacked strong Fe FHILs. Normal type 2 AGN were also selected with property 3, suggesting that this emission line is important in type 2 AGN \citep{mignoli13}. Finally property 6 selected both typical quasars and quiescent galaxies. These observations were confirmed with spot checks of their spectra. 

Searches based on pairs of the criteria also failed to constrain the search. Therefore additional criteria were needed in order to constrain the search. 

We then searched for objects using all four properties (2 AND 3 AND 5 AND 6). This search produced just two CLiF AGN. One of these was the previously detected Mrk 1388, and one new CLiF AGN: J1641+43. 

As [NeV]$\lambda$3425 limited the search to $z$ $>$ 0.109 we then scaled back the search to the three criteria excluding 3. This search found one additional CLiF AGN candidate: J1241+44. 

These search criteria failed to find the known CLiF AGN III Zw 77, although it is in the SDSS database. We obtained the SDSS spectrum for III Zw 77 separately

While somewhat unsuccessful, this exercise serves to highlight both the apparent rarity of CLiF AGN in the AGN population, and the complexity of selecting such objects.  

In this paper we make use of SDSS spectra (Figure \ref{fig:spec1}). The basic properties of the CLiF AGN sample studied in this paper are given in Table \ref{tab:prop}. Note that the outputs of the SDSS pipeline are used only for the sample selection. Detailed measurements of emission line parameters such as the flux and velocity widths are measured using our own methods ($\S$\ref{sect:measurments}). The redshifts were determined using single Gaussian fits to the [OIII]$\lambda$5007 emission line. This line was chosen because it is the most prominent emission line in the optical spectra of these and most other AGN. 

\section{Emission Spectra}\label{sect:measurments}

\begin{center}
\begin{table*}
\centering
\caption{Key results from the optical spectra of the CLiF AGN. `N$_{Lines}$' refers to the total number of emission lines with S/N $>$ 3, N$_{FHIL}$ is the number of detected FHILs, N$_{[FeVII]}$ is the number of [FeVII] lines, N$_{Fe}$ is the total number of iron emission lines, [FeVII]/H$\beta$ is the H$\beta$ flux ratio for the [FeVII]$\lambda$6087 emission line, [FeX]/H$\beta$ is the H$\beta$ flux ratio for the [FeX]$\lambda$6375 emission line, [NeV]/H$\beta$ is the H$\beta$ flux ratio for the [NeV]$\lambda$3425 emission line, N$_{HI}$ is the number of detected HI lines and N$_{?}$ is the number of unidentified lines. The column `BL' indicates whether broad emission lines were detected in the optical spectrum. `Y' means yes, `?' indicates tentative evidence and `N' indicates a non-detection. All results were obtained using the Starlink package {\sc dipso}.}
\begin{tabular}{lcccccccccc}
\hline
Name	&	N$_{Lines}$	& N$_{FHIL}$	&	N$_{[FeVII]}$	&	N$_{Fe}$	&	 [FeVII]/H$\beta$ 	&	 [FeX]/H$\beta$ 	&	 [NeV]/H$\beta$ 	&	N$_{HI}$	&	N$_{?}$	&	BL	\\
	&	&	 I$_P$ $>$ 54.4 eV &		& 	& 	& 	& 	&	&	\\
\hline																			
Q1131+16	&	99	&	 19	&	7	&	23	&	 0.50	&	 0.23	&	 1.01	&	10	&	30$^a$	&	?	\\
ESO 138 G1	&	71	&	 24	&	7	&	26	&	 0.24	&	 0.08	&	 1.22	&	9	&	2	&	N	\\
Mrk 1388	&	62	&	 13	&	5	&	11	&	 0.34	&	 0.19	&	{\it  b}	&	7	&	17	&	?	\\
III Zw 77	&	51	&	 13	&	6	&	11	&	0.39	&	 0.21	&	{\it  b}	&	8	&	6	&	Y$^{c}$	\\
J1241+44	&	34	&	 12	&	4	&	9	&	1.11	&	 1.88	&	{\it  b}	&	3	&	3	&	N	\\
Tololo 0109-383	&	31	&	 11	&	5	&	8	&	 0.28	&	 0.26	&	 0.57	&	6	&	d	&	N	\\
J1641+43	&	30	&	 8	        &	4	&	7	&	 0.50	&	 0.21	&	 1.73	&	4	&	1	&	N	\\
\hline																			
\end{tabular}
\begin{tablenotes}
       \item[a]$^a$ A few `unknown' lines have been identified since \citet{rose11}.
	\item[b]$^b$ [NeV]$\lambda$3425 is shifted out of the observed frame of the spectra for these objects.
       \item[c]$^c$ The BLR emission of III Zw 77 is significantly blueshifted (-675$\pm$25 km s$^{-1}$) with respect to the NLR emission. 
	\item[d]$^d$ There were no `unknown' lines indicated in \citet{fosbury83}.
     \end{tablenotes}
\label{tab:elprop}
\end{table*}
\end{center}

In Figure \ref{fig:spec1} we plot the spectra for the four CLiF AGN with SDSS spectra. There is a remarkable number of emission lines. We performed a uniform line detection process by making fits to the lines and identifying them as described below.

\subsection{Line detection}\label{sect:gfits}

\begin{figure*}
\centering
\includegraphics[scale=0.62, trim=0.0cm 0cm 0.00cm 0.5cm]{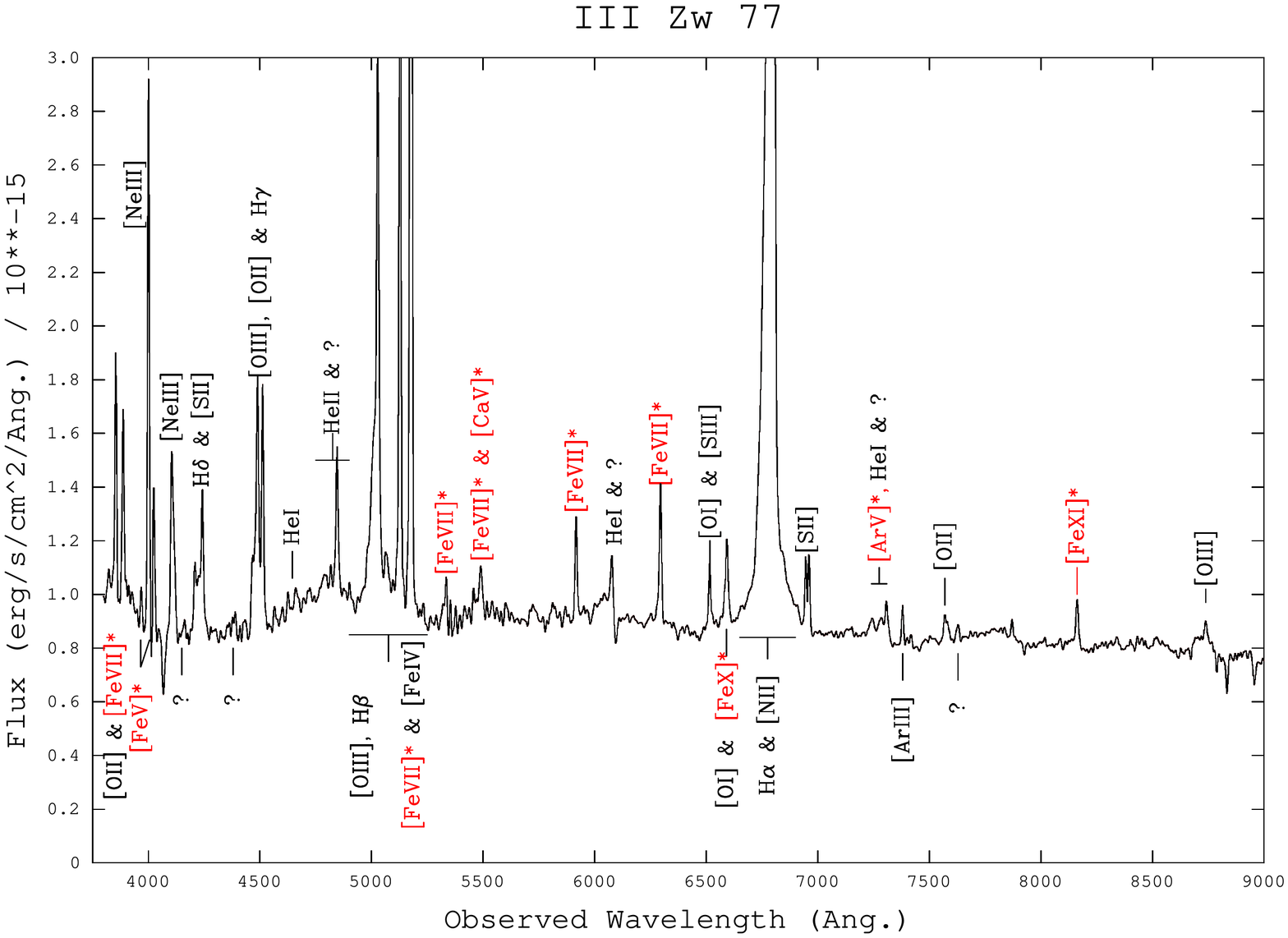}
\includegraphics[scale=0.62, trim=0.0cm 2cm 0.00cm 3.5cm]{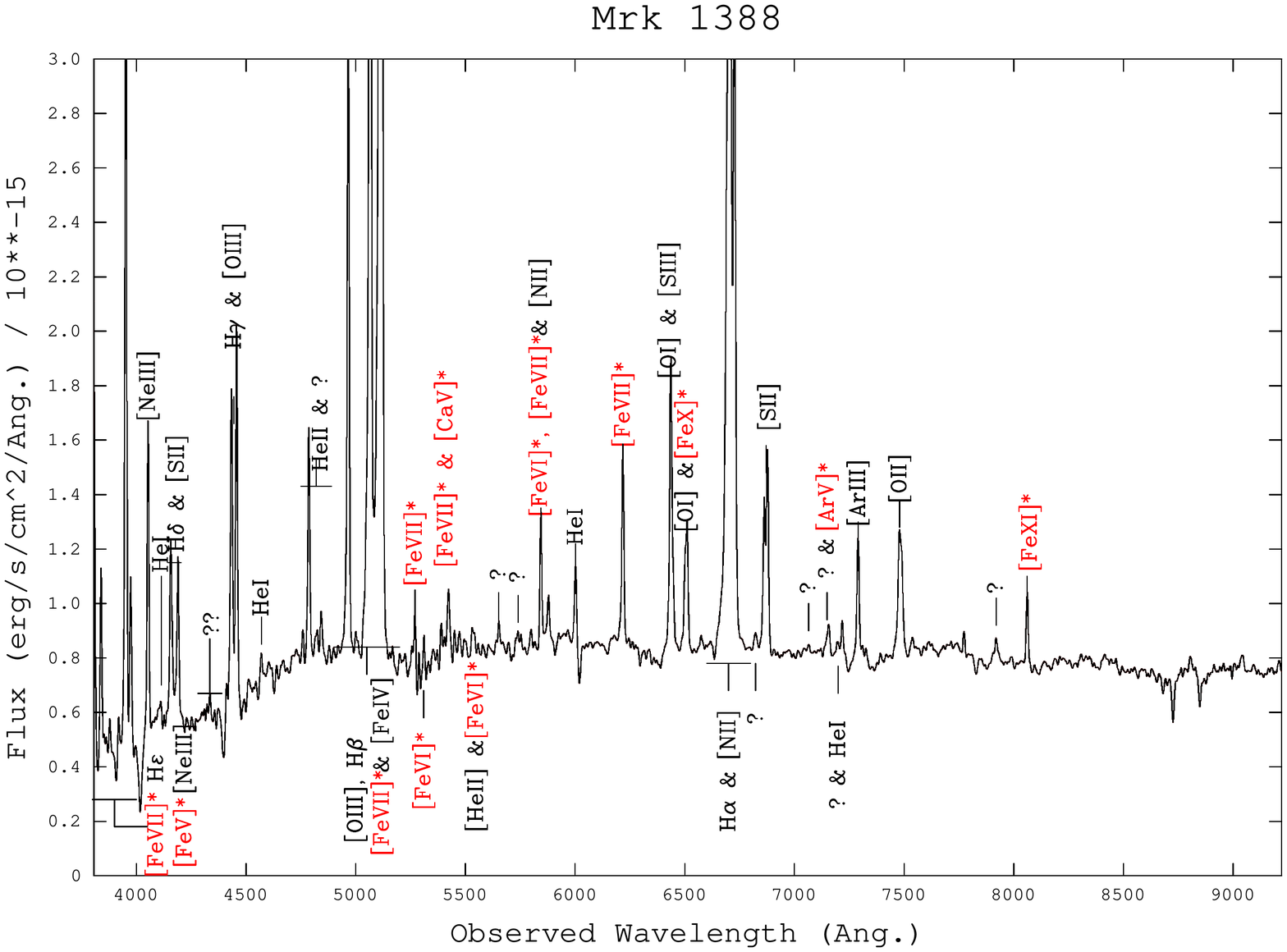}
\caption{Labeled SDSS spectra for III Zw 77 and Mrk 1388. FHILs are labeled in red and marked with *. Note that we could not label all the emission lines on the spectra due to space constraints. The flux scale is measured in units of 10$^{-15}$ erg s$^{-1}$ \AA $^{-1}$ cm$^{-2}$, the Observed Wavelength is measured in units of \AA . {\sc continued on next page}.}
\end{figure*}

\begin{figure*}
\centering
\ContinuedFloat
\includegraphics[scale=0.62, trim=0.0cm 0cm 0.00cm 0.5cm]{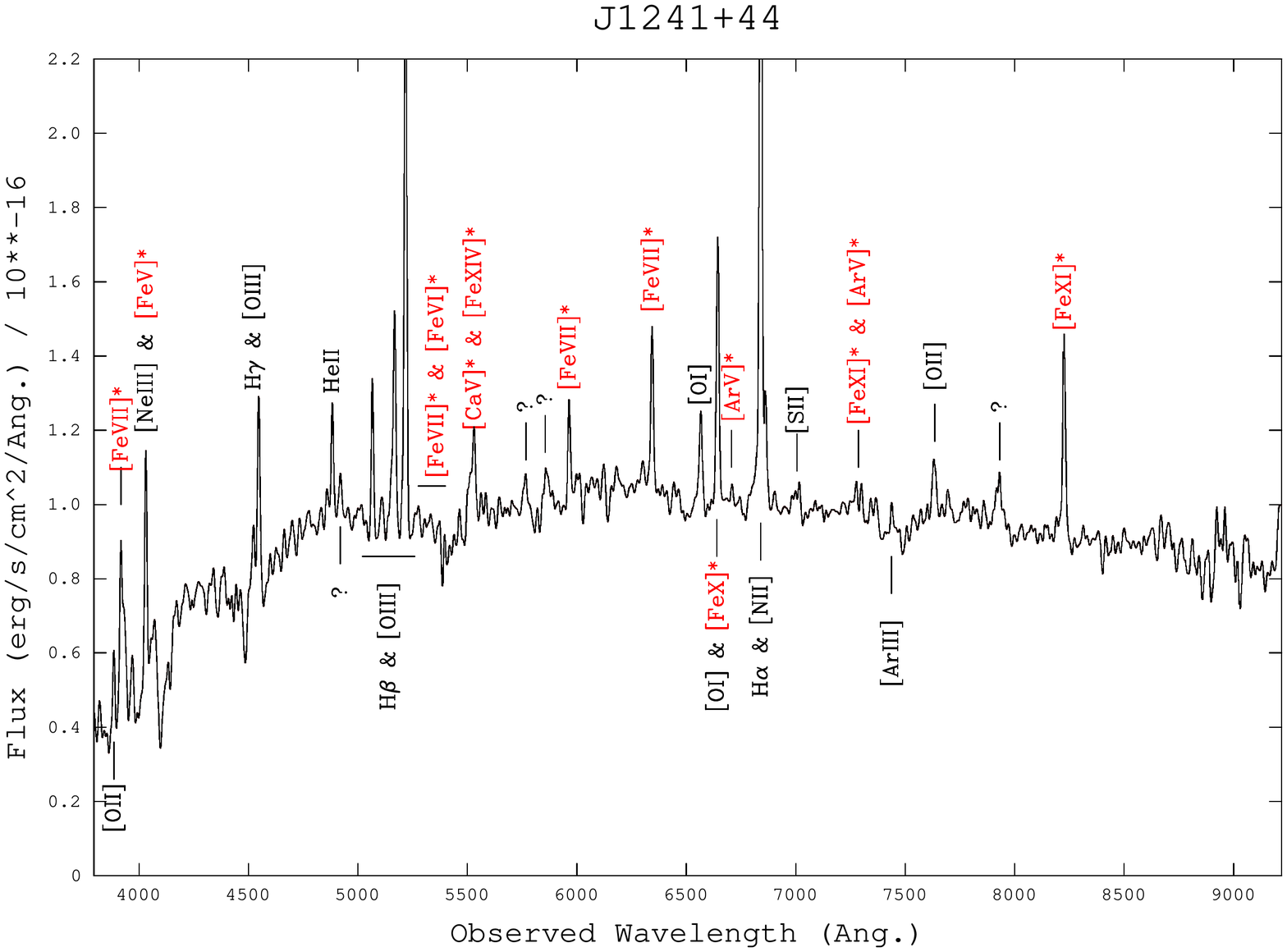}
\includegraphics[scale=0.62, trim=0.0cm 2cm 0.00cm 3.5cm]{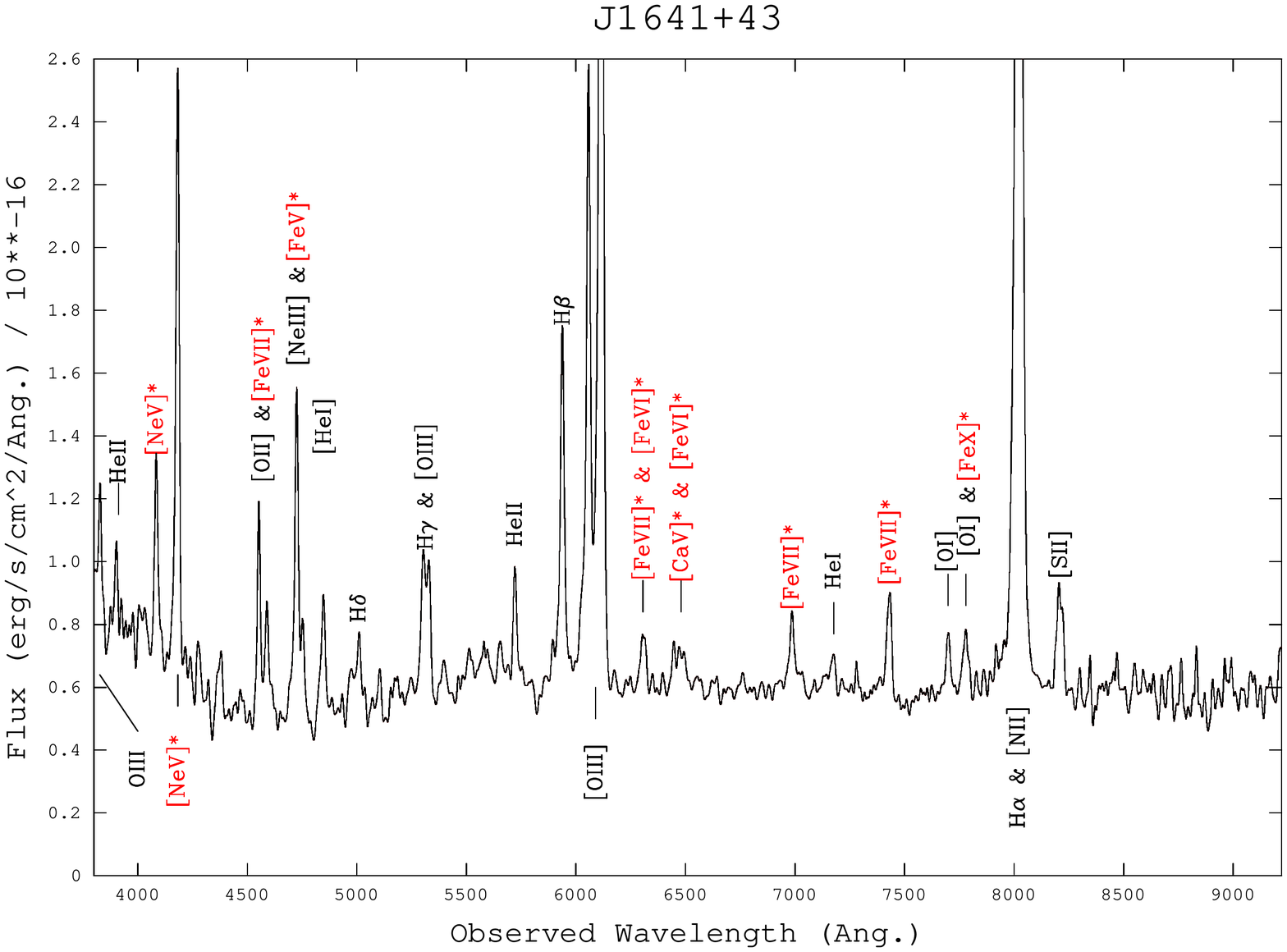}
\caption{Labeled SDSS spectra for J1241+44 and 1641+43. The flux scale is measured in units of 10$^{-16}$ erg s$^{-1}$ \AA $^{-1}$ cm$^{-2}$.}
\label{fig:spec1}
\end{figure*}

All line detections were determined by fitting single Gaussians to the emission features in the CLiF AGN spectra, using the Starlink package {\sc dipso}\footnote{http://starlink.jach.hawaii.edu/starlink}, accepting  those with S/N $>$ 3. We define S/N as the total flux measured using the Gaussian profile divided by the uncertainty of this flux from the fit as provided by {\sc dipso}. For the three CLiF AGN without SDSS spectra (Q1131+16, ESO 138 G1 and Tololo 0109-373), we use the spectra and data presented in Table \ref{tab:prop}. 30-99 lines were found in the CLiF AGN (Table \ref{tab:elprop}).

\begin{figure}
\centering
\includegraphics[scale=0.45, trim=1.5cm 3cm 1.5cm 3.5cm]{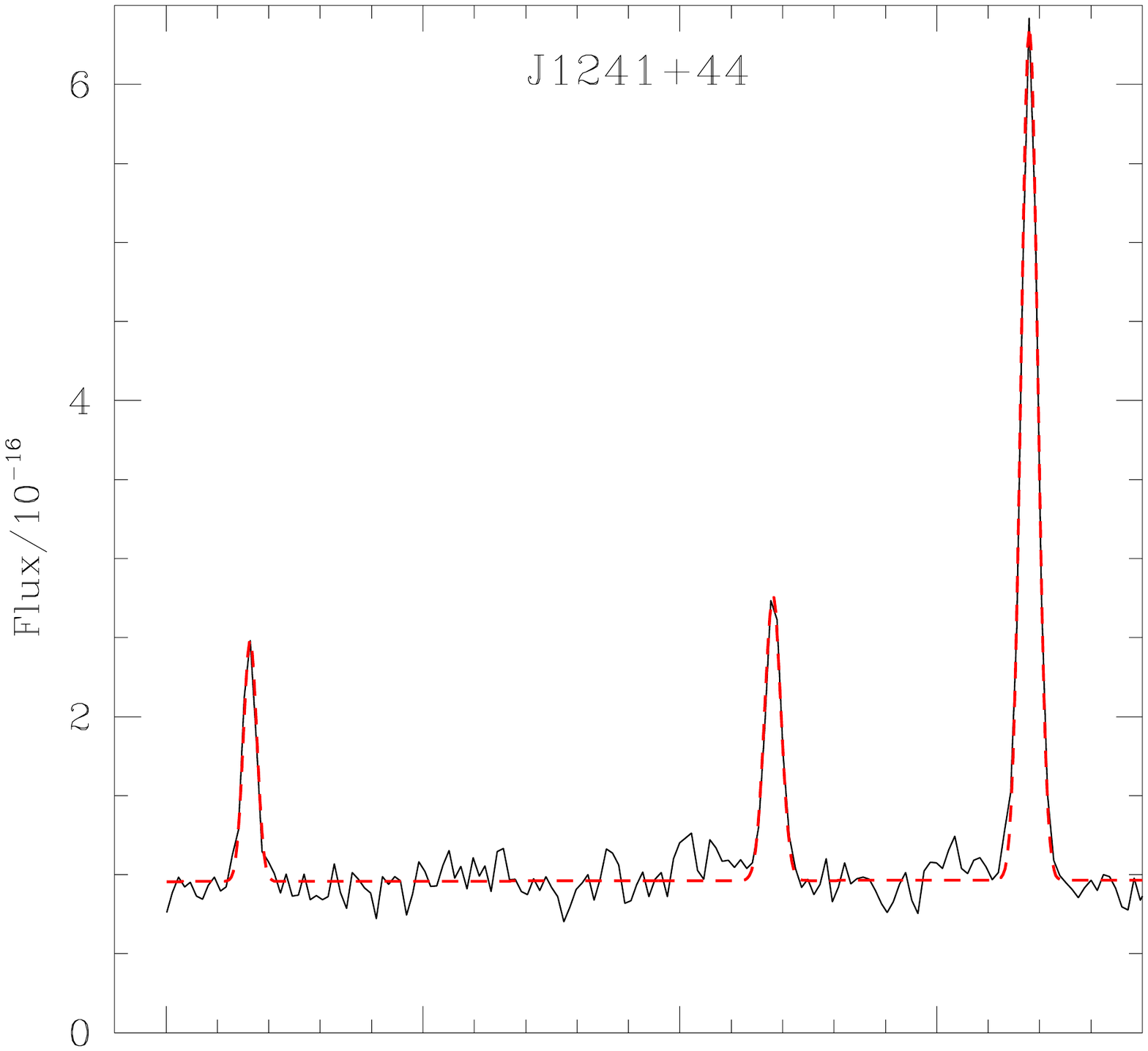}\\
\includegraphics[scale=0.45, trim=1.5cm 3cm 1.5cm 18.1cm]{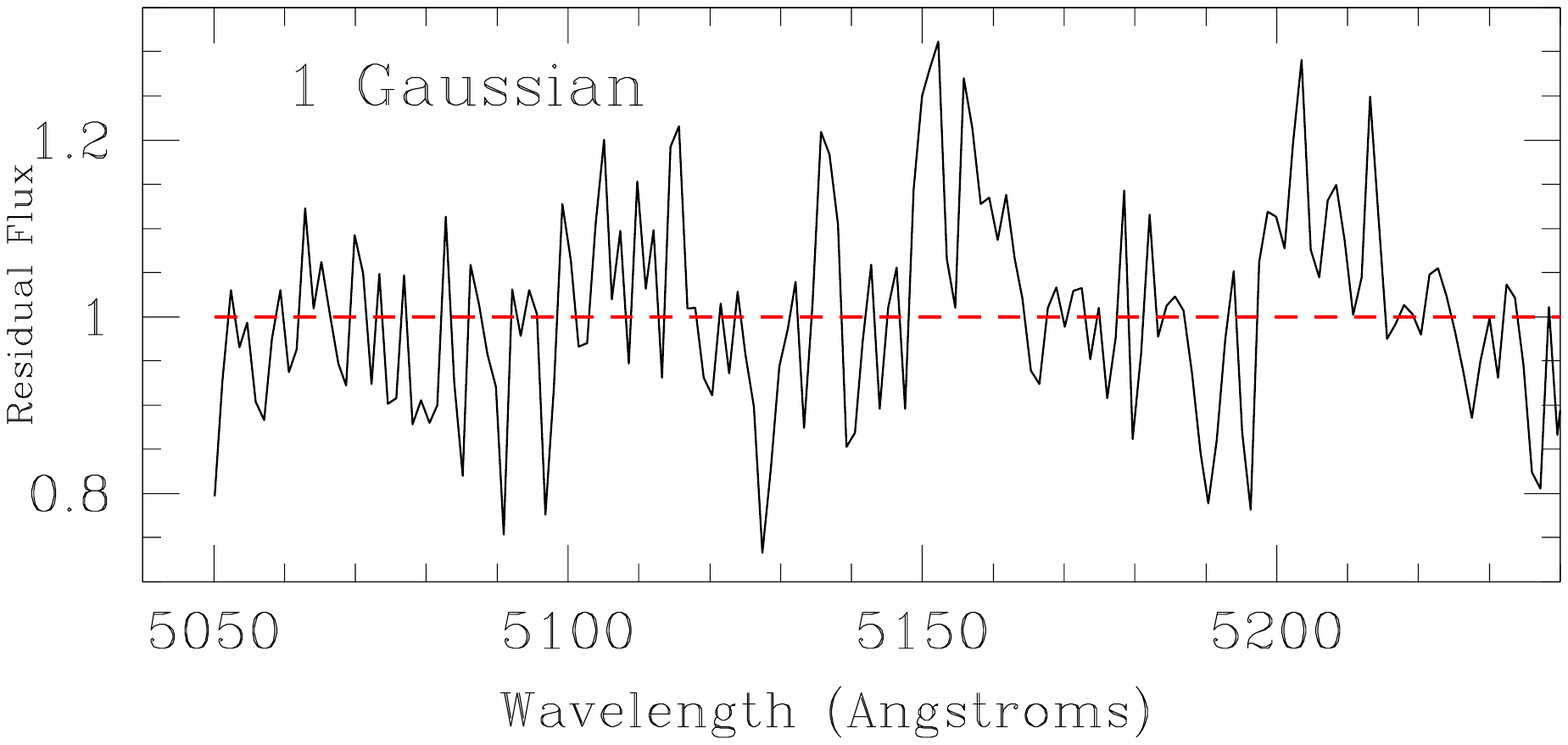}
\caption{{\bf Top.} Multi-component model fits to the H$\beta$ and [OIII]$\lambda\lambda$4959,5007 emission lines for 1241+44. The model fits are drawn with the dashed red line. The flux is measured in units of erg s$^{-1}$ \AA $^{-1}$ cm$^{-2}$ and the wavelength is in units of \AA . {\bf Bottom.} The residuals from the single Gaussian fits.}
\label{fig:12fit}
\end{figure}

\begin{figure}
\centering
\includegraphics[scale=0.45, trim=1.5cm 3cm 1.5cm 3.5cm]{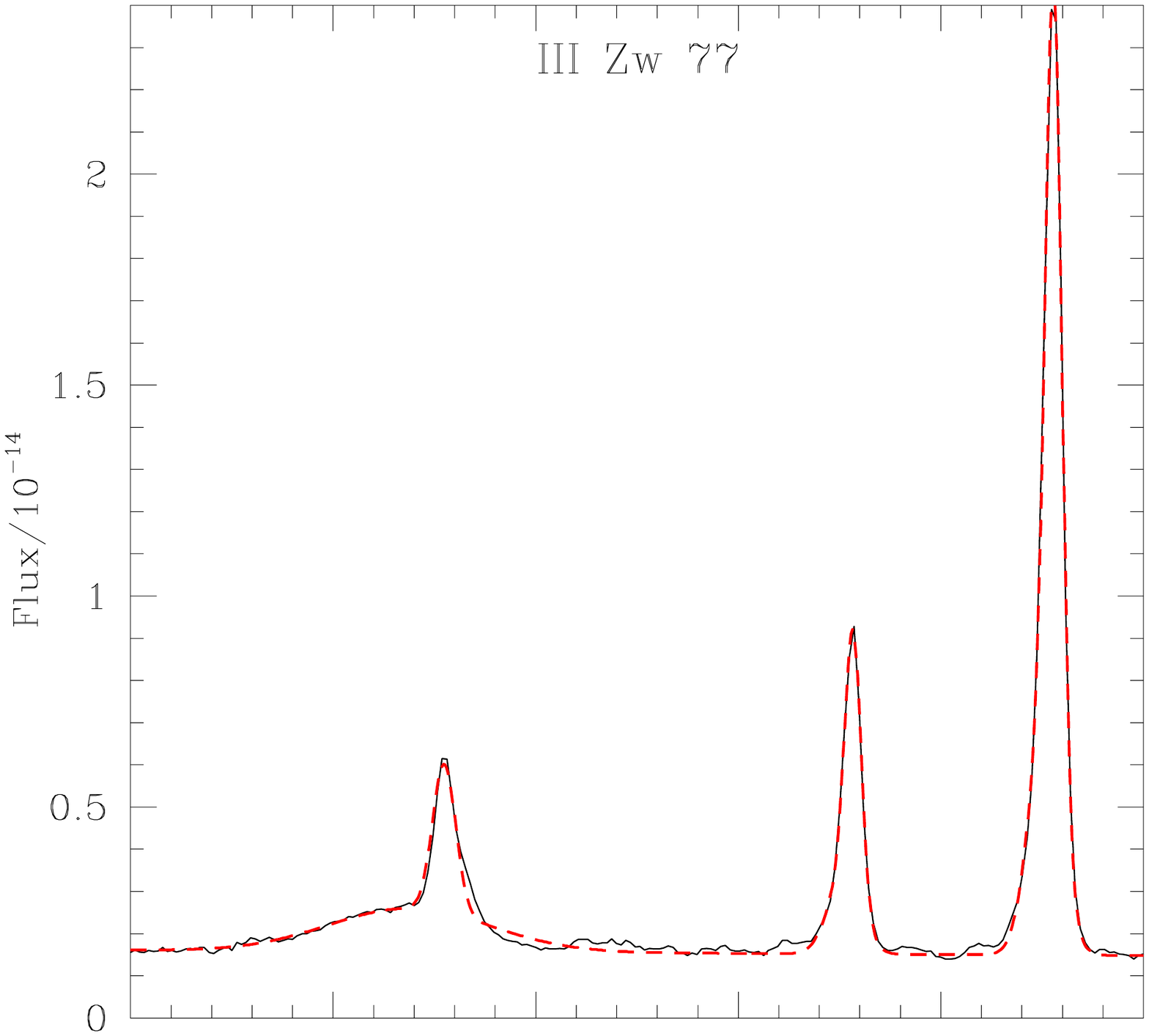}\\
\includegraphics[scale=0.45, trim=1.5cm 3cm 1.5cm 18.1cm]{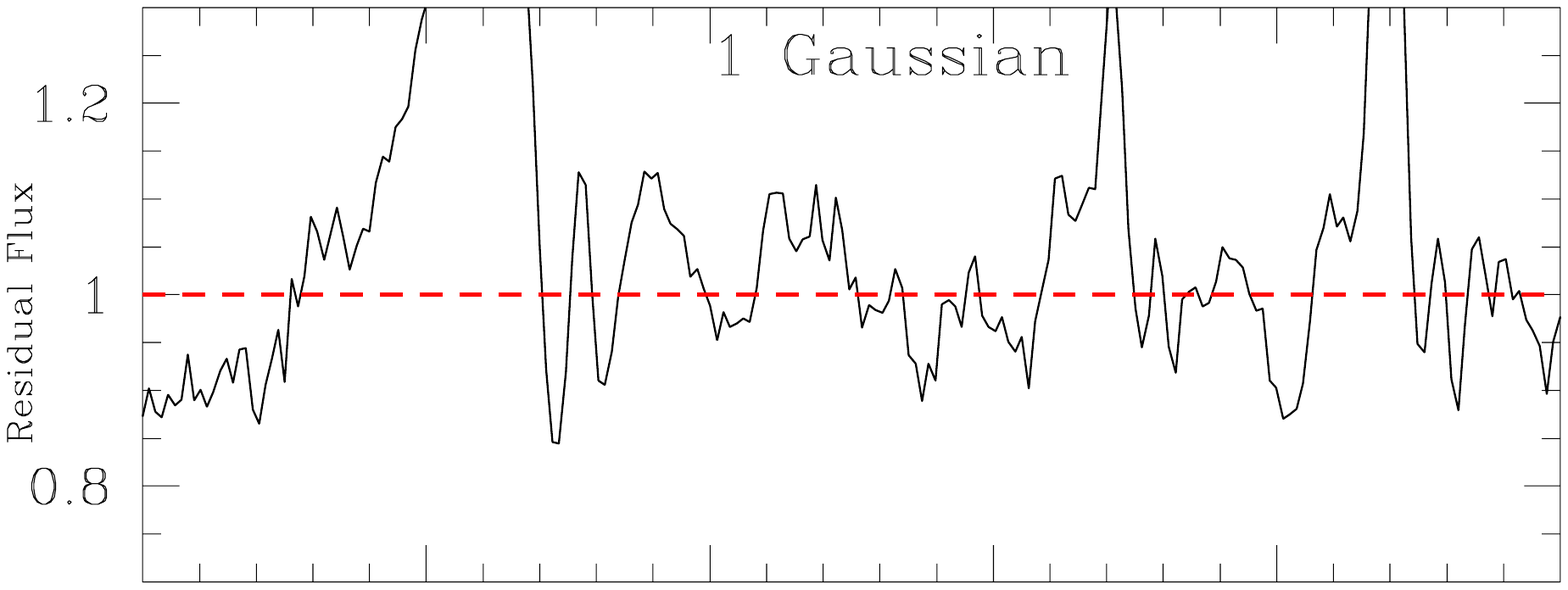}
\includegraphics[scale=0.45, trim=1.5cm 5cm 1.5cm 18.1cm]{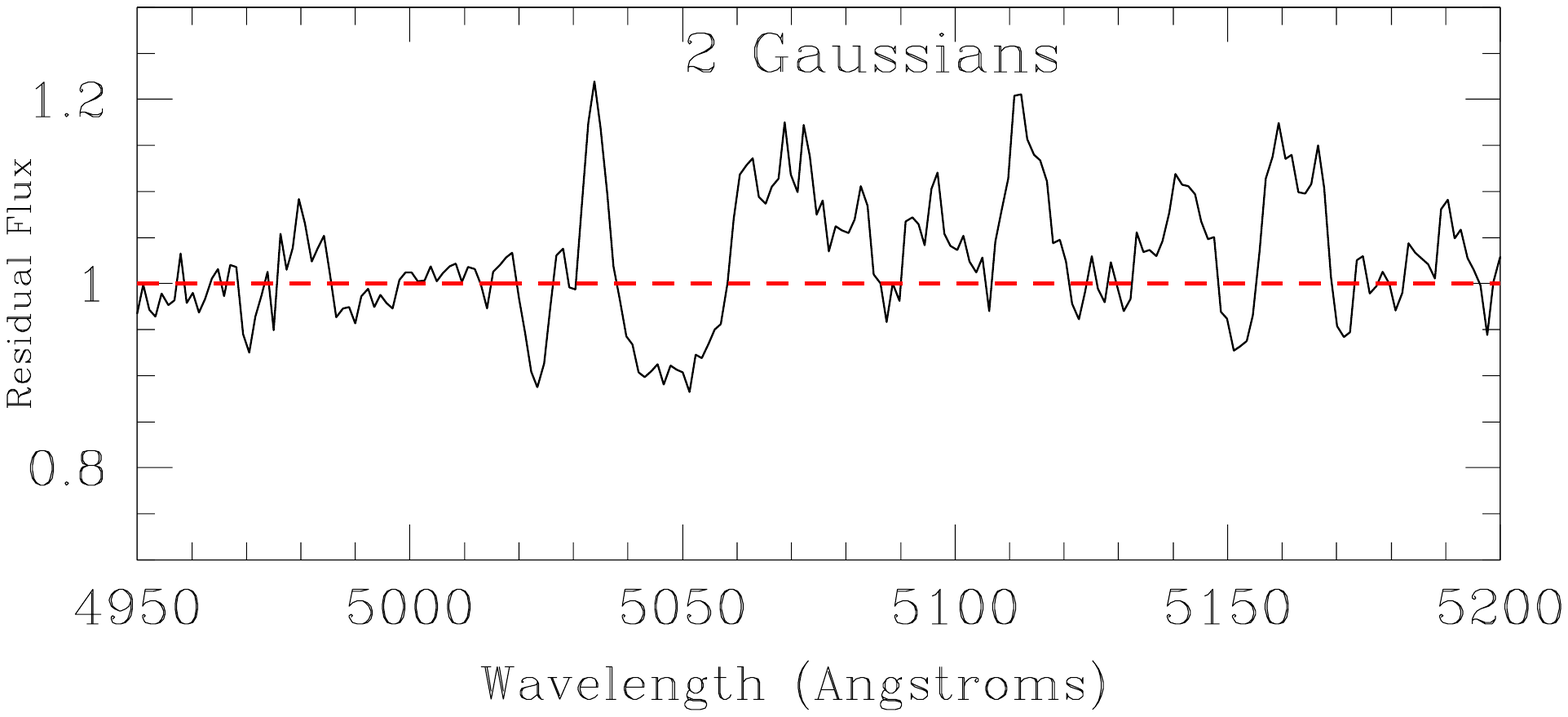}
\caption{{\bf Top.} Multi-component model fits to the H$\beta$ and [OIII]$\lambda\lambda$4959,5007 emission lines for III Zw 77. The model fits are drawn with the dashed red line. The flux is measured in units of erg s$^{-1}$ \AA $^{-1}$ cm$^{-2}$ and the wavelength is in units of \AA . {\bf Middle.} The residuals from the single Gaussian fits. {\bf Bottom.} The residuals from the double Gaussian fits.}
\label{fig:zwfit}
\end{figure} 

\begin{figure}
\centering
\includegraphics[scale=0.45, trim=1.5cm 3cm 1.5cm 3.5cm]{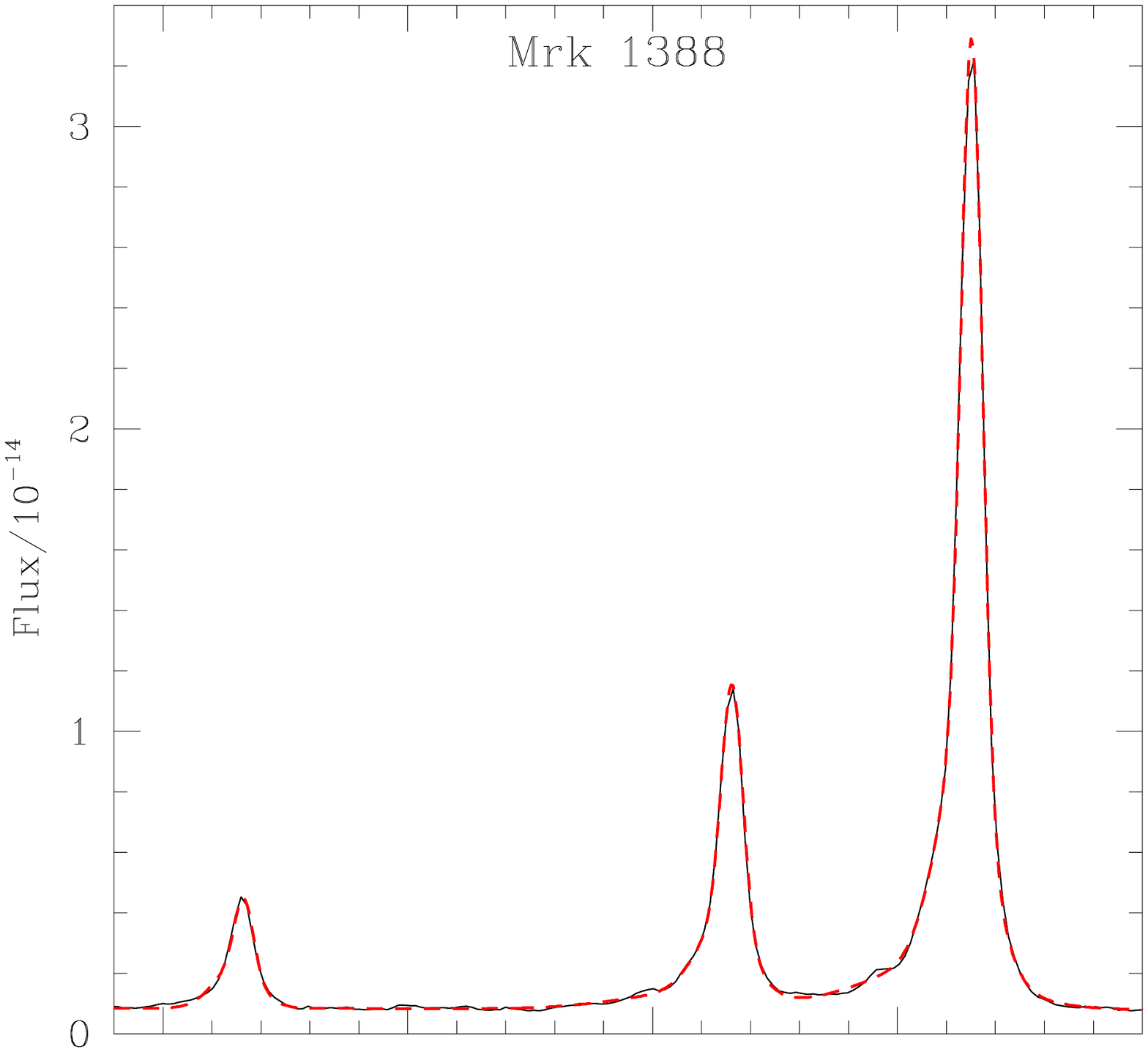}\\
\includegraphics[scale=0.45, trim=1.5cm 3cm 1.5cm 18.1cm]{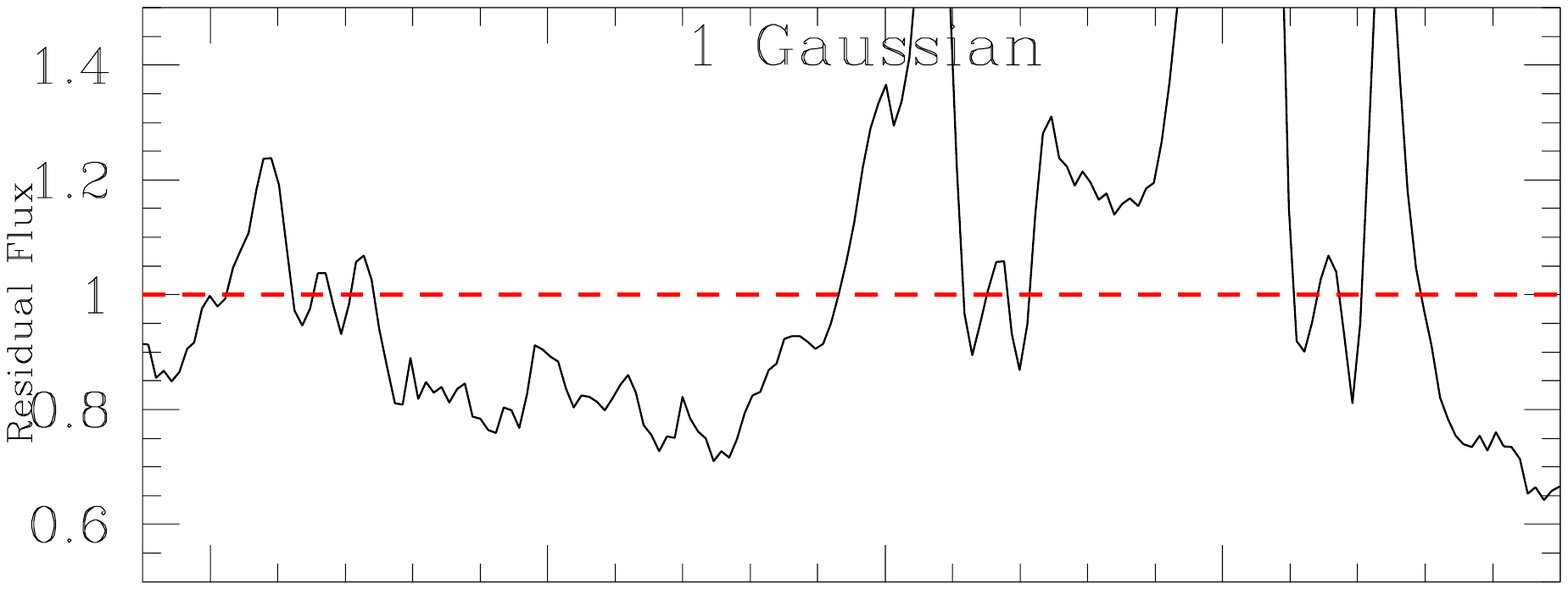}
\includegraphics[scale=0.45, trim=1.5cm 5cm 1.5cm 18.1cm]{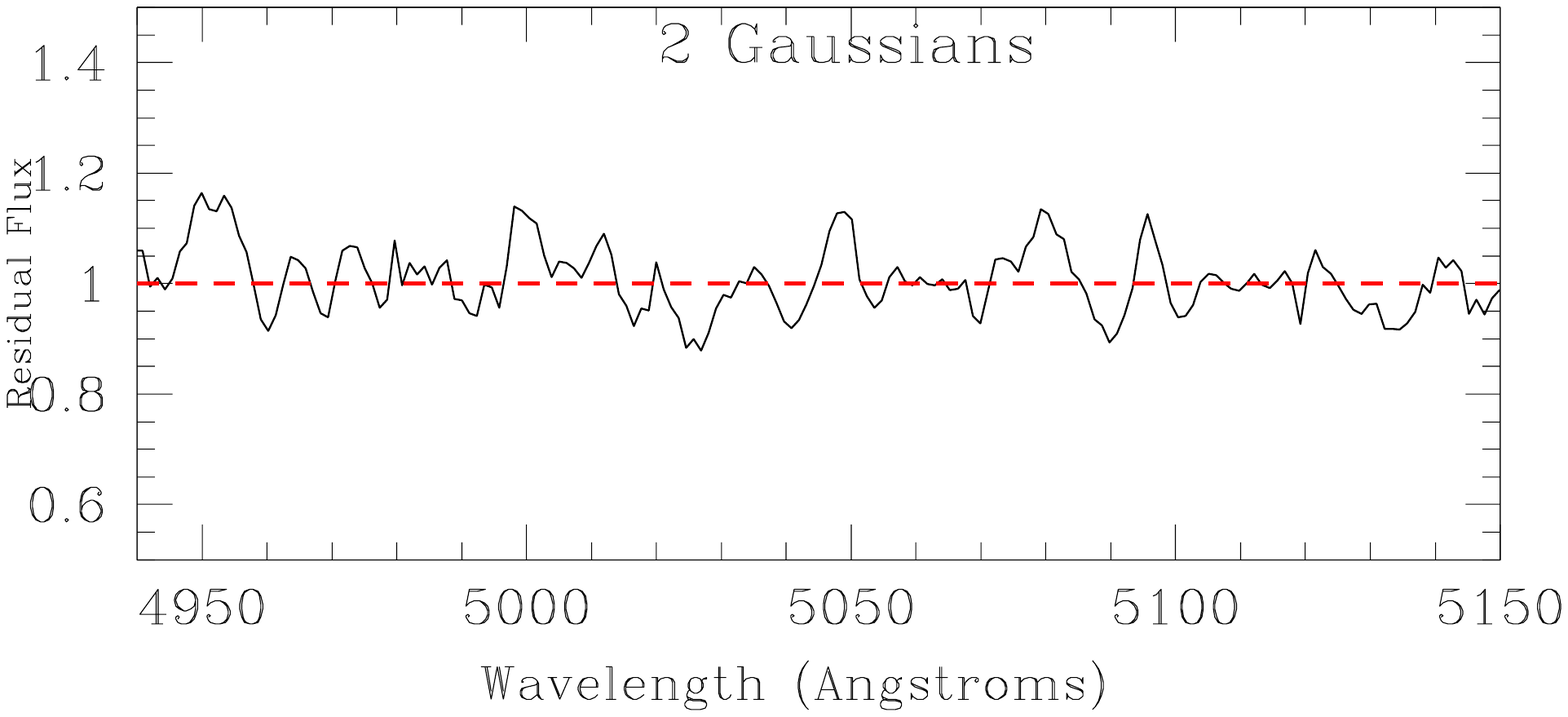}
\caption{{\bf Top.} Multi-component model fits to the H$\beta$ and [OIII]$\lambda\lambda$4959,5007 emission lines for Mrk 1388. The model fits are drawn with the dashed red line. The flux is measured in units of erg s$^{-1}$ \AA $^{-1}$ cm$^{-2}$ and the wavelength is in units of \AA . {\bf Middle.} The residuals from the single Gaussian fits. {\bf Bottom.} The residuals from the double Gaussian fits.}
\label{fig:mrkfit}
\end{figure}

\begin{figure}
\centering
\includegraphics[scale=0.45, trim=1.5cm 3cm 1.5cm 3.5cm]{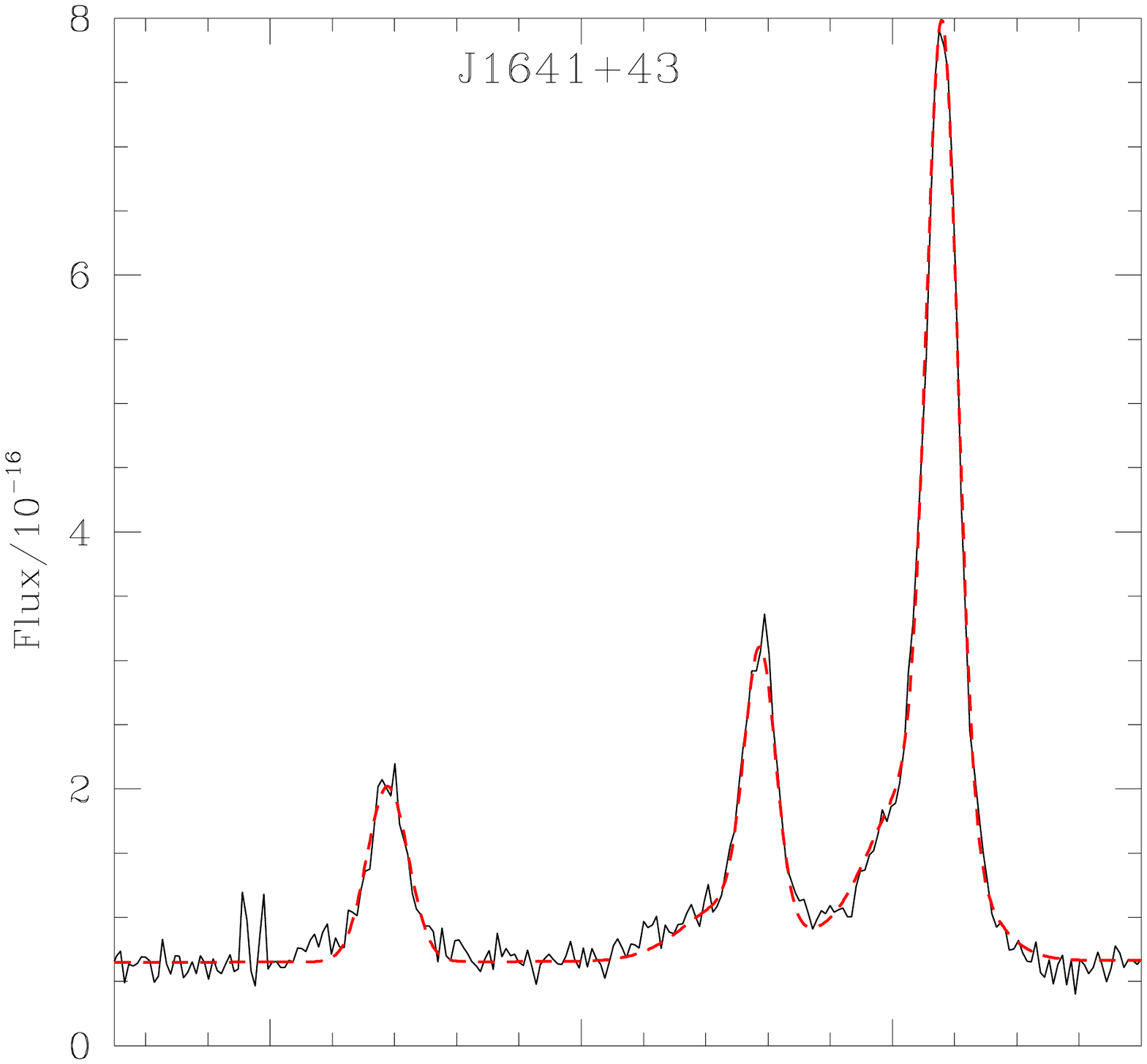}\\
\includegraphics[scale=0.45, trim=1.5cm 3cm 1.5cm 18.1cm]{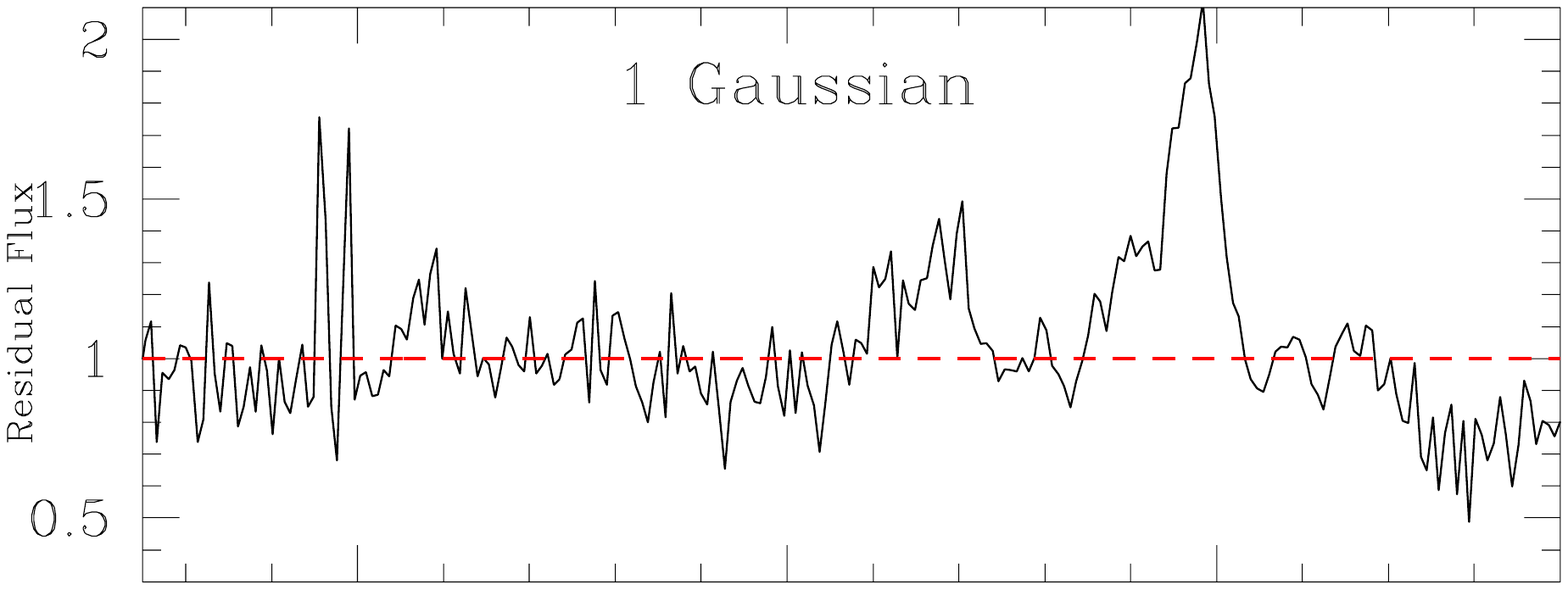}
\includegraphics[scale=0.45, trim=1.5cm 5cm 1.5cm 18.1cm]{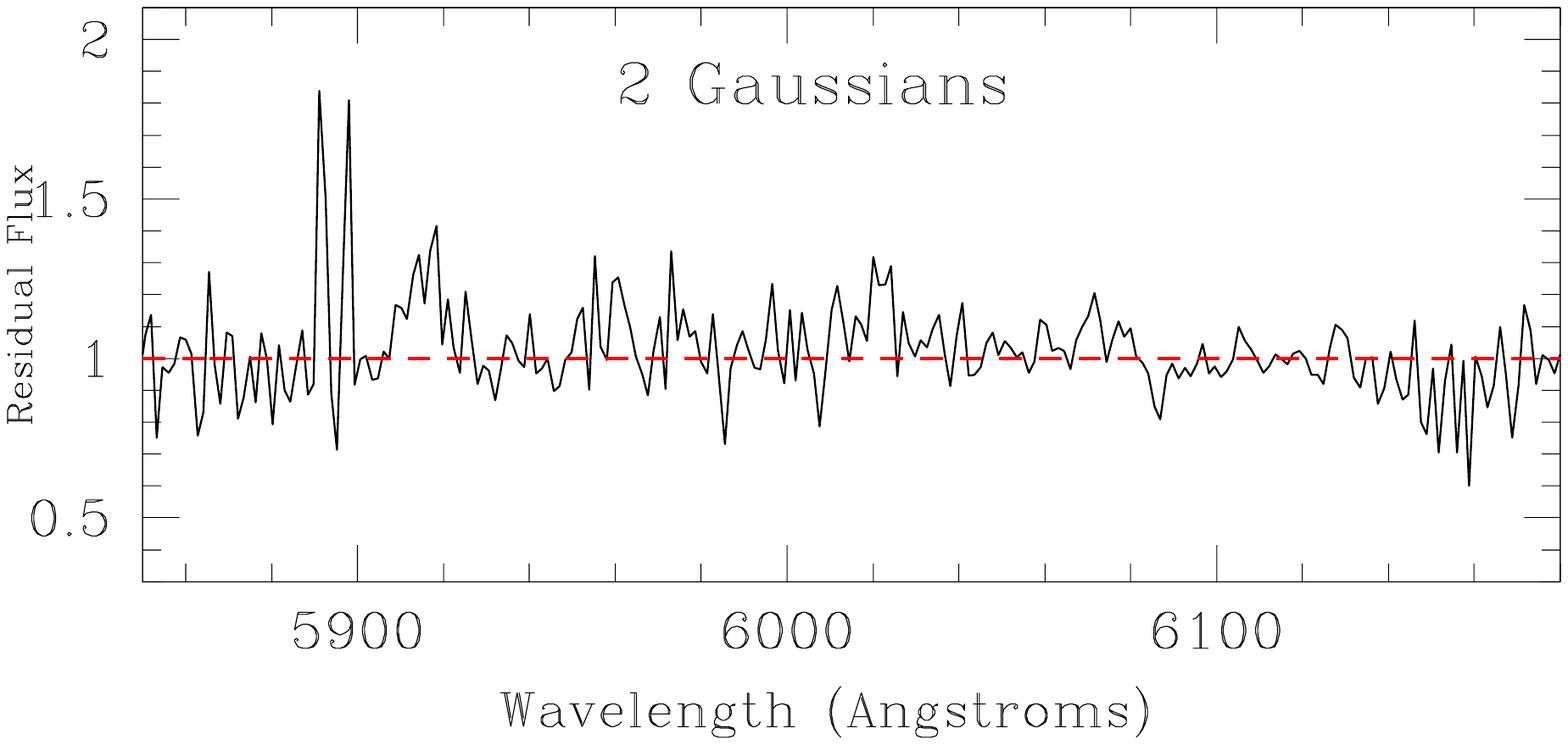}
\caption{{\bf Top.} Multi-component model fits to the H$\beta$ and [OIII]$\lambda\lambda$4959,5007 emission lines for 1641+43. The model fits are drawn with the dashed red line. The flux is measured in units of erg s$^{-1}$ \AA $^{-1}$ cm$^{-2}$ and the wavelength is in units of \AA . {\bf Middle.} The residuals from the single Gaussian fits. {\bf Bottom.} The residuals from the double Gaussian fits.}
\label{fig:16fit}
\end{figure}

Single-Gaussian profiles, did not always provide adequate fits to the wings or peaks of the strongest lines (EW $>$ 10 \AA)\footnote{These include H$\alpha$, H$\beta$, [OII]$\lambda$$\lambda$3726,3729 [OIII]$\lambda$$\lambda$$\lambda$4363,4959,5007, [NII]$\lambda$6583, and [FeVII]$\lambda$6087.} It is clear that these fits are inadequate for most of the CLiF AGN as the fitting residuals from the single Gaussian fits are large for the strong lines (Figures \ref{fig:zwfit}-\ref{fig:16fit}). 

To address this problem, double-Gaussian models were fitted to the strong spectral lines. To produce the models, both broad and narrow Gaussians were fitted to the H$\beta$ and [OIII] line profiles, where the narrow component widths for the H$\beta$ and [OIII] lines are linked, yet the line centers, intensities and broad H$\beta$ width are free parameters. All but J1241+44 required a double-Gaussian model to fit these emission lines. Figures \ref{fig:12fit}-\ref{fig:16fit} show the residuals for the one and two-Gaussian fits to the [OIII]$\lambda\lambda$4959,5007 emission lines. The variations shown in the residual plots of the fitting models used to obtain the emission line fluxes are dominated by noise in the continua of the spectra. Intrinsic velocity widths for the Gaussian components were obtained by correcting each of the components' FWHM in quadrature using the SDSS instrumental width of $\sim$3\AA . The intrinsic kinematics of the narrow (N) and broad (B) components are given in Table \ref{tab:gmods}. 
 
The parameters from the double-Gaussian fits were then used to fit all the strong (EW $>$ 3 \AA ) emission lines throughout the CLiF AGN spectra (FWHM, velocity shift and relative amplitudes). 

The fits for the weakest lines (EW $\leq$ 3 \AA ) were overestimated with the double-Gaussian model because the broader component required by the double-Gaussian model tended to fit the local continuum. In these cases a single-Gaussian fit was sufficient. We indicate the number of Gaussians required to fit each individual emission line in Tables \ref{tab:mrk1388}-\ref{tab:1641}. 

We also take in to account constraints that are fixed by atomic physics. The [OIII]$\lambda$$\lambda$4959,5007 doublet was modeled using the constraint that they have the same FWHM, a 1:3 intensity ratio and the line center of $\lambda$4959 fixed relative to the center of $\lambda$5007. The same approach was used for the [OI]$\lambda$$\lambda$6300,6364 (3:1 ratio) doublet blend (blended with [SIII]$\lambda$6312 and [FeX]$\lambda$6375) and the [NII]$\lambda$$\lambda$6548,6584 doublet (1:3 ratio), which is also often blended with the H$\alpha$ emission line.

Throughout this paper, we use the summed fluxes of the two Gaussians. We do not use the data for weak emission lines (EW $\leq$ 3 \AA ) in any of the following investigation. The measured fluxes for the emission lines are presented for the SDSS CLiF AGN in Tables \ref{tab:mrk1388}-\ref{tab:1641} in the Appendix.	

\begin{center}
\begin{table}
\centering
\caption{The double-Gaussian model parameters derived from the [OIII]$\lambda$$\lambda$4959,5007 and H$\beta$ emission lines of the CLiF AGN spectra. The `FWHM$_N$' and `FWHM$_B$' columns present the rest-frame widths of the narrow and broad components of the model, which have been corrected for instrumental broadening. The `$\Delta$$v$' column gives the rest-frame velocity separation of the broad relative to the narrow kinematic components. The `Ratio' column give the amplitude ratio for the intensities of the broad to narrow components.}
\begin{tabular}{lcccc}
\hline
Name	&	FWHM$_N$			&	FWHM$_B$	&	$\Delta$$v$ & Ratio.\\
	&	 (km s$^{-1}$)		&	 (km s$^{-1}$)	& (km s$^{-1}$)	& 	 \\
\hline								
III Zw 77	&	216$\pm$11	&	512$\pm$32	&	-127$\pm$11 & 0.29	\\	
Mrk 1388	&	250$\pm$21	&	723$\pm$57	&	-106$\pm$10 & 0.79	\\	
J1241+44	&	145$\pm$9	&	-	&	-106$\pm$10 & -	\\	
J1641+43	&	563$\pm$47	&	1770$\pm$148	&	-412$\pm$38 & 0.30	\\	
\hline
\end{tabular}								
\label{tab:gmods}
\end{table}
\end{center}	

\subsection{Line identification}\label{sect:lineids} 

We regarded an identification to be secure for an emission line if its line center was within 1 $\sigma$ of the wavelength predicted for that particular emission line based on the redshift (as measured using the [OIII]$\lambda$5007 emission line). We used \citet{osterbrock81} to identify the majority of the emission lines in the CLiF AGN spectra. For the rest of the peculiar emission lines we used \citet{meinel75} and \citet{kaler76}. 

\subsection{Emission line properties}

The emission lines are of a wide variety including 9-26 FHILs and 1-30 unidentified lines (Table \ref{tab:elprop}). A full list of the measured wavelengths, velocity widths (FWHM) and velocity shifts ($\Delta$v) for the emission lines are given in Tables \ref{tab:mrk1388}-\ref{tab:1641} in the Appendix. Below we discuss each group of emission lines.

\subsubsection{Unidentified lines}\label{sect:?}

CLiF AGN host unknown emission lines in their spectra. Such emission lines are not reported in typical AGN. Four CLiF AGN have 1-3, but III Zw 77, Mrk 1388 and Q1131+16 have 6, 17 and 30 unknown lines respectively (Table \ref{tab:elprop}). Only Q1131+16 and ESO 138 G1 share the same unknown emission lines. Q1131+16 has 2 and 3 in common unknown emission lines with J1241+44 and Mrk 1388 respectively.  Mrk 1388 has 1 and 3 in common unknown emission lines with J1241+44 and III Zw 77 respectively. Finally, III Zw 77 and J1641+43 share 1 common unknown emission line.

\subsubsection{FHILs}

Each CLiF AGN has {\it at least} 10 FHILs. These FHILs are dominated by Fe ions, particularly [FeVII] (Table \ref{tab:elprop}; Tables \ref{tab:mrk1388}-\ref{tab:1641}). All CLiF AGN show emission from [FeVII]$\lambda\lambda\lambda\lambda$3759,5159,5720,6087, [FeX]$\lambda$6375 and [CaV]$\lambda$5307. The [FeVII]$\lambda$6087 line is particularly prominent, with flux ratios [FeVII]/H$\beta$ of 0.24-1.11 (EW of 5-27 \AA ). However, in the spectra that cover [NeV]$\lambda$3426, this line has flux ratios [NeV]/H$\beta$ of 0.57-1.73. 

In addition, J1241+44 has [ArV]$\lambda$6435 and [FeXI]$\lambda$6985 that are not detected even in Q1131+16 and ESO 138 G1, the two CLiF AGN with the most FHILs. 

\subsubsection{Balmer emission lines}\label{sect:bls}

Strong (EW $>$ 3 \AA ) higher level HI Balmer emission lines (H$\epsilon$ - H$_{16}$) are detected in 5/7 of the CLiF AGN (Table \ref{tab:elprop}). Hydrogen emission from transitions with energy levels greater than H$\epsilon$ is not typically seen in AGN, implying some unusual physics in CLiF AGN, far from Case B. We investigate the HI emission of CLiF AGN in more detail in $\S$\ref{sect:halpha}. 

\subsubsection{BLR emission lines}

In this work line width was not a selection criterion yet broad emission lines are not clearly detected in the spectra of the CLiF AGN suggesting a possible CLiF-type 2 AGN link. The exception is III Zw 77 in which the BLR Balmer emission lines are peculiarly blueshifted by -675$\pm$25 km s$^{-1}$ with respect to the NLR Balmer emission lines. This is an expected signature of a black hole recoil event, where after the merger of two SMBHs, anisotropic emission of gravitational waves gives a `kick' to the resulting SMBH after the merger (e.g. see \citealt{favata04}; \citealt{civano12}). While this phenomenon is interesting in its own right, III Zw 77 is the only recoil candidate to date that is a CLiF AGN. This subject is beyond the scope of this work.   

\citet{rose11} reported tentative evidence for broad emission at the base of the H$\alpha$+[NII] blend for Q1131+16. In this work we assume that there is tentative evidence for a BLR if the Balmer recombination lines show broad wings that cannot be accounted for by using [OIII]$\lambda$5007 as an NLR template. Only Mrk 1388 shows such evidence. The velocity width (FWHM) for this emission component is 1070$\pm$160 km s$^{-1}$, however it is weak (represents 8\% of the flux of the blend) compared with the narrow H$\alpha$+[NII] blend and is not present for higher order Balmer emission.   

\section{Results}

\subsection{A unique CLiF region on the BPT diagram} \label{sect:halpha}

\begin{figure}
\centering
\includegraphics[scale=0.38, trim=0.1cm 4.6cm 0.1cm 3.6cm]{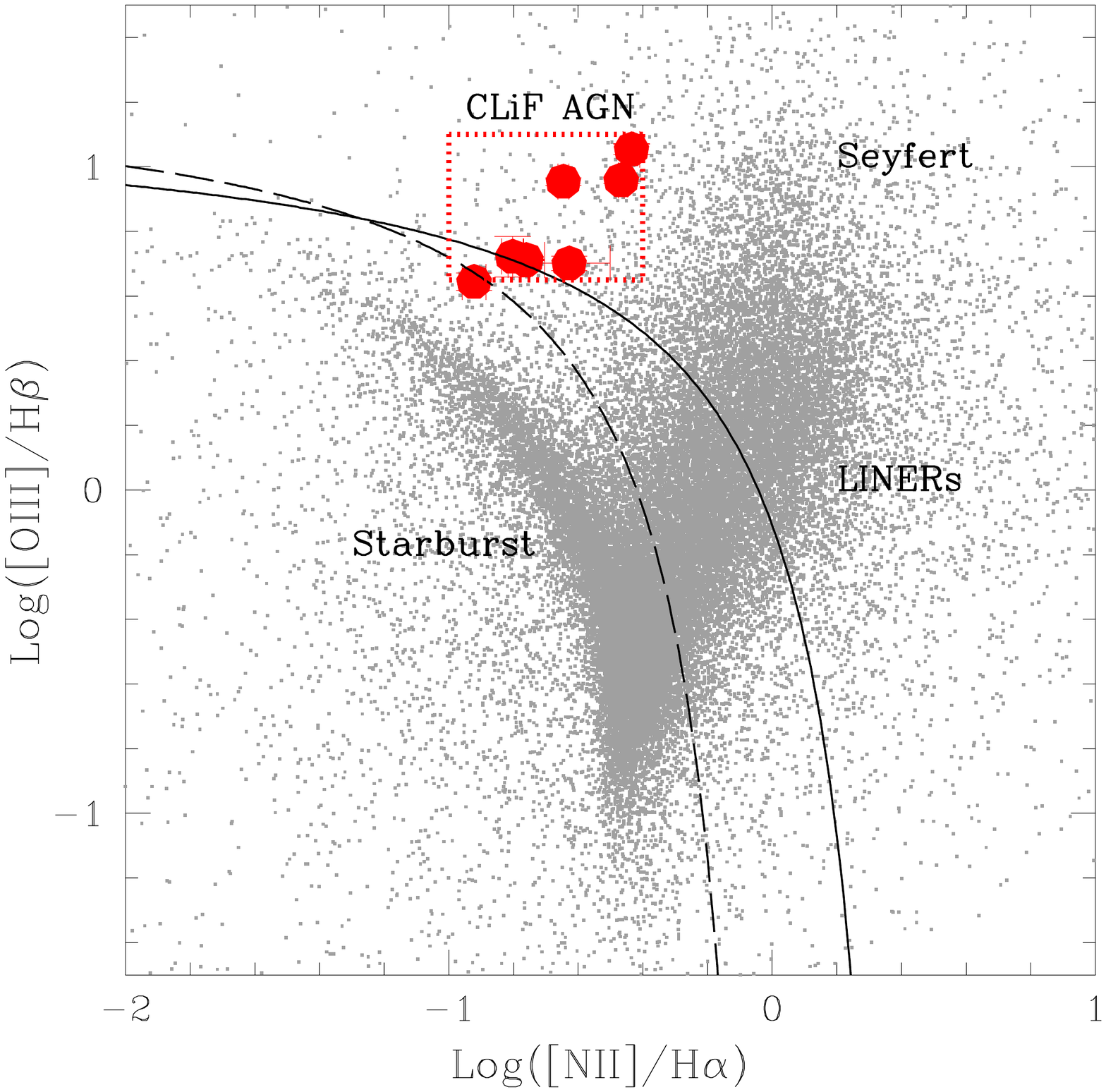}
\caption{Diagnostic plot of Log$_{10}$([OIII]/H$\beta$) vs Log$_{10}$([NII]/H$\alpha$). AGN are defined to lie above the solid line, HII-region like galaxies below the dashed line, and composite galaxies between these boundaries. The red circles represent the CLiF AGN. The small points represent objects detected by the SDSS and are taken from the SDSS-DR 7 release. The region outlined by the red dotted line is a `CLiF AGN Region'.}
\label{fig:bpt}
\end{figure}

The Baldwin, Phillips \& Terlevich (BPT) diagnostic diagrams use emission line ratios to determine whether the emission region of an object is photoionized by an AGN, or by a starburst (\citealt{bpt}; \citealt{veilleux87}; \citealt{kewley06}). In Figure \ref{fig:bpt} we plot log$_{10}$([OIII]/H$\beta$) vs log$_{10}$([NII]/H$\alpha$) for the CLiF AGN and SDSS emission line objects which include both AGN and starburst galaxies. The measurements for the SDSS objects were a subset (6000 objects) of the MPA-JHU SDSS DR7 release of spectrum measurements\footnote{http://www.mpa-garching.mpg.de/SDSS/DR7/}. This subset is composed of measurements from objects with spectra which were reported to have the highest S/N, as well as having [OIII]/H$\beta$ and [NII]/H$\alpha$ ratios which were consistent with the ranges shown in Figure \ref{fig:bpt}.      

The SDSS AGN divide cleanly in the right-hand branch from star-forming galaxies in the left-hand branch. The solid line represents the extreme upper starburst line determined using the upper limit of theoretical pure stellar photoionization models \citep{kewley01}. The dashed line indicates the empirically derived boundary between HII region like galaxies and AGN \citep{kauffmann03}. 

All the CLiF AGN fall in the cleft between AGN and the starburst galaxy branches. They are clearly removed from the main body of the SDSS AGN (grey dots) along the x-axis by log$_{10}$ [NII]/H$\alpha$ $\sim$ 0.2-1. Either the [NII] emission is weaker, or H$\alpha$ is enhanced in CLiF AGN compared to typical AGN. 

Balmer line recombination ratios are the most common tool to correct AGN spectra for intrinsic dust extinction assuming conditions close to Case B. The H$\alpha$/H$\beta$ ratio is the favored Balmer decrement for this task because these emission lines are the strongest and have the widest wavelength separation. However, the large number of Balmer lines from high energy transitions shows that Case B does not apply to CLiF AGN (\citealt{rose11}).

\begin{figure}
\centering
\includegraphics[scale=0.38, trim=0.1cm 4.6cm 0.1cm 3.6cm]{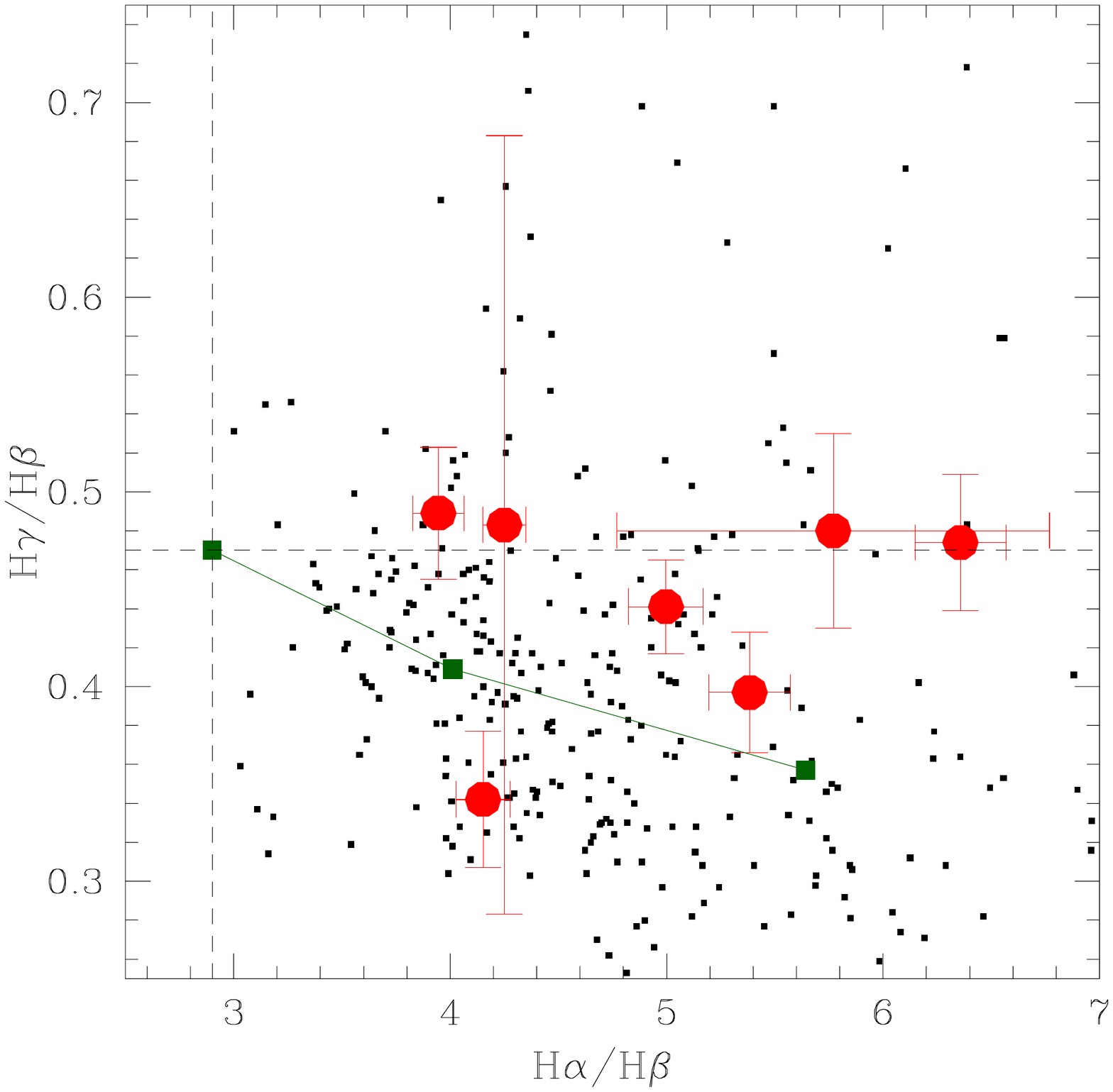}
\caption{Plot showing H$\gamma$/H$\beta$ versus H$\alpha$/H$\beta$. The CLiF AGN are indicated by red circles and the type 2 AGN of \citet{reyes08} are indicated by black crosses. The theoretical Case B values for the decrements are indicated by the dashed lines. The dark green squares indicate the Balmer decrement values for reddening of A$_V$ = 1 and 2 (labelled).}
\label{fig:baldec}
\end{figure}

In Figure \ref{fig:baldec} we plot H$\gamma$/H$\beta$ versus H$\alpha$/H$\beta$ for the CLiF AGN (red circles) and the type 2 AGN of Reyes et al. (2008; black crosses). We limit the \citet{reyes08} sample to $z$ $<$ 0.42 so that H$\alpha$ lies in the SDSS spectrum, and use Balmer lines which have EW $>$ 3 \AA, leaving 325 out of the original 887 type 2 AGN. We use the SDSS pipeline measurements for the Balmer decrements of the type 2 AGN of \citet{reyes08}.

A reddening line for \citet{calzetti94} dust is shown with A$_V$= 0, 1 and 2 marked. Most of the type 2 AGN of \citet{reyes08} roughly follow the reddening trend indicated by the reddening line. Within the 1$\sigma$ uncertainties 83\% have values consistent with the reddening line.    

It is clear that while the H$\alpha$/H$\beta$ ratios for the CLiF AGN all exceed the Case B values of 2.9 by up to a factor $\sim$2, the majority of the H$\gamma$/H$\beta$ agree with the Case B value of 0.47 within the uncertainties (Table \ref{tab:bals}). These Case B values apply to the low density (n$_{e}$ = 10$^{2}$-10$^{4}$ cm$^{-3}$) limit, with a temperature of 10000K \citep{osterbrock06}.  

\begin{table}
\centering
\caption{The H$\alpha$/H$\beta$ and H$\gamma$/H$\beta$ Balmer decrements for the CLiF AGN.}
\begin{tabular}{lcc}
\hline
Name	&	H$\alpha$/H$\beta$	&	H$\gamma$/H$\beta$	\\
\hline
Q1131+16	&	5.00$\pm$0.17	&	0.44$\pm$0.02	\\
ESO 138 G1	&	4.25$\pm$0.10	&	0.48$\pm$0.20	\\
Mrk 1388	&	3.95$\pm$0.12	&	0.49$\pm$0.03	\\
III Zw 77	&	4.15$\pm$0.13	&	0.34$\pm$0.04	\\
J1241+44	&	6.36$\pm$0.21	&	0.47$\pm$0.04	\\
Tololo 0109-383	&	5.77$\pm$1.00	&	0.48$\pm$0.05	\\
J1641+43	&	5.38$\pm$0.19	&	0.40$\pm$0.03	\\
Case B		&	2.9		&	0.47		\\
\hline
\end{tabular}								
\label{tab:bals}
\end{table}

Instead, 5/7 of the CLiF AGN are inconsistent with the reddening line. These objects all have H$\alpha$/H$\beta$ ratios exceeding Case B, yet their H$\gamma$/H$\beta$ ratios are consistent with the Case B value. Only III Zw 77 and J1641+44 lie close to the reddening line and of these only III Zw 77 $>$ 2$\sigma$ from case B H$\gamma$/H$\beta$. It is likely that reddening does not dominate the CLiF AGN Balmer decrements.

A visual inspection of the spectra of the remaining 17\% of the type 2 AGN of \citet{reyes08} that do not follow the reddening line show that they are not CLiF AGN. Their positions on Figure \ref{fig:bpt} are possibly due to uncertainties with the SDSS pipeline fits e.g. degeneracies with the H$\gamma$+[OIII]$\lambda$4363 blend.    
 
\subsection{Kinematics} \label{sect:kine}

\begin{figure*}
\centering
\includegraphics[scale=0.38, trim=0.1cm 4.6cm 0.1cm 3.6cm]{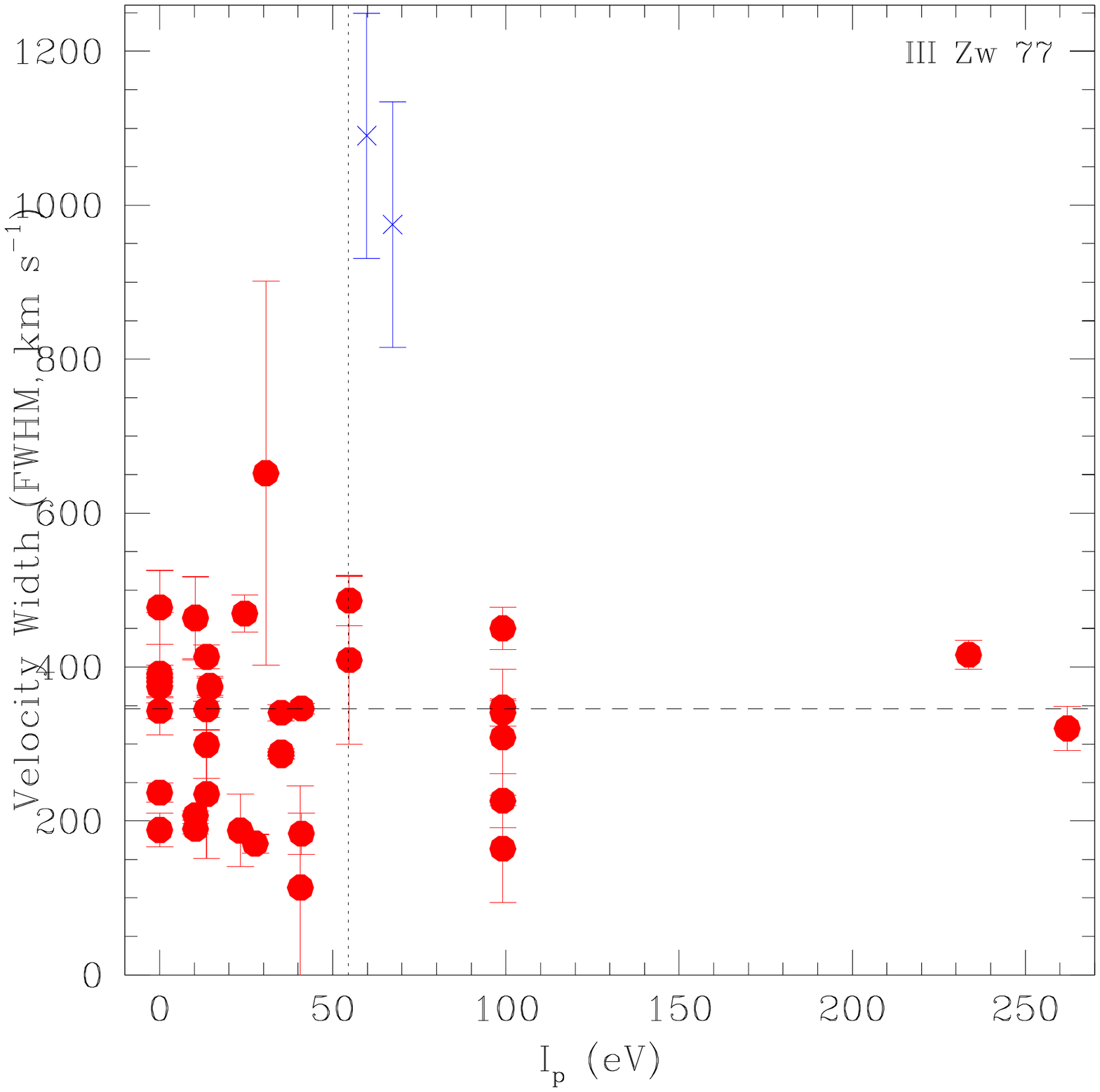}
\includegraphics[scale=0.38, trim=0.1cm 4.6cm 0.1cm 3.6cm]{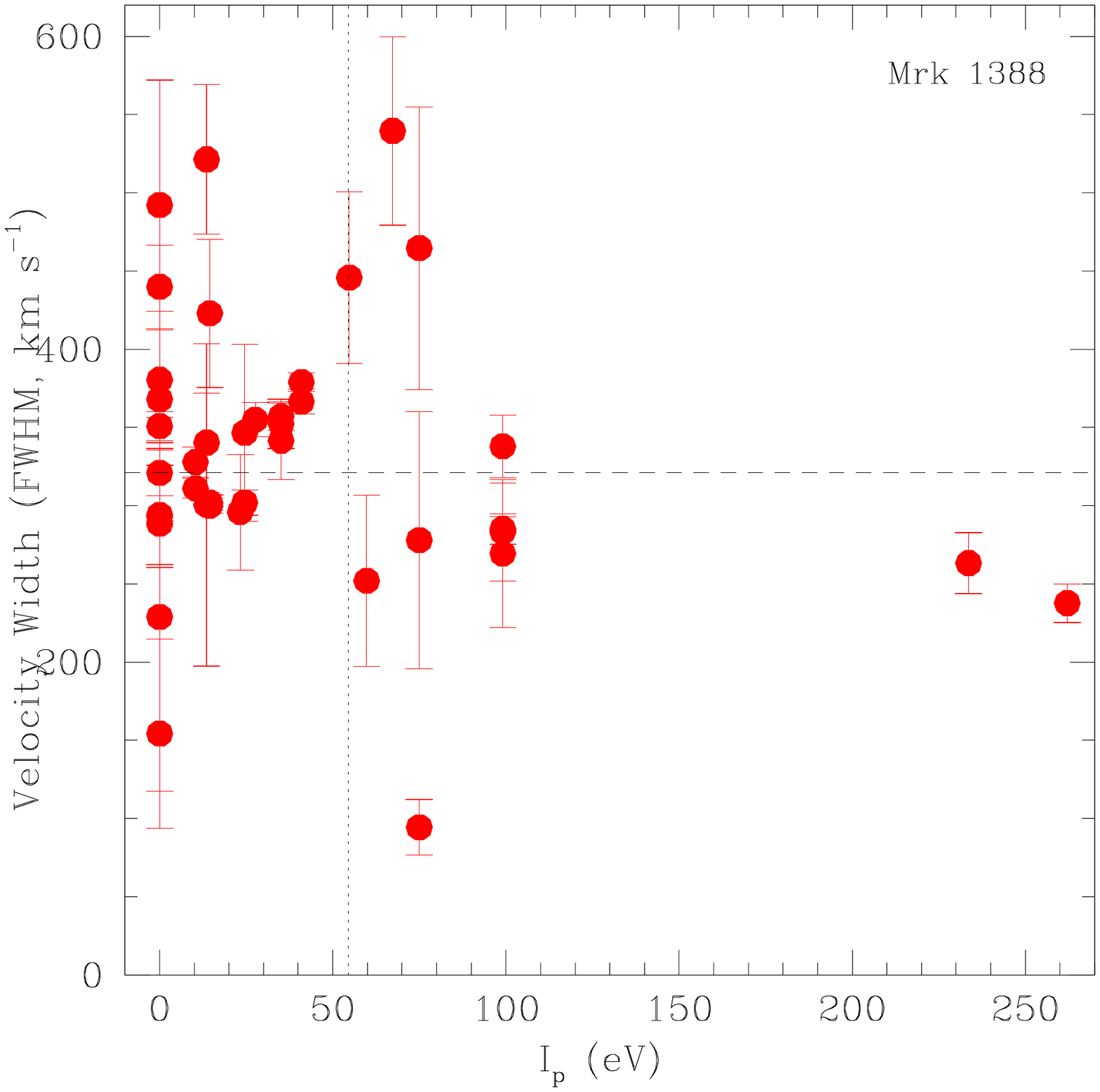}\\
\includegraphics[scale=0.38, trim=0.1cm 4.6cm 0.1cm 3.6cm]{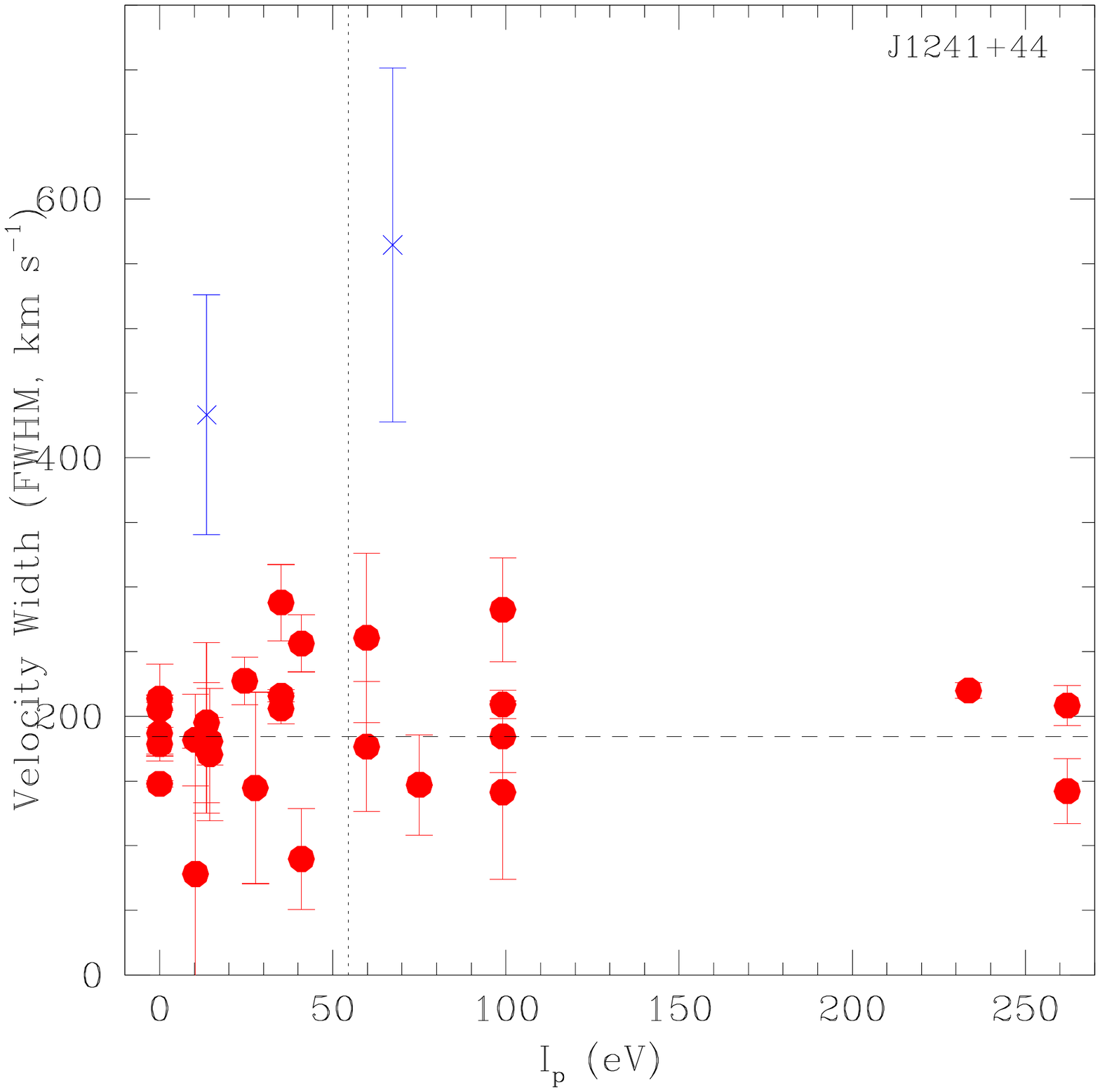}
\includegraphics[scale=0.38, trim=0.1cm 4.6cm 0.1cm 3.6cm]{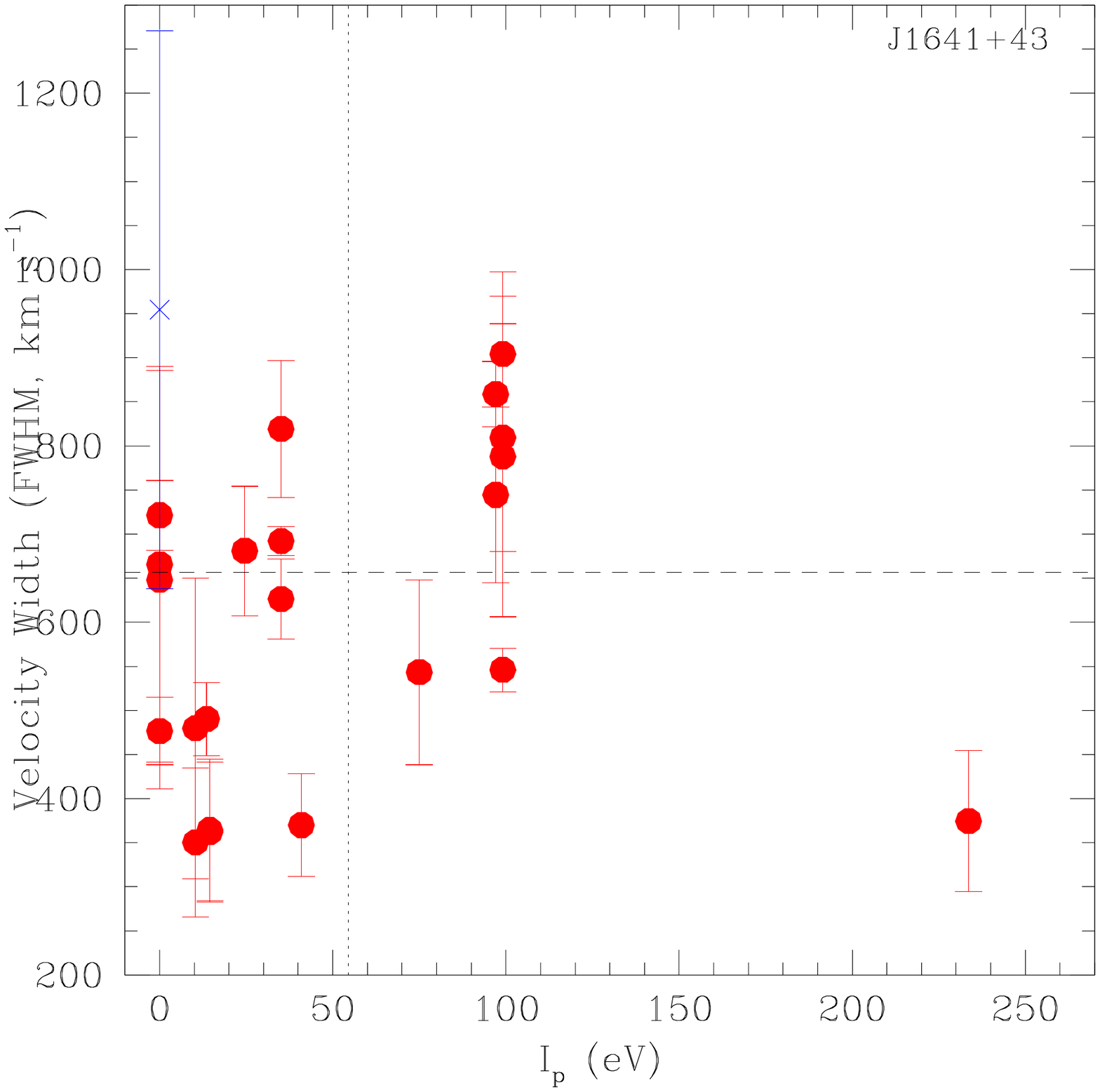}
\caption{Velocity widths FWHM (km s$^{-1}$) for the single Gaussian fits to the emission lines versus I$_P$ (eV) for the CLiF AGN (red circles). The blue crosses are for the emission lines which have a much higher FWHM than the median of all the emission lines of the CLiF AGN. The median velocity widths are indicated by the dashed lines. The dotted line indicates the boundary between the FLILs and FHILs (I$_P$ = 54.4 ev; the HeII edge).}
\label{fig:fwhm}
\end{figure*}

\begin{figure*}
\centering
\includegraphics[scale=0.38, trim=0.1cm 4.6cm 0.1cm 3.6cm]{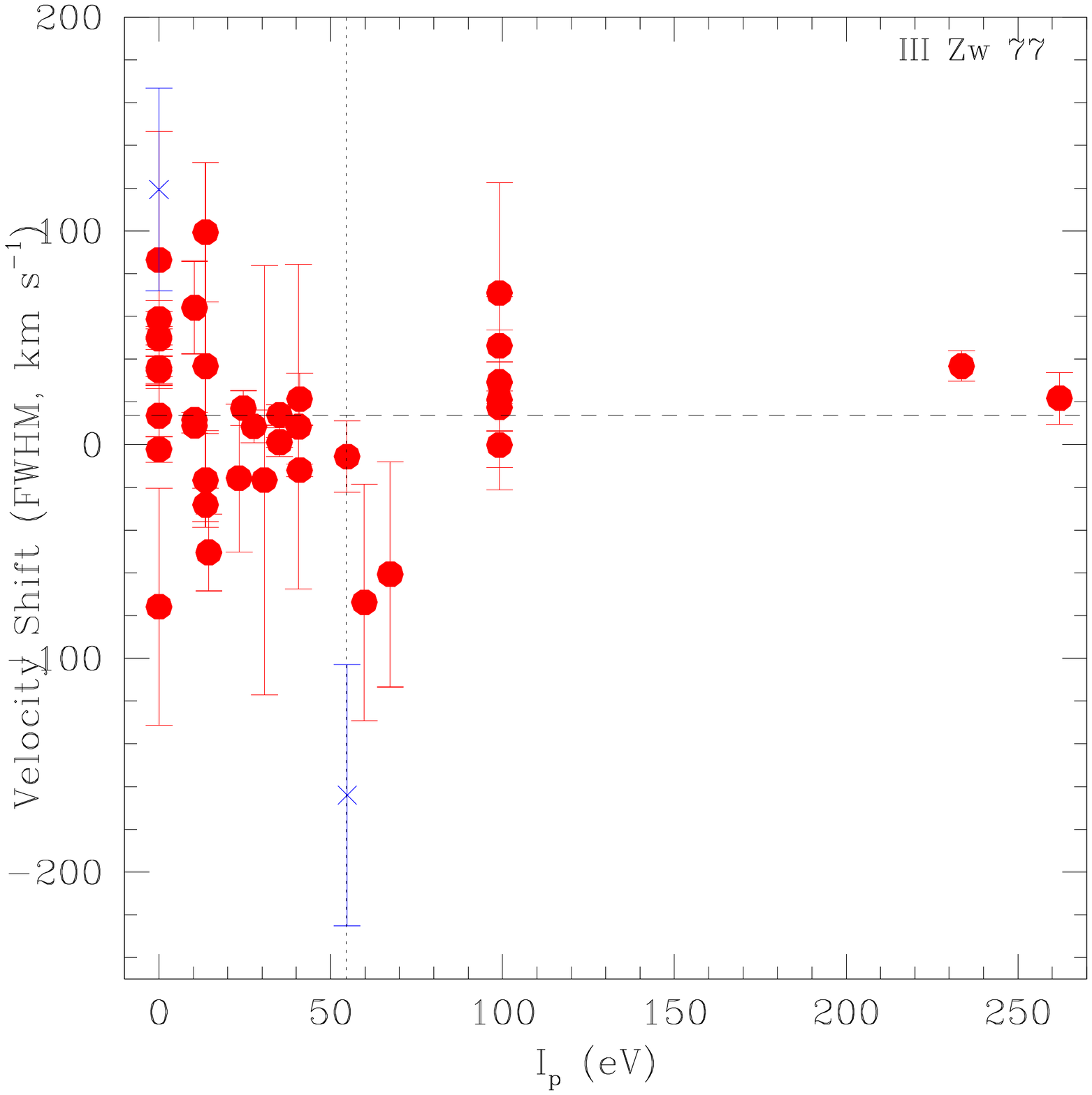}
\includegraphics[scale=0.38, trim=0.1cm 4.6cm 0.1cm 3.6cm]{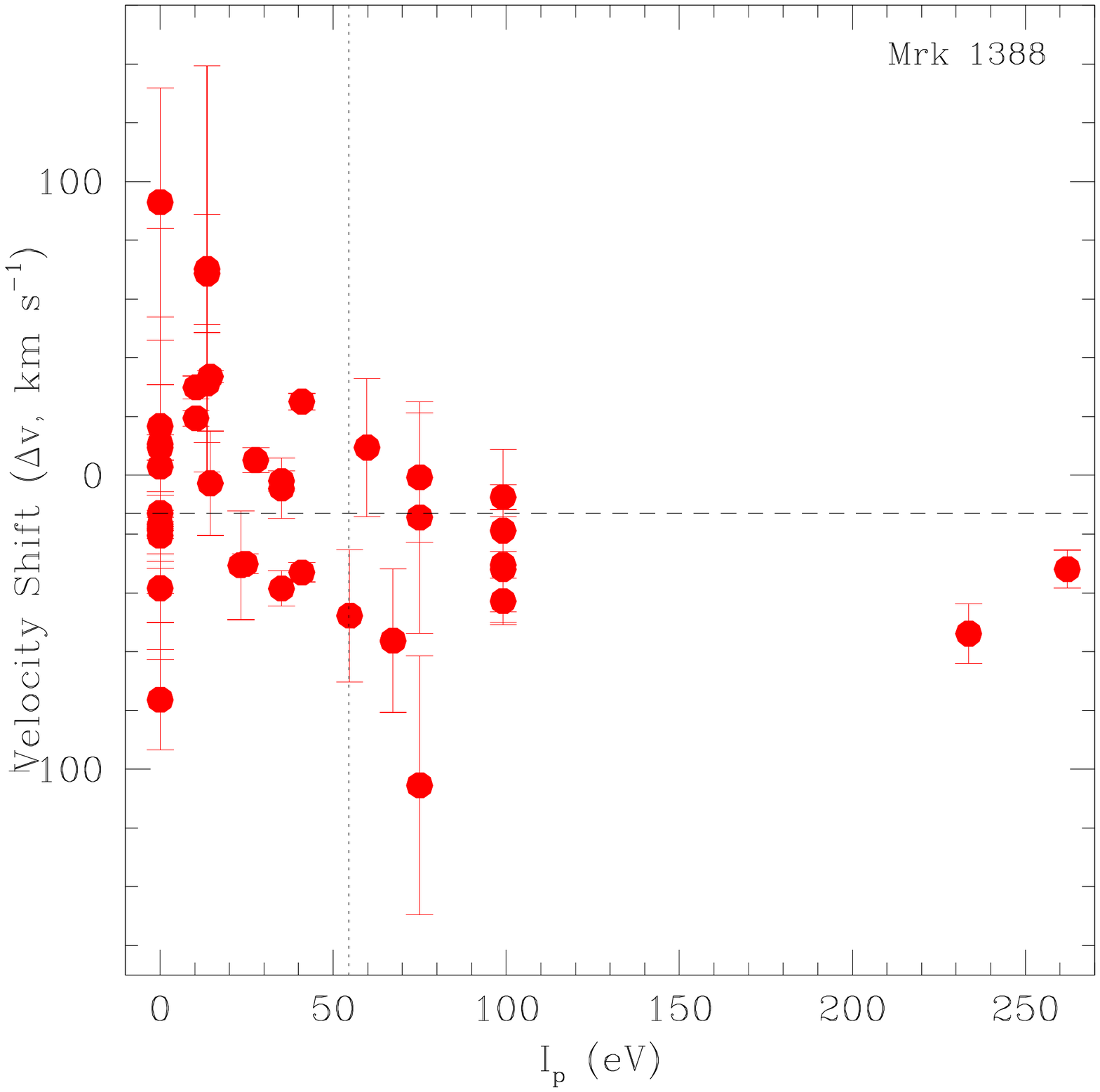}\\
\includegraphics[scale=0.38, trim=0.1cm 4.6cm 0.1cm 3.6cm]{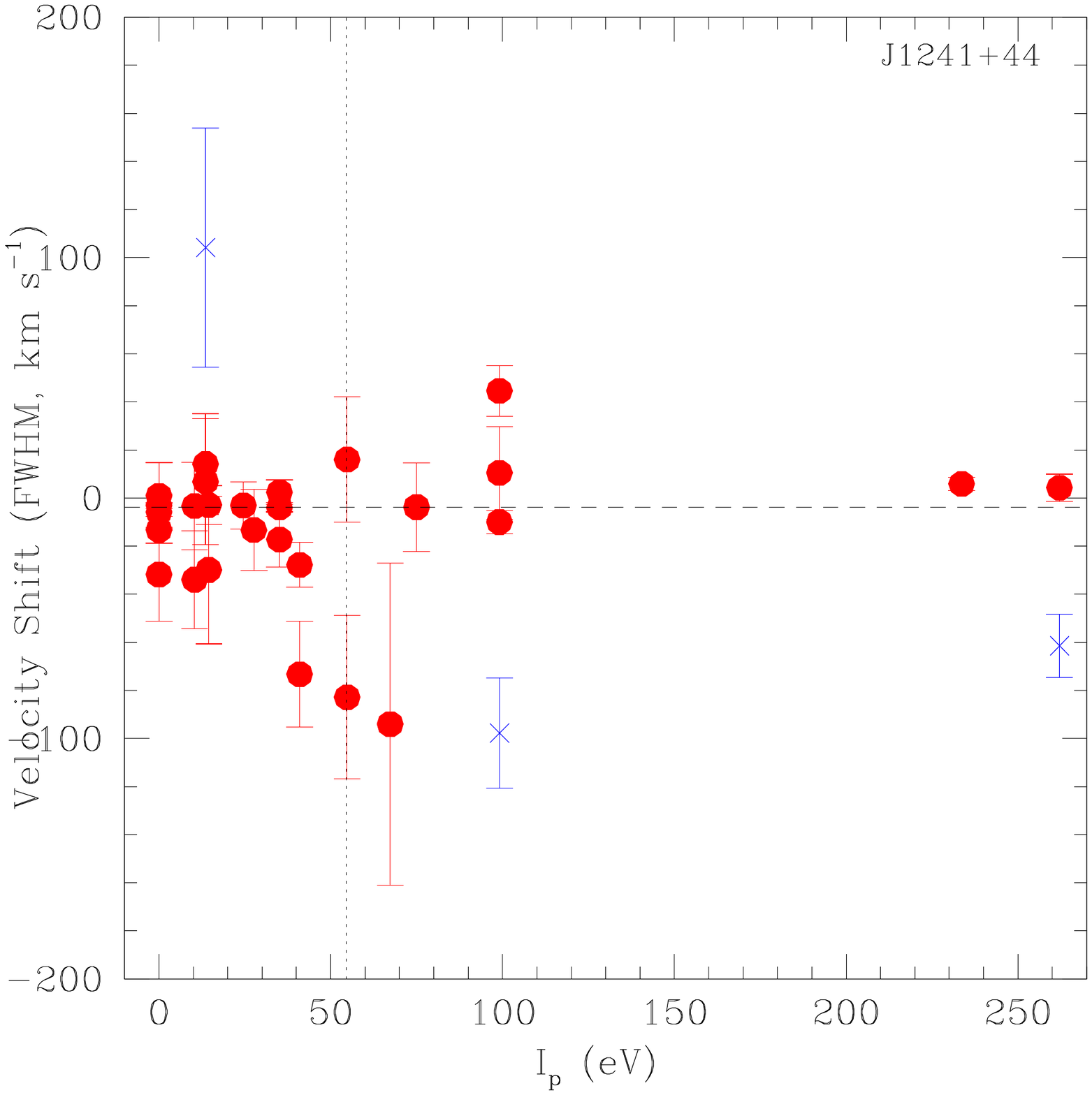}
\includegraphics[scale=0.38, trim=0.1cm 4.6cm 0.1cm 3.6cm]{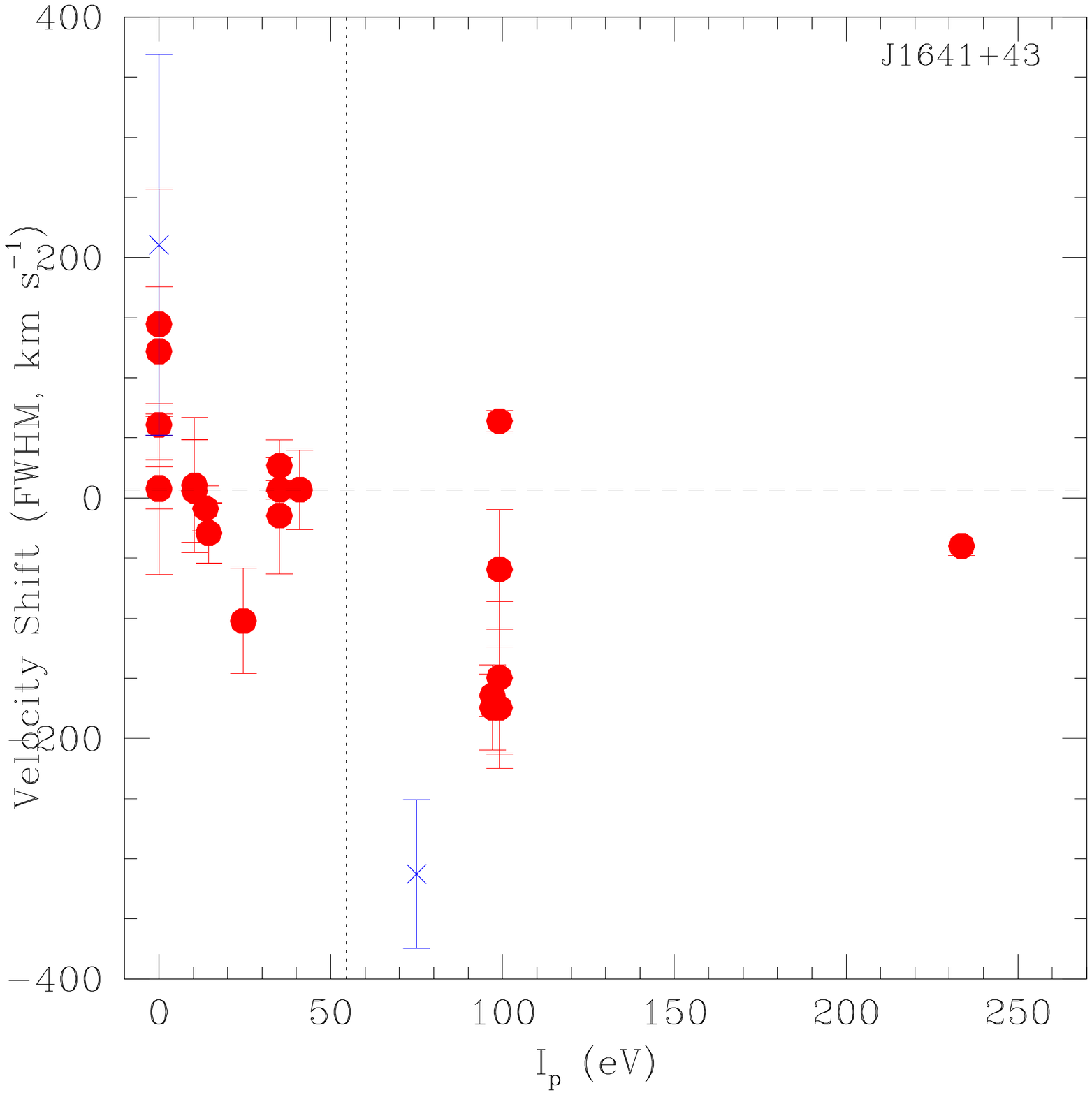}
\caption{Velocity shift (km s$^{-1}$) with respect to the redshift of the CLiF AGN for the single Gaussian fits to the emission lines versus I$_P$ (eV) for the CLiF AGN (red circles). The blue crosses are for the emission lines which have a much much larger $\Delta$v than the median of all of the emission lines of the CLiF AGN. The median velocity shifts are indicated by the dashed lines. The dotted line indicates the boundary between the FLILs and FHILs (I$_P$ = 54.4 ev; the HeII edge).}
\label{fig:shift}
\end{figure*}

We investigated the kinematics of the CLiF AGN with SDSS spectroscopy using the velocity shifts ($\Delta$v, and FWHM from the single Gaussian fits to the emission lines of the CLiF AGN because there are not enough lines for a broad and narrow component separation ($\S$\ref{sect:gfits}).

\subsubsection{FWHM vs I$_P$}

In Figure \ref{fig:fwhm} we plot the velocity width (km s$^{-1}$) versus I$_P$ (eV) for the emission lines detected in each of the CLiF AGN.  There is significant scatter in the measured widths of a few hundred km s$^{-1}$. However there is no trend for FWHM to increase with I$_P$. 

There are four lines (blue crosses) which have much higher FWHM than the median of all the emission lines of the CLiF AGN (horizontal dashed lines). These emission lines are:

\begin{itemize}

\item $[$ArV]$\lambda$7004 (I$_P$ = 59.81 eV) in III Zw 77 which may be intrinsically broader, but could always be blended with an unknown emission line.

\item $[$CaV]$\lambda$5309 (I$_P$ = 67.30 eV) in III Zw 77 and J1241+44 possibly blended with [FeXIV]$\lambda$5303. 

\item H$\gamma$ blended with [OII]$\lambda$4415 in J1641+43. The low S/N of the blending companions leads to Gaussian fitting degeneracies which are not resolved. 

\end{itemize} 

The 3 CLiF AGN without SDSS spectra also show no trend for FWHM with I$_P$ (\citealt{rose11}, \citealt{fosbury83} and \citealt{alloin92}). However, \citet{alloin92} found a median velocity width for all the emission lines (FLILs and FHILs) around 1000 km s$^{-1}$ for ESO 138 G1, 2-3 times higher than found for the SDSS CLiF AGN sample. 

\subsubsection{$\Delta$v vs I$_P$} 

In Figure \ref{fig:shift} we plot the velocity shifts ($\Delta$v; km s$^{-1}$) relative to the [OIII]$\lambda$5007 line, versus I$_p$ (eV). Again there is significant scatter of a few hundred km s$^{-1}$. There is no evidence for a trend with I$_P$. 

There are 7 lines which have much larger $\Delta$v (blue crosses) than the median of all of the emission lines of the CLiF AGN (horizontal dashed lines). These emission lines are:

\begin{itemize}

\item $[$OII]$\lambda\lambda$3726,3729 doublet (I$_P$ = 13.61 eV) in III Zw 77, Mrk 1388 and J1241+44 whose blended line center (assumed to be 3727.4 \AA ) is dictated by the electron density of the emission region.

\item $[$OI]$\lambda$6364 (I$_P$ = 0 eV) in Mrk 1388 which is in a blend with [FeX]$\lambda$6375.

\item $[$FeXI]$\lambda$6984 (I$_P$ = 262.1 eV) in J1241+44 an uncommon [FeXI] emission line.

\item $[$FeVII]$\lambda$5159 (I$_P$ = 99.1 eV) in J1241+44 and [FeVI]$\lambda$5176 (I$_P$ = 75.0 eV) in J1641+43 both of which are blended with each other.

\end{itemize}

The outliers in blends are likely the result of fitting degeneracies. For the [FeXI]$\lambda$6984 line, it is possible that this emission line has been misidentified.     

Q1131+16 also shows no trend for $\Delta$v with I$_P$ \citep{rose11}. Unfortunately \citet{fosbury83} and \citet{alloin92} do not comment on $\Delta$v for the FHILs.

\subsection{The [OIII]$\lambda$4363 Emitting Region} \label{sect:4363}

The emission-line flux ratio of the FLILs [OIII] 4363/5007 is a widely used diagnostic for the ionization mechanism and temperature of NLR in AGN \citep{osterbrock06} as it has little or no contamination from star formation compared to other NLR emission lines (e.g. [NII], [SII] and [OI]). However, [OIII]$\lambda$4363 has a 50 times higher critical density (n$_{C}$ = 10$^{7.6}$ cm$^{-3}$; \citealt{osterbrock06}) than [OIII]$\lambda$5007 (n$_{C}$ = 10$^{5.8}$ cm$^{-3}$), comparable to those of the FHILs. 

\citet{nagao01} showed that, like [FeVII], the [OIII] 4363/5007 ratio increases across Seyfert galaxy types from type 2 to type 1. This suggests that a significant portion of the [OIII]$\lambda$4363 flux may be emitted from the [FeVII] emitting clouds in the CLiF region.

\begin{figure}
\centering
\includegraphics[scale=0.38, trim=0.1cm 4.6cm 0.1cm 3.6cm]{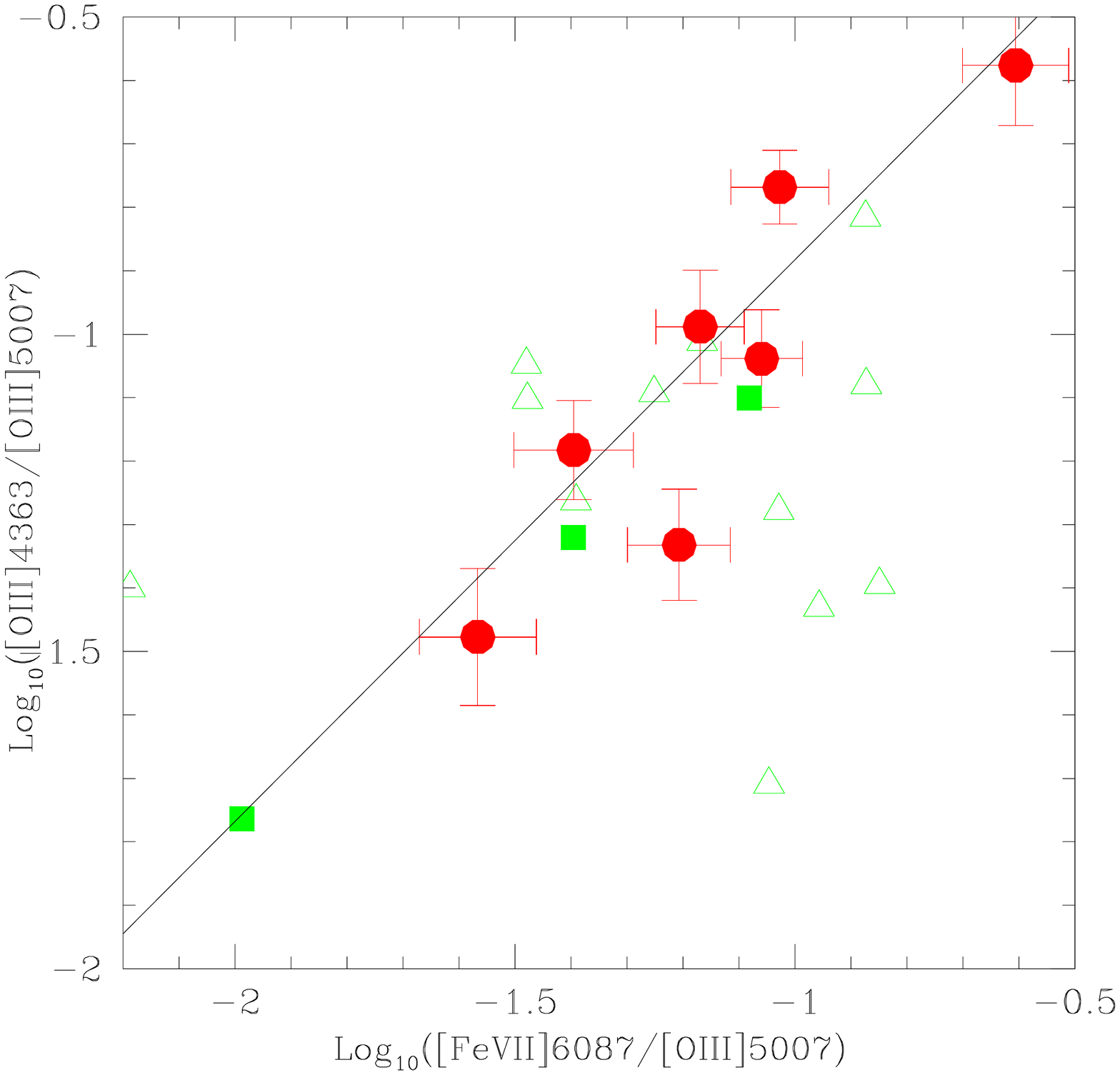}
\caption{The strength of [OIII]$\lambda$4363 versus [FeVII]$\lambda$6087, normalized to [OIII]$\lambda$5007. The CLiF AGN are indicated by red circles, the narrow-line \citet{erkens97} objects by green squares, and the broad-line \citet{erkens97} objects by open green triangles. The solid line represents the best-fit slope to the CLiF AGN.}
\label{fig:4363}
\end{figure}

We address this possibility directly in Figure \ref{fig:4363} where we plot the strength of [OIII]$\lambda$4363 against [FeVII]$\lambda$6087. Both values are normalized to [OIII]$\lambda$5007 to eliminate AGN luminosity as a variable. Figure \ref{fig:4363} includes both the CLiF AGN and the \citet{erkens97} FHIL sample. \citet{erkens97} observed a sample of 15 Seyfert galaxies and 2 emission line radio galaxies based on the presence of FHILs in previous work. For both CLiF AGN, and the narrow-line \citet{erkens97} samples, there is a clear correlation of [OIII]$\lambda$4363 with [FeVII]$\lambda$6087. The correlation coefficient for these objects is R = 0.92. Using Spearman's rank correlation test, this correlation is statistically significant, with a p-value = 0.05\%.

This correlation clearly indicates that both the [FeVII] and [OIII]$\lambda$4363 fluxes are emitted from the same emission region in both CLiF AGN and in the narrow-line \citet{erkens97} AGN i.e. the CLiF region. Given that the FHILs in both the CLiF AGN and \citet{erkens97} narrow-line AGN samples share the same kinematics as the FLILs, this correlation suggests a strong link between the FHIL emitting clouds and the [OIII]$\lambda$4363 emitting clouds.

The broad-line \citet{erkens97} AGN show no such correlation (R=-0.01, p=0.98). In the broad-line \citet{erkens97} objects FHILs do not share the same kinematics as the FLILs, but have higher FWHMs, and are blueshifted with respect to FLILs. The magnitude of the blueshift is correlated with the FWHM. This is another difference between the CLiF AGN and typical type 1 AGN. 

The correlation of [OIII]$\lambda$4363 and [FeVII]$\lambda$6087 fluxes suggests that using the [OIII] 4363/5007 emission-line flux ratio is not a reliable way to estimate temperature, at least for the CLiF AGN. Whether this is a problem for typical AGN requires the determination of the level of contamination of [OIII]$\lambda$4363 from the CLiF region.   

\subsection{Emission Region Diagnostics} \label{sect:params} 

In order to investigate the physical conditions of both the FLIL and FHIL emission clouds, we used the photoionization code CLOUDY\footnote{The version of CLOUDY used in this paper is c13.03.} \citep{ferland98} to create plane-parallel, single-slab photoionization models for the CLiF AGN emission lines. We assumed that the clouds were radiation bounded and had a solar composition. 

We varied the ionization parameter ($U$) over the range -2.5 $\leq$ log$_{10}$ [$U$] $\leq$ 0, in increments of log$_{10}$ [$U$] = 0.5, and the hydrogen density (n$_H$) over the range 3.0 $\leq$ log$_{10}$ [n$_H$ cm$^{-3}$] $\leq$ 7.0 in steps of log [n$_H$ cm$^{-3}$] = 0.5. The models were run assuming ionizing continuum power-law indices $\alpha$ -1.5, -1.2 and -0.8\footnote{The emission is described by F$_{\nu}$ $\propto$ $\nu$$^{+\alpha}$, where $\alpha$ is the power law spectral index.}. We used the `table power law' command of CLOUDY to describe the input ionizing continuum SED. This has a default spectral range from 10$\mu$m to 50 keV \citep{ferland98}. The results are shown in Figure \ref{fig:highdiag}. We do not show grids which assumes a power-law of $\alpha$=-1.5 or -0.8 because these models failed to reproduce the observed emission line ratios for the CLiF AGN. 

We did not include dust grains when modeling the FHIL and FLIL clouds. This assumption is supported by the work of \citet{ferguson97}, \citet{nagao03} and \citet{nagao06} who demonstrated that for a large sample of AGNs the observed FHIL strengths could not be reproduced in CLOUDY models that contain dust grains.

\subsubsection{FHILs}\label{sect:fhilc}

\begin{figure*}
\centering
\includegraphics[scale=0.38, trim=0.1cm 4.6cm 0.1cm 3.6cm]{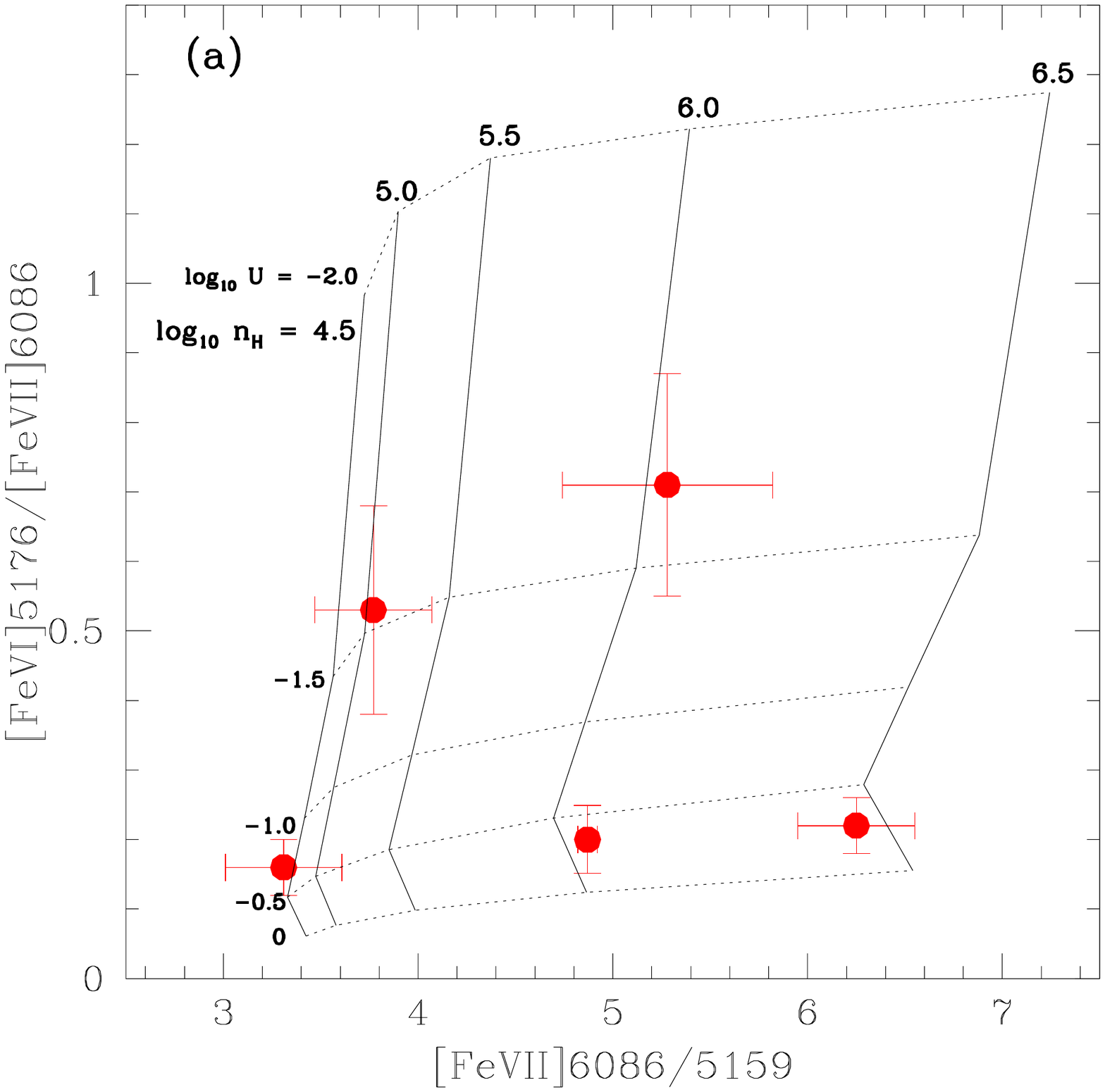}
\includegraphics[scale=0.38, trim=0.1cm 4.6cm 0.1cm 3.6cm]{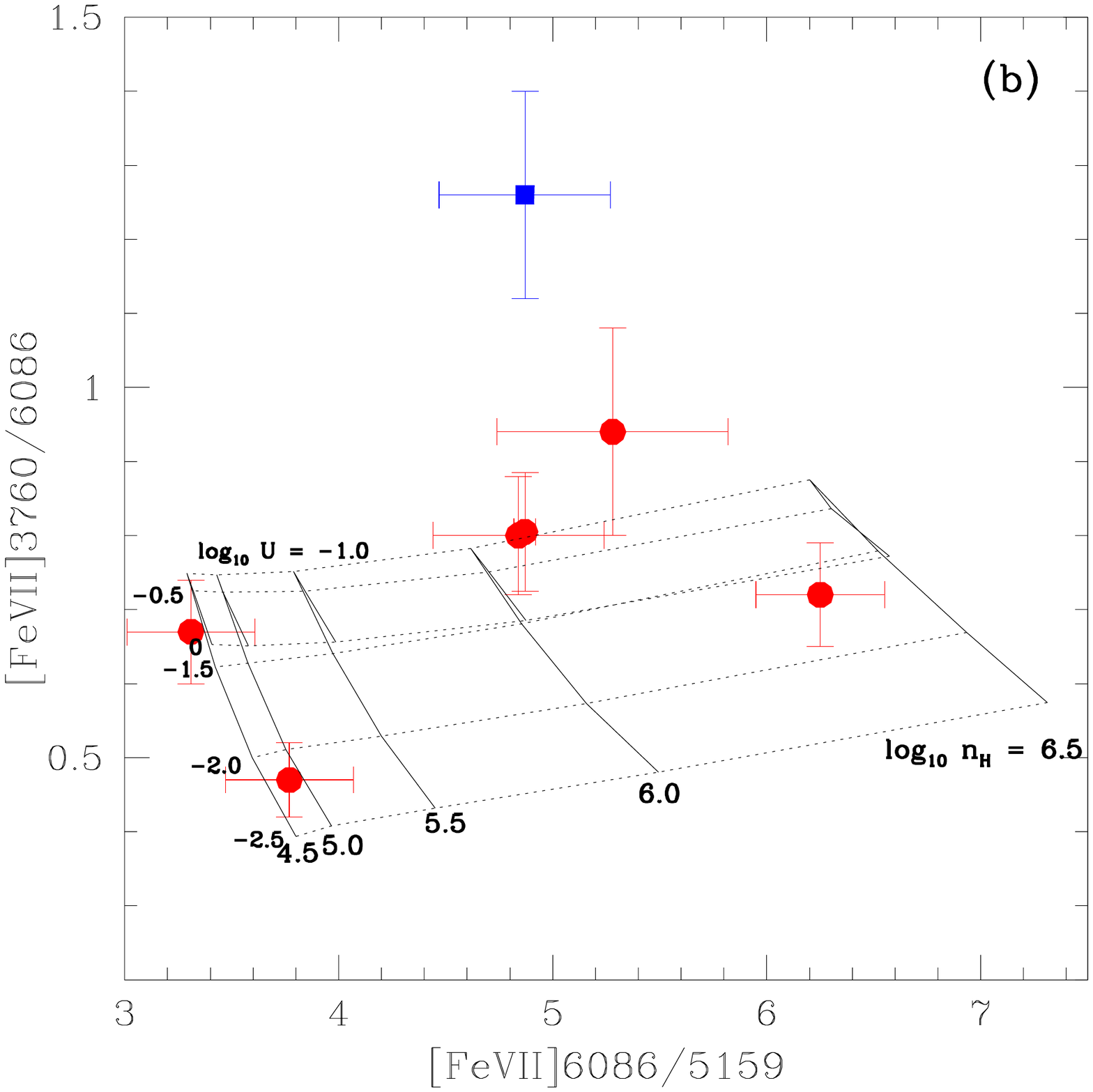}\\
\includegraphics[scale=0.38, trim=0.1cm 4.6cm 0.1cm 3.6cm]{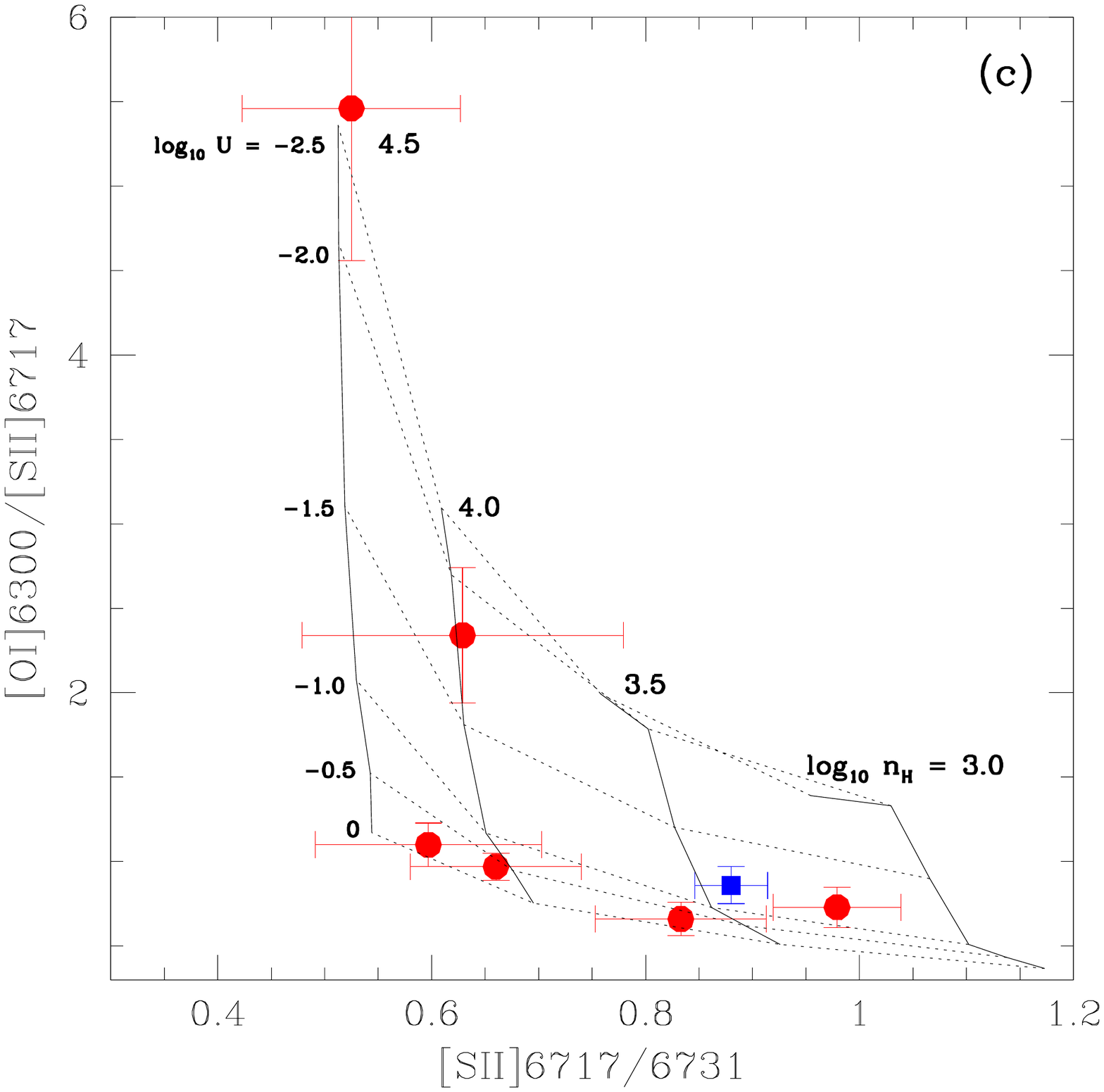}
\includegraphics[scale=0.38, trim=0.1cm 4.6cm 0.1cm 3.6cm]{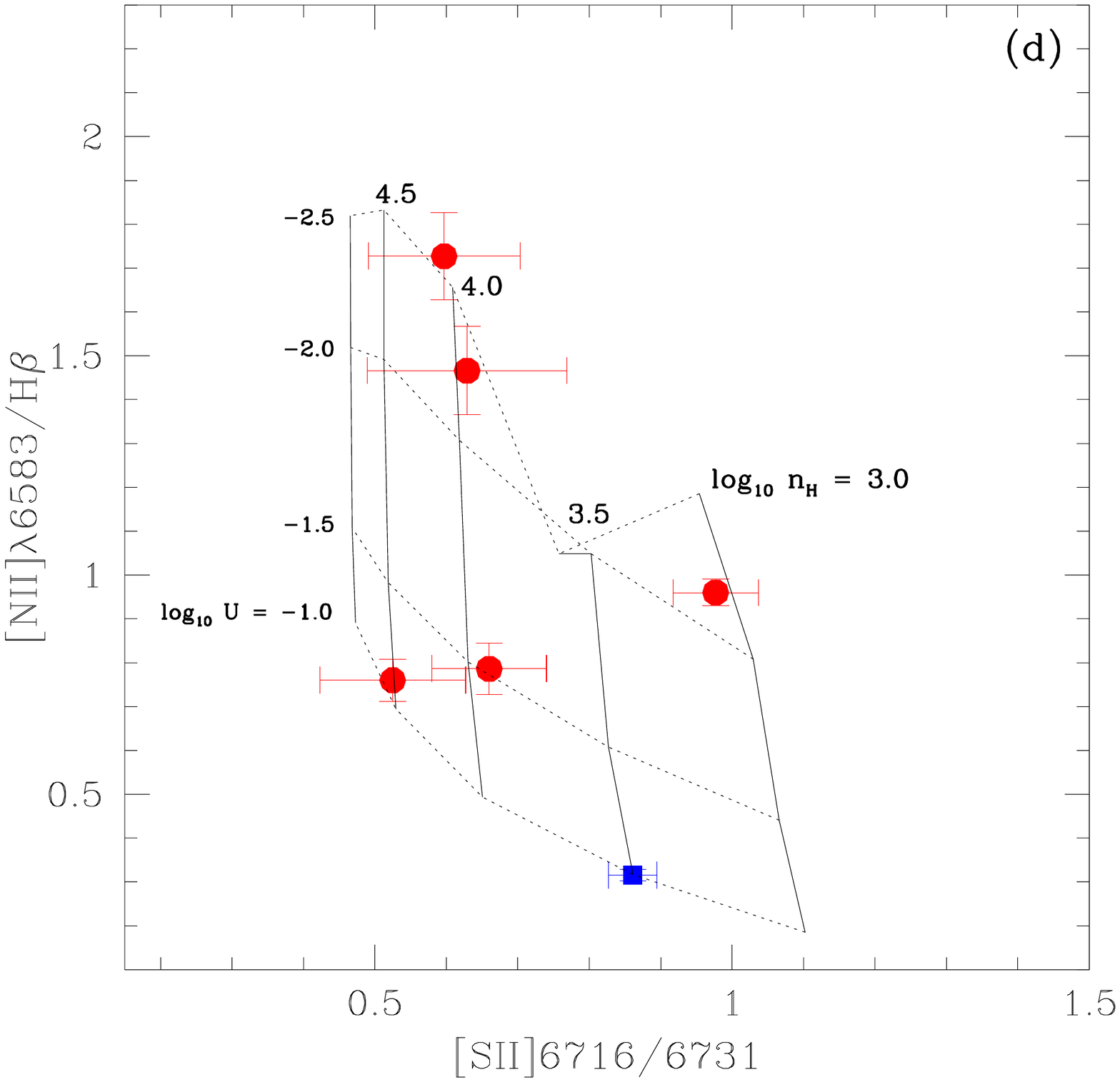}\\
\caption{{\bf (a)} Diagnostic plot for the FLIL region showing the measured [FeVI]$\lambda$5176/[FeVII]$\lambda$6087 versus [FeVII]6087/5159 ratios for the CLiF AGN (red circles). III Zw 77 is indicated separately with a blue square. Iso density lines are solid and iso ionization parameter lines are dotted. {\bf (b)}. Diagnostic plot for the FLIL region showing the measured [FeVII]3760/6087 versus [FeVII]6087/5159 ratios. {\bf (c)} Diagnostic plot for the FLIL region showing the measured [OI]$\lambda$6300/[SII]$\lambda$6717 versus [SII]6717/6731 ratios. {\bf (d)} Diagnostic plot for the FLIL region showing the measured [NII]$\lambda$6583/H$\beta$ versus [SII]6717/6731 ratios.}
\label{fig:highdiag}
\end{figure*}

First we consider the FHILs. In Figure \ref{fig:highdiag}(a) we plot a grid of [FeVI]$\lambda$5176/[FeVII]$\lambda$6086 versus [FeVII]6086/5159 contours (iso-n$_H$ and iso-U). We chose the [FeVI]$\lambda$5176/[FeVII]$\lambda$6086 ratio because it is sensitive to both $U$ and the ionizing power-law, and we chose [FeVII]6086/5159 because this is the only FHIL ratio whose emission lines were detected in all CLiF AGN, that is sensitive to density yet not as sensitive to variations in $U$. Note that [FeVI]$\lambda$5176 was not detected in III Zw 77 and therefore it is not included on Figure \ref{fig:highdiag}(a).

In Figure \ref{fig:highdiag}(b) we plot a grid of [FeVII]3760/6086 versus [FeVII]6086/5159 contours. We chose [FeVII]3760/6086 because it is sensitive to $U$.

All the ratios studied in Figures \ref{fig:highdiag}(a)\&(b) consist of Fe lines. Therefore, they are completely free from the uncertainty in both the relative elemental abundance ratio, and the dust depletion factor.

Comparing the line ratios for the CLiF AGN with the grid models in Figures \ref{fig:highdiag}(a)\&(b) we find the ratios are consistent with n$_H$ in the range 4.5 $<$ log$_{10}$ (n$_H$ cm$^{-3}$) $<$ 6.5 and $U$ in the range -2.0 $<$ log [$U$] $<$ 0. 

We cannot fit a $U$ value for the FHIL emitting clouds of III Zw 77 (blue square). This is true for all assumed power-laws for the ionizing continuum. 

\subsubsection{H$\alpha$ emitting region}\label{sect:fex}

The observed H$\alpha$/H$\beta$ ratio is very large (3.95-6.36 with the Case B value of 2.9) for a given H$\gamma$/H$\beta$ ratio in CLiF AGN. A plausible explanation is that the high density of the CLiF region in these objects, coupled with an extended partially ionized zone (\citealt{gaskell84}; \citealt{rose11}), will lead to significant collisional excitation of H$\alpha$ in the partially ionized zones of clouds in the CLiF region. The other Balmer series emission lines, will not be enhanced as they have lower cross-sections for collisional excitation, as well as higher excitation temperatures (\citealt{gaskell84}; \citealt{osterbrock06}; \citealt{rose11}).

\begin{figure}
\centering
\includegraphics[scale=0.38, trim=0.1cm 4.6cm 0.1cm 3.6cm]{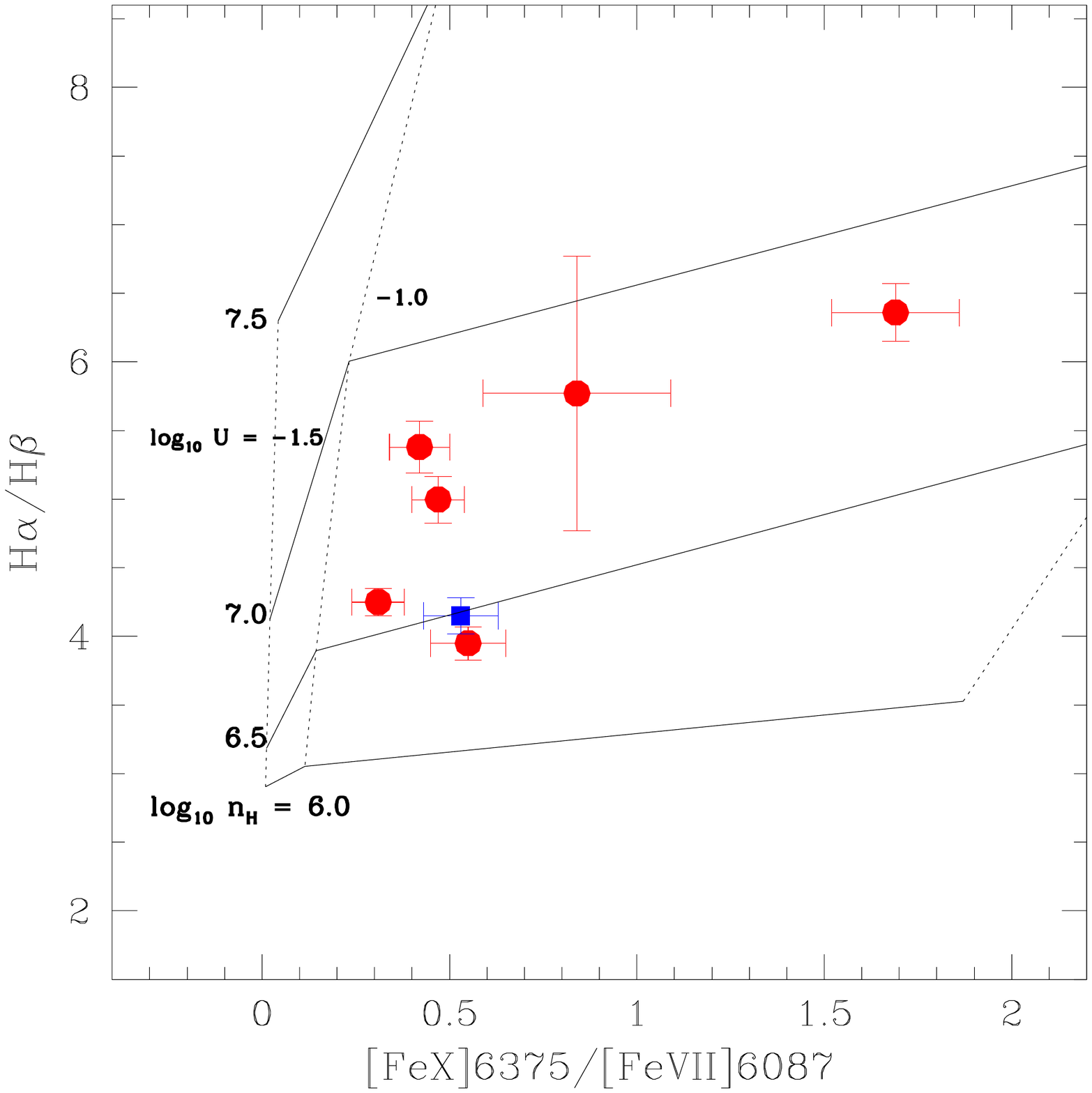}
\caption{Diagnostic plot for the CLiF region showing the measured H$\alpha$/H$\beta$ versus [FeX]$\lambda$6375/[FeVII]$\lambda$6087 ratios for the CLiF AGN (red circles). III Zw 77 is indicated separately with a blue square. Iso density lines are solid and iso ionization parameter lines are dotted.}
\label{fig:fex}
\end{figure}

The conditions needed to produce the high observed H$\alpha$/H$\beta$ flux ratios were investigated using CLOUDY. In Figure \ref{fig:fex} we plot a grid of H$\alpha$/H$\beta$ versus [FeX]$\lambda$6375/[FeVII]$\lambda$6087 contours (iso-n$_H$ and iso-U). We chose [FeX]$\lambda$6375/[FeVII]$\lambda$6087 because this ratio is sensitive to both $U$ and the ionizing power-law, and because both [FeVII]$\lambda$6087 and [FeX]$\lambda$6375 have relatively high n$_{crit}$ (10$^{7.6}$ and 10$^{9.7}$ cm$^{-3}$ respectively) and are therefore not so sensitive to variations in the density.

Comparing the line ratios with the grid model we find that the H$\alpha$/H$\beta$ is consistent with densities in the range 6.0 $<$ log$_{10}$ (n$_H$ cm$^{-3}$) $<$ 7.5. These densities are higher than found in typical NLR (log (n$_H$ cm$^{-3}$) $<$ 4.0; \citealt{osterbrock06}), and 0.5-1 dex higher than the FHIL clouds. 

A high H$\alpha$/H$\beta$ ratios could also be produced by lower metal abundances and/or a harder ionizing continuum (\citealt{gaskell84}; \citealt{rose11}). However, extremely low metallicities (Z $<$ 10$^{-15}$ Z$_{\sun}$) are required in order to reproduce these ratios. At these metallicities it is questionable whether strong Fe emission lines would be observed. In addition, altering the slope of the ionizing continuum results in the models being unable to reproduce the observed flux ratios of the FHILs.

The [FeX]$\lambda$6375/[FeVII]$\lambda$6087 ratios imply $U$ in the range -1.0 $<$ log [$U$] $<$ -0.5. This falls in to the range in $U$ found for the FHILs in $\S$\ref{sect:fhilc}.

The ranges for $U$ and n$_H$ for the FHIL and H$\alpha$ emitting clouds of each CLiF AGN are given in Table \ref{tab:clfhil}.

\begin{table*}
\centering
\caption{Log$_{10}$ $U$ and log$_{10}$ n$_H$ ranges for the FHIL and H$\alpha$ emitting clouds of the CLiF AGN. Both $lower$ and $upper$ subscripts indicate the lower and upper limits of the ranges respectively.}
\begin{tabular}{lcccccccc}
\hline
	&		&	FHIL	&		&	& & H$\alpha$ & &	\\
Name	&	U$_{lower}$	&	U$_{upper}$	&	n$_{H,lower}$	&	n$_{H,upper}$ & U$_{lower}$	&	U$_{upper}$	&	n$_{H,lower}$	&	n$_{H,upper}$	\\
\hline									
Q1131+16	&	-1.5	&	0	&	5.5	&	6.5 & -1.0 & -0.5 & 6.5 & 7.0	\\
III Zw 77	&	-	&	-	&	-	&	- & -1.0 & -0.5 & 6.0 & 6.5	\\
Mrk 1388	&	-2.0	&	-1.5	&	4.5	&	5.5 & -1.0 & -0.5 & 6.0 & 6.5	\\
ESO 138 G1	&	-1.5	&	0	&	4.5	&	5.5 & -1.0 & -0.5 & 6.5 & 7.0	\\
Tololo 0109-383	&	-1.5	&	0	&	5.5	&	6.5 & -1.0 & -0.5 & 6.5 & 7.5	\\
J1241+44	&	-2.0	&	-1.0	&	5.5	&	6.5 & -1.0 & -0.5 & 6.5 & 7.0	\\
J1641+43	&	-2.0	&	0	&	6.0	&	6.5 & -1.0 & -0.5 & 6.5 & 7.0	\\
\hline
\end{tabular}
\label{tab:clfhil}
\end{table*}

\subsubsection{FLILs}\label{sect:flilc}

Next we consider the FLILs, i.e. the normal NLR lines. In Figure \ref{fig:highdiag}(c) we plot a grid of [OI]$\lambda$6300/[SII]$\lambda$6317 versus [SII]6716/6731 contours. We chose [OI]$\lambda$6300/[SII]$\lambda$6317 because this ratio is sensitive to both the ionizing power-law and $U$, and we chose [SII]6716/6731 because it is sensitive to n$_e$ \citep{osterbrock06}.

In Figure \ref{fig:highdiag}(d) we plot a grid of [NII]$\lambda$6583/H$\beta$ versus [SII]6716/6731 contours. We chose [NII]$\lambda$6583/H$\beta$ because this ratio is sensitive to both the ionizing power-law and $U$. The [NII]$\lambda$6583 emission is sensitive to the metallicity (\citealt{thiasa91}; \citealt{thiasa98}). Decreasing the metallicity will increase the required $U$ needed to reproduce the [NII]$\lambda$6583/H$\beta$ ratio. However this is a prominent FLIL ratio detected in all the CLiF AGN spectra. 

The ratios for the CLiF AGN (red circles) are consistent for n$_H$ in the range 3.0 $<$ log$_{10}$ (n$_H$ cm$^{-3}$) $<$ 4.5 and $U$ -2.5 $<$ log [$U$] $<$ 0. The ranges for $U$ and n$_H$ for the FLIL region of each CLiF AGN are given in Table \ref{tab:clflil}. These densities are consistent with those found for the NLRs of typical AGNs (e.g. \citealt{osterbrock06}). The lower end of the $U$ range is consistent with those found for the NLR of AGN (-2.5 $<$ log [$U$] $<$ -2.0; \citealt{bennert05}). The FHIL densities exceed those found for the FLILs by 2 dex. While there is some overlap for the $U$ required by the FLIL and FHIL ratios, overall the $U$ required by the FHILs is higher by 0.5 dex assuming a solar metallicity. 

\begin{table}
\centering
\caption{Log$_{10}$ $U$ and log$_{10}$ n$_H$ ranges for the FLIL regions of the CLiF AGN. Both $lower$ and $upper$ subscripts indicate the lower and upper limits of the ranges respectively.}
\begin{tabular}{lcccc}
\hline
Name	&	U$_{lower}$	&	U$_{upper}$	&	n$_{H,lower}$	&	n$_{H,upper}$	\\
\hline									
Q1131+16	&	-2.0	&	0	&	3.5	&	4.5	\\
III Zw 77	&	-1.5	&	-1.0	&	3.0	&	4.0	\\
Mrk 1388	&	-2.5	&	-1.5	&	3.5	&	4.5	\\
ESO 138 G1	&	-2.5	&	-1.0	&	3.0	&	3.5	\\
Tololo 0109-383	&	-1.5	&	-1.0	&	3.0	&	4.0	\\
J1241+44	&	-2.5	&	-1.0	&	4.0	&	4.5	\\
J1641+43	&	-2.5	&	0	&	3.5	&	4.5	\\
\hline
\end{tabular}
\label{tab:clflil}
\end{table}

\subsubsection{Comparisons of FHIL and FLIL region conditions}\label{sect:compare}

Overall, the range of both $U$ and n$_H$ increase from the FLIL emitting region (-2.5 $<$ log [$U$] $<$ 0, 3.0 $<$ log$_{10}$ (n$_H$ cm$^{-3}$) $<$ 4.5) to the FHIL (-2.0 $<$ log [$U$] $<$ 0, 4.5 $<$ log$_{10}$ (n$_H$ cm$^{-3}$) $<$ 6.5) and boosted H$\alpha$ emitting clouds (-1.0 $<$ log [$U$] $<$ -0.5, 6.0 $<$ log$_{10}$ (n$_H$ cm$^{-3}$) $<$ 7.5) for all the CLiF AGN. The median increase in $U$ from the FLIL to the FHIL and boosted H$\alpha$ emitting clouds is $\sim$ 0.5 dex for the lower end of the $U$ range, and $\sim$ 1.0 dex for the upper end. The median increase for n$_H$ is $\sim$2 dex at lower end of the n$_H$ range, and 3 dex for the upper end (see Tables \ref{tab:clfhil} \& \ref{tab:clflil}).

\begin{figure}
\centering
\includegraphics[scale=0.38, trim=0.1cm 4.6cm 0.1cm 3.6cm]{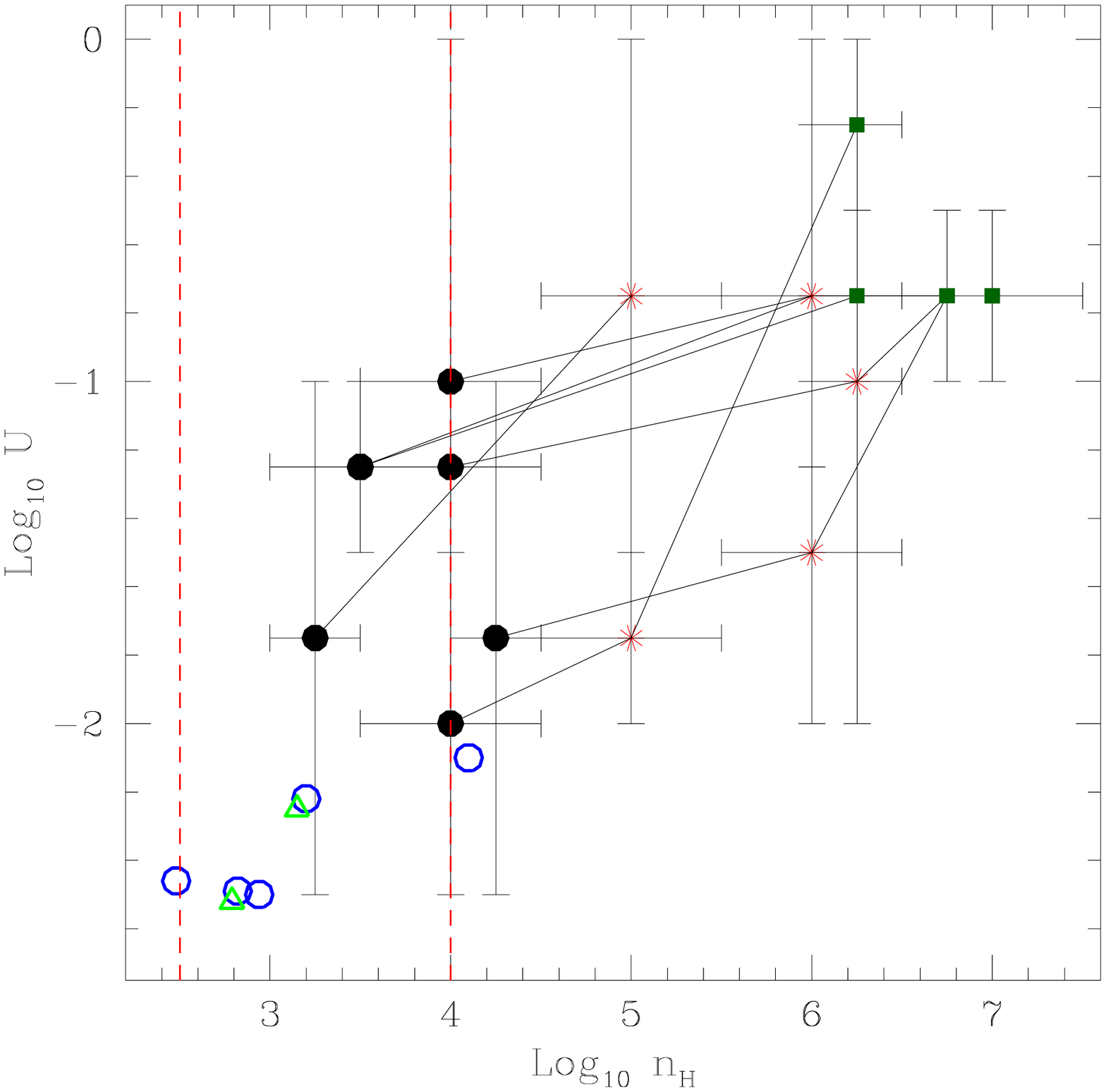}
\caption{A comparison of the CLOUDY model results for the FLIL, FHIL and H$\alpha$ emitting clouds for the CLiF AGN. We plot log$_{10}$ $U$ versus log$_{10}$ n$_H$. The results for the FLILs are indicated by black circles, the results for the FHILs are indicated by red asterisks and the results for the H$\alpha$ emitting clouds are indicated by green squares. The error bars present the $U$ and n$_H$ ranges. The results for individual objects are connected by solid lines. The results from the NLRs of the \citet{bennert05} sample are indicated by open blue circles for the type 1 AGN, and open green triangles for the type 2 AGN. The expected range in density for the NLR is indicated by the red dotted lines \citep{osterbrock06}.}
\label{fig:unplot}
\end{figure}

In Figure \ref{fig:unplot} we map the results of $\S$\ref{sect:fhilc}-\ref{sect:flilc} on to log$_{10}$ $U$ versus log$_{10}$ n$_H$ space. The lines connect the FLIL (black circles), FHIL (red asterisks) and H$\alpha$ (green squares) values for individual objects. We use the median $U$ and n$_H$ values to mark the positions of the CLiF AGN emission regions on Figure \ref{fig:unplot}, the error bars give the full ranges for these values. III Zw 77 and Tololo 0109-383 have the same median $U$ and n$_H$ values for the FLIL emitting region and therefore overlap. Also ESO 138 G1, J1241+44 and J1641+43 have the same median $U$ and n$_H$ values for the FHIL emitting clouds. Finally, Q1131+16 and Tololo 0109-383 have the same median $U$ and n$_H$ values for the boosted H$\alpha$ emitting clouds. Both the median $U$ and n$_H$ values increase from the FLIL emitting region to the FHIL emitting clouds and then to the H$\alpha$ emitting clouds for all the CLiF AGN. However, we note that the range in $U$ for the FHIL and boosted H$\alpha$ emitting clouds overlap in all but two CLiF AGN: Mrk 1388 and J1241+44.

For comparison, we use the n$_H$ and $U$ measurements from \citet{bennert05} for the NLR of normal AGN from spatially resolved VLT/FORS1 and NTT/EMMI spectroscopy for 7 objects (5/7 type 1 AGN, open blue circles; 2/7 type 2 AGN, open green triangles). This work shows that the range of n$_H$ (2.5 $<$ log$_{10}$ n$_H$ $<$ 4.1) measured for these NLRs are consistent with the range found for the FLILs of 5/7 of the CLiF AGN, as well as the typical density of the NLR (2.5 $<$ log$_{10}$ n$_H$ $<$ 4.0; \citealt{osterbrock06}). However, the $U$ values for the \citet{bennert05} measurements are consistent with only the low values found for the CLiF AGN (-2.6 $<$ log$_{10}$ $U$ $<$ -2.0). The difference could be due to aperture effects, where we are sampling emission regions which are closer to the SMBH in the CLiF AGN using the SDSS spectroscopy, when compared with the spatially resolved spectroscopy of Bennert (2005). 

\subsection{The CLiF region radial distance}\label{sect:wall}

Given the computed $U$ and n$_H$ for the FHIL and boosted H$\alpha$ emitting clouds CLiF AGN from $\S$\ref{sect:params} we can calculate a range of distances (R$_{CLiF}$) to the CLiF regions given the ionizing luminosity (L$_{ION}$):  

\begin{equation}
R_{CLiF} =(\frac{ L_{ION} }{4 \pi U n_H c\left \langle h\nu  \right \rangle } )^{1/2}  ,
\label{eqn:lion}
\end{equation}

\noindent where $U$ is the ionization parameter, n$_{H}$ is the hydrogen density, $\left \langle h\nu  \right \rangle_{}$ is the mean ionizing photon energy\footnote{We assume a value of 56 eV for $\left \langle h\nu  \right \rangle_{}$, which is based on an ionizing power law of -1.2, with the photon energy limits of 13.6 eV and 5 keV.}, and c is the speed of light. Note that we assume the CLiF emitting region includes both the boosted H$\alpha$ emitting clouds and the strong FHIL emitting clouds, as both these unusual emission phenomena are seen for all CLiF AGN.   

L$_{ION}$ is estimated from L$_{[OIII]}$ using both the bolometric correction of \citet{heckman04} (L$_{BOL}$= 3500 L$_{[OIII]}$, with a variance of 0.38 dex), and the \citet{elvis94} SED (L$_{BOL}$ $\approx$ 3.1$\pm$0.2 L$_{ION}$). Given that L$_{ION}$ is proportional to L$_{[OIII]}$, and that we could not confidently correct L$_{[OIII]}$ for dust extinction, the computed radial distances to the CLiF emitting region must be considered lower limits.

\begin{table*}
\centering
\caption{Radial distance (R$_{CLiF}$) and sublimation distance (R$_{SUB}$) ranges for the CLiF AGN.}
\begin{tabular}{lccccc}
\hline
Name	&	R$_{CLiF}$	&	R$_{CLiF}$	&	R$_{sub}$	&	R$_{sub}$	&	R$_{sub}$	\\
	&	$lower$	&	$upper$	&	$1800K$	&	$1500K$	&	$1200K$	\\
\hline											
Q1131+16	&	0.52$\pm$0.01	&	16.49$\pm$0.38	&	0.85$\pm$0.05	&	1.51$\pm$0.09	&	4.34$\pm$0.25	\\
III Zw 77	&	0.41$\pm$0.01	&	2.33$\pm$0.06	&	0.68$\pm$0.04	&	1.09$\pm$0.06	&	1.94$\pm$0.11	\\
Mrk 1388	&	0.33$\pm$0.01	&	32.52$\pm$0.74	&	0.16$\pm$0.01	&	0.53$\pm$0.03	&	1.52$\pm$0.09	\\
Tololo 0109-383	&	0.04$\pm$0.01	&	2.03$\pm$0.18	&	0.19$\pm$0.01	&	0.53$\pm$0.02	&	2.38$\pm$0.14	\\
J1241+44	&	0.09$\pm$0.01	&	1.98$\pm$0.13	&	0.10$\pm$0.01	&	0.16$\pm$0.02	&	0.29$\pm$0.02	\\
J1641+43	&	0.51$\pm$0.02	&	28.65$\pm$1.29	&	1.48$\pm$0.09	&	2.43$\pm$0.14	&	4.24$\pm$0.24	\\
\hline
\end{tabular}
\label{tab:dist}
\end{table*}

Using this method we find a range of radial distances for the CLiF emitting region: 0.04 $<$ R$_{CLiF}$ $<$ 32.5 pc (Table \ref{tab:dist}). The uncertainties given were computed using the uncertainties for the [OIII]$\lambda$5007 flux because they were the dominant source of uncertainty in the calculation. We could not calculate distance range for ESO 138 G1, because the spectrum in \citet{alloin92} was not flux calibrated. In addition, the distance range calculations for III Zw 77 relies solely on the CLOUDY grid presented in Figure \ref{fig:fex}. 

The lower end of this distance range is consistent with the inner torus MAGNUM reverberation results of \citet{yoshii14}, who find AGN-torus distances of up to 0.1 pc. The higher end is consistent with disk radii measurements based on IR interferometry and high-resolution molecular line observations (e.g. \citealt{chou07}; \citealt{tristram07}), as well as results from fitting IR SEDs and spectroscopy \citep{alonso11}. 

We cannot make a direct comparison with the expected BLR radius for the CLiF AGN because we do not detect the BLR in the CLiF AGN spectra. 

\begin{figure}
\centering
\includegraphics[scale=0.38, trim=0.1cm 4.6cm 0.1cm 3.6cm]{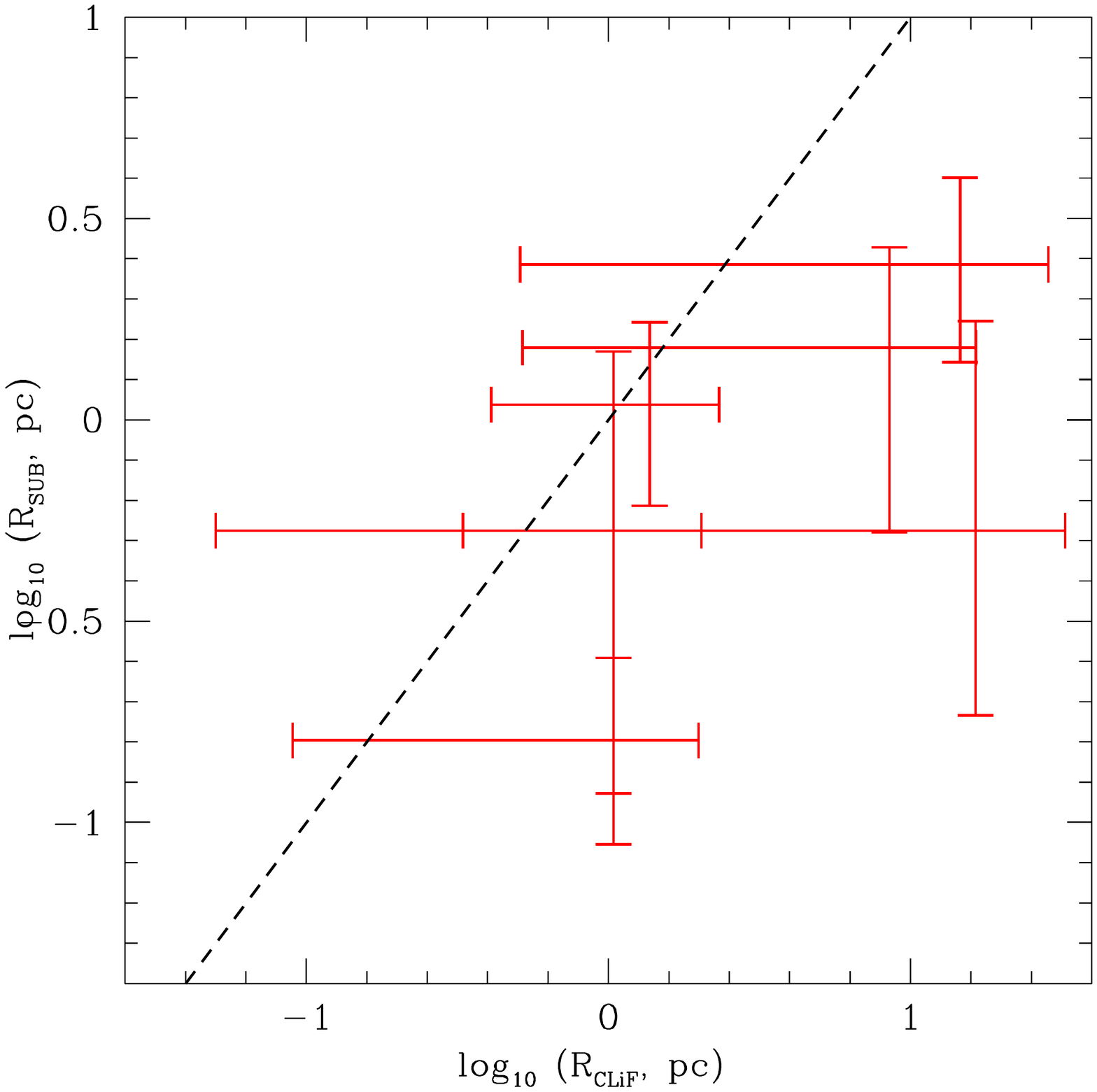}
\caption{Log$_{10}$ R$_{SUB}$ ranges versus log$_{10}$ R$_{CLiF}$ ranges for the CLiF AGN (red lines). The dashed black line indicates a one-to-one ratio of R$_{SUB}$ and R$_{CLiF}$.}
\label{fig:hotdist}
\end{figure}

We also determine dust sublimation distances (R$_{SUB}$) for the CLiF AGN using the relationship given by \citet{elitzur08}. We assume dust sublimation temperatures in the range 1200 $<$ T$_{SUB}$ $<$ 1800 K, consistent with the expected range for torus dust grains (\citealt{barvainis87}; \citealt{rod06}). The uncertainties were computed using both the uncertainty on the bolometric corrections and the uncertainties for the [OIII]$\lambda$5007 flux. 

Figure \ref{fig:hotdist} plots the log$_{10}$ R$_{SUB}$ ranges versus the log$_{10}$ R$_{CLiF}$ ranges (pc) for the CLiF AGN (red lines) given in Table \ref{tab:dist}. The x-axis error bars represent the R$_{CLiF}$ distance ranges. These ranges intersect with the R$_{SUB}$ ranges at the median R$_{CLiF}$ value. The y-axis error bars represent the R$_{SUB}$ distance ranges. The lowest R$_{SUB}$ values are for R$_{SUB}$ at a sublimation temperature of 1800 K, and the highest values are at a temperature of 1200 K. The R$_{SUB}$ ranges intersect with the R$_{CLiF}$ ranges at distances which assume a sublimation temperature of 1500 K. The dashed black line indicates the one-to-one relationship between R$_{CLiF}$ and R$_{SUB}$. Note that all R$_{CLiF}$ ranges intersect with the one-to-one line.

However there is a caveat, L$_{[OIII]}$ has not been corrected for aperture effects. Due to the limited coverage of the SDSS fiber diameter (3 arcseconds), a fraction of the total [OIII] flux will be lost. Therefore, L$_{[OIII]}$ could be stronger than observed, which will subsequently increase both L$_{ION}$ and L$_{BOL}$, and therefore increase both R$_{CLiF}$ and R$_{SUB}$. But the increase in R$_{CLiF}$ and R$_{SUB}$ will be the same, and therefore the comparison of both parameters remain valid. 

We find a good agreement between the ranges for R$_{CLiF}$ and R$_{SUB}$ as both ranges overlap for the CLiF AGN. In addition the R$_{CLiF}$ ranges encompass the one-to-one ratios of R$_{SUB}$ and R$_{CLiF}$. Although we use L$_{[OIII]}$ to calculate both R$_{CLiF}$ and R$_{SUB}$, the ratios between these two parameters do not have to agree.

\section{Discussion}

\subsection{The CLiF Region - the Inner Wall of the Torus?}\label{sect:innert}

Given that both the dust grain sublimation and CLiF emitting region distance ranges are consistent ($\S$\ref{sect:wall}), it is likely that the CLiF emitting region is closely associated with the inner wall of the torus, as suggested by \citet{murayami98} and \citet{rose11}.    

This CLiF location is supported by the observation that FHIL emission is rich in iron lines particularly [FeVII], which can be enhanced relative to the dusty NLR \citep{mor09} by the evaporation of dust grains in the inner torus wall, releasing the iron locked up in the grains \citep{pier95}. 

\citet{rose11} suggested that Q1131+16 is observed with our line of sight at a relatively large angle to the torus axis, so we have a maximal view of the far wall of the torus, while the AGN continuum is still obscured by the near side of the torus. 

We emphasize that such a geometry is consistent with the relatively narrow FWHM and small velocity shifts of the FHILs, since the range of circular velocities associated with the torus will be restricted largely to transverse motions in the sky plane, and any vertical gas motions would be directed close to perpendicular to the line of sight \citep{rose11}.

\subsection{Collisional excitation of H$\alpha$}

In $\S$\ref{sect:halpha} we showed that the CLiF AGN occupy their own region on the BPT diagram due to a factor $\sim$1.4-2.2 boosted H$\alpha$ emission.

This boosting may be due to collisional excitation in the high density CLiF region coupled with an extended partially ionized zone, so that Hydrogen-like ions can be ionized (\citealt{gaskell84}; \citealt{rose11}). The other Balmer series emission lines, are not enhanced as they have lower cross-sections for collisional excitation, as well as higher excitation temperatures (\citealt{gaskell84}; \citealt{osterbrock06}; \citealt{rose11}). To explain the high H$\alpha$/H$\beta$ ratio measured in the spectra of the CLiF AGN, requires a major ($>$35\%) contribution to the line flux of H$\alpha$ from the CLiF region. Indeed, in Figure \ref{fig:fex} we investigated this possibility using CLOUDY modeling and found a required density range of 6.0 $<$ log$_{10}$ (n$_H$ cm$^{-3}$) $<$ 7.5, which is higher than expected for the NLR ($<$ 10$^{4.5}$ cm$^{-3}$) and the FHIL emitting clouds (10$^{4.5}$ $<$ n$_H$ $<$ 10$^{6.5}$ cm$^{-3}$). 

As there is a difference between the densities of the boosted H$\alpha$ and FHIL emitting clouds, but no major difference in $U$ ($\S$\ref{sect:compare}), we suggest that both sets of emission clouds are located in the CLiF region (possibly the inner torus wall, see $\S$\ref{sect:innert}), but there is a range of densities in the CLiF region, where the highest density clouds produce the boosted H$\alpha$ flux.

In Figure \ref{fig:bpt} we defined a box (red dotted line) to indicate the `CLiF AGN Region' of the BPT diagram (limits: -1.0 $<$ log$_{10}$([NII]/H$\alpha$) $<$ -0.4, and 0.65 $<$ log$_{10}$([OIII]/H$\beta$) $<$ 1.1). Within these limits, the BOSS catalog \citep{ahn12} has 1284 CLiF AGN candidates. This population is just 0.5\% of the 259008 objects in the AGN region of the BPT diagram (limits: -0.4 $<$ log$_{10}$([NII]/H$\alpha$) $<$ 1, and -0.1 $<$ log$_{10}$([OIII]/H$\beta$) $<$ 1.5).  

This region is based solely on the CLiF AGN studied in this paper. It is possible that higher levels of collisional excitation of H$\alpha$ could push log$_{10}$([NII]/H$\alpha$) values lower than -1.0. These CLiF AGN candidates will be investigated in a further paper.

\section{Conclusion}

When compared to `typical' AGN, the optical spectra of CLiF AGN show unusual properties in the sense that they are rich in emission lines, and have a remarkable variety of species.

Our analysis of seven CLiF AGN has shown that:

\begin{enumerate}

\item[1.] The H$\alpha$ Balmer emission line is boosted by a factor $\sim$1.4-2.2 when compared to higher energy Balmer transitions, putting them in a distinct location on the BPT diagram. 

\item[2.] There is no difference between the velocity widths of the FLILs and FHILs.

\item[3.] There is no evidence that the FHILs are blueshifted with respect to the FLILs. 

\item[4.] The [OIII]$\lambda$4363 emission line is emitted from the same region as the FHILs in CLiF AGN as its emission flux is correlated with that of [FeVII]$\lambda$6087.

\item[5.] The FHILs of CLiF AGN are emitted from dense clouds (10$^{4.5}$ $<$ n$_H$ $<$ 10$^{6.5}$ cm$^{-3}$) when compared to the NLR.

\item[6.] The boosted H$\alpha$ emission lines of CLiF AGN are emitted from denser clouds (10$^{6.0}$ $<$ n$_H$ $<$ 10$^{7.5}$ cm$^{-3}$) when compared to the FHIL emitting clouds.

\item[7.] We calculate small radial distances (0.04 $<$ R$_{CLiF}$ $<$ 32.5 pc) to the CLiF emitting region, which are comparable to the sublimation distance for dust grains (0.10 $<$ R$_{SUB}$ $<$ 4.3 pc).  

\end{enumerate}

A plausible model is that these objects are being viewed with a specific geometry which gives a maximal view of the inner wall on the far side of the torus, not at such a large angle from the disk plane that the BLR and continuum are exposed. In this scenario the inner torus wall is the FHIL and boosted H$\alpha$ emitting region, which consists of gas with a range of densities that are needed to produce both strong FHILs and boosted H$\alpha$ emission.

We plan to follow up the CLiF AGN by (a) obtaining both X-ray and UV observations for the CLiF AGN, (b) modeling both the strength of the emission lines and kinematics of the CLiF emitting region, (c) using the BPT cleft to find more (Rose et al. 2015), and (d) modeling the geometry (Glidden et al. 2015).

\section*{Acknowledgments}

We gratefully acknowledge Eugene Avrett, Thaisa Storchi Bergmann and Hagai Netzer for their insightful discussions. This work was supported in part by NASA grant NNX13AE88G. We thank the referee for useful comments and suggestions which improved this work substantially. The authors acknowledge the data analysis facilities provided by the Starlink Project, which is run by CCLRC on behalf of PPARC. This publication makes use of data products from the Two Micron All Sky Survey, which is a joint project of the University of Massachusetts and the Infrared Processing and Analysis Center/California Institute of Technology, funded by the National Aeronautics and Space Administration and the National Science Foundation. This publication also makes use of data products from NEOWISE, which is a project of the Jet Propulsion Laboratory/California Institute of Technology, funded by the Planetary Science Division of the National Aeronautics and Space Administration. This research has made use of the NASA/IPAC Extragalactic Database (NED) which is operated by the Jet Propulsion Laboratory, California Institute of Technology, under contract with the National Aeronautics and Space Administration. Funding for SDSS-III has been provided by the Alfred P. Sloan Foundation, the Participating Institutions, the National Science Foundation, and the U.S. Department of Energy Office of Science.

\appendix

\section{Emission line tables}

\begin{table*}
\centering
\caption{Line identifications for the SDSS spectrum of Mrk 1388.  `I$_{P}$ presents the ionization potentials for the securely identified emission species. Flux ratios are relative to H$\beta$ and are not corrected for reddening. Note that the fluxes presented here are from the double-Gaussian fitting model. Line IDs, $\Delta$v and FWHM are determined by fitting a single Gaussian to the emission lines. The total flux of the H$\beta$ emission line is (2.65$\pm$0.19) x 10$^{-15}$ erg s$^{-1}$ cm$^{-2}$ \AA$^{-1}$. The column `Gaussians' indicates the number of Gaussians required to fit the profile of the emission line.}
\begin{tabular}{lcccccccc}
\hline
Line ID	&	I$_{P}$ (eV)	&	$\lambda$ (\AA )	&	$\lambda$$_{SDSS}$  (\AA ) 	&	F/F$_{H\beta}$	&	EW (\AA )	&	$\Delta$v (km s$^{-1}$)	&	FWHM (km s$^{-1}$)	&	Gaussians	\\
\hline																	
$[$FeVII]	&	99.1	&	3758.9	&	3839.5$\pm$0.1	&	0.16$\pm$0.02	&	7.4	&	-42.8$\pm$7.9	&	337.9$\pm$20.1	&	2	\\
HI	&	0	&	3770.6	&	3852.1$\pm$0.5	&	0.03$\pm$0.01	&	1.5	&	3.0$\pm$43.0	&	154.2$\pm$60.5	&	1	\\
HI	&	0	&	3797.9	&	3878.9$\pm$0.2	&	0.06$\pm$0.01	&	3	&	-76.4$\pm$17.0	&	380.4$\pm$43.8	&	1	\\
HI	&	0	&	3835.4	&	3918.4$\pm$0.9	&	0.04$\pm$0.01	&	2	&	10.7$\pm$73.3	&	228.8$\pm$111.6	&	1	\\
$[$NeIII]	&	41.0	&	3868.8	&	3951.9$\pm$0.0	&	1.30$\pm$0.28	&	74.2	&	-33.0$\pm$3.3	&	366.4$\pm$8.1	&	2	\\
$[$FeV]	&	54.8	&	3891.3	&	3974.7$\pm$0.3	&	0.28$\pm$0.08	&	16	&	-47.8$\pm$22.6	&	445.7$\pm$54.9	&	2	\\
$[$NeIII]	&	41.0	&	3967.5	&	4053.5$\pm$0.0	&	0.44$\pm$0.10	&	25.8	&	25.1$\pm$2.7	&	379.0$\pm$6.1	&	2	\\
$[$SII]	&	10.4	&	4068.6	&	4155.8$\pm$0.1	&	0.09$\pm$0.02	&	3.9	&	-82.5$\pm$10.2	&	382.4$\pm$6.8	&	2	\\
$[$SII]	&	10.4	&	4076.3	&	4163.3$\pm$0.1	&	0.11$\pm$0.03	&	4.8	&	-75.6$\pm$12.5	&	382.4$\pm$6.8	&	2	\\
H$\delta$	&	13.6	&	4101.7	&	4189.3$\pm$0.1	&	0.24$\pm$0.05	&	10.5	&	-71.4$\pm$25.1	&	382.0$\pm$6.1	&	2	\\
?	&		&	-	&	4310.5$\pm$0.5	&	0.03$\pm$0.01	&	1.6	&	-	&	194.2$\pm$43.3	&	1	\\
?	&		&	-	&	4322.0$\pm$0.3	&	0.03$\pm$0.01	&	1.6	&	-	&	206.3$\pm$32.2	&	1	\\
?	&		&	-	&	4335.5$\pm$0.4	&	0.05$\pm$0.01	&	2.6	&	-	&	376.7$\pm$55.4	&	1	\\
$[$OII]	&	13.6	&	4317.2	&	4411.5$\pm$1.0	&	0.04$\pm$0.01	&	2.1	&	70.2$\pm$69.1	&	300.5$\pm$103.0	&	1	\\
H$\gamma$	&	0	&	4340.5	&	4434.0$\pm$0.1	&	0.49$\pm$0.03	&	24.7	&	-20.6$\pm$8.6	&	439.9$\pm$26.7	&	2	\\
$[$OIII]	&	35.1	&	4363.2	&	4456.9$\pm$0.1	&	0.56$\pm$0.10	&	28.3	&	-38.5$\pm$6.0	&	352.3$\pm$15.7	&	2	\\
HeI	&	0	&	4471.4	&	4569.4$\pm$0.6	&	0.06$\pm$0.02	&	2.9	&	92.9$\pm$39.0	&	492.2$\pm$79.8	&	1	\\
?	&		&	-	&	4759.4$\pm$0.5	&	0.04$\pm$0.01	&	2	&	-	&	260.0$\pm$50.6	&	1	\\
HeII	&	24.6	&	4685.7	&	4786.4$\pm$0.1	&	0.34$\pm$0.08	&	17	&	-30.2$\pm$3.4	&	301.9$\pm$8.1	&	2	\\
H$\beta$	&	0	&	4861.3	&	4966.0$\pm$0.1	&	-	&	30.8	&	-17.9$\pm$3.8	&	350.8$\pm$9.2	&	2	\\
$[$OIII]	&	35.1	&	4958.9	&	5065.9$\pm$0.2	&	2.86$\pm$0.19	&	88.2	&	-4.5$\pm$10.3	&	341.5$\pm$24.9	&	2	\\
$[$OIII]	&	35.1	&	5006.8	&	5114.9$\pm$0.1	&	8.57$\pm$0.21	&	264.5	&	-	&	357.0$\pm$8.7	&	2	\\
$[$FeVII]	&	99.1	&	5159.0	&	5269.9$\pm$0.3	&	0.09$\pm$0.02	&	3	&	-32.0$\pm$17.9	&	283.2$\pm$31.3	&	2	\\
$[$FeVI]	&	75.0	&	5176.4	&	5288.2$\pm$0.4	&	0.18$\pm$0.05	&	6.1	&	-0.7$\pm$22.0	&	94.4$\pm$17.9	&	2	\\
$[$NII]	&	14.5	&	5199.5	&	5311.5$\pm$0.4	&	0.04$\pm$0.01	&	1.5	&	17.3$\pm$11.2	&	243.5$\pm$38.3	&	1	\\
$[$FeVII]	&	99.1	&	5276.4	&	5390.0$\pm$0.5	&	0.07$\pm$0.02	&	3.5	&	-18.8$\pm$27.6	&	269.4$\pm$47.3	&	2	\\
$[$CaV]	&	67.3	&	5309.1	&	5422.8$\pm$0.4	&	0.15$\pm$0.04	&	7.5	&	-56.3$\pm$24.4	&	539.6$\pm$60.1	&	2	\\
HeII	&	24.6	&	-	&	5529.9$\pm$0.5	&	0.06$\pm$0.02	&	3	&	-	&	346.5$\pm$56.5	&	1	\\
$[$FeVI]	&	75.0	&	5425.7	&	5540.9$\pm$0.8	&	0.06$\pm$0.02	&	3	&	-105.5$\pm$44.0	&	464.6$\pm$90.2	&	2	\\
?	&		&	-	&	5557.9$\pm$0.5	&	0.03$\pm$0.01	&	1.6	&	-	&	250.3$\pm$41.2	&		2\\
?	&		&	-	&	5578.2$\pm$0.3	&	0.03$\pm$0.01	&	1.6	&	-	&	109.4$\pm$23.4	&		2\\
?	&		&	-	&	5739.7$\pm$0.5	&	0.03$\pm$0.01	&	1.6	&	-	&	213.2$\pm$58.9	&		2\\
$[$FeVI]	&	75.0	&	5677.0	&	5799.3$\pm$0.7	&	0.04$\pm$0.01	&	1.5	&	-14.3$\pm$39.4	&	278.0$\pm$82.0	&	1	\\
$[$FeVII]	&	99.1	&	5720.7	&	5844.1$\pm$0.1	&	0.20$\pm$0.03	&	6.2	&	-7.4$\pm$4.2	&	285.2$\pm$9.5	&	2	\\
$[$NII]	&	14.5	&	5754.6	&	5878.8$\pm$0.3	&	0.09$\pm$0.02	&	3	&	-2.7$\pm$17.8	&	423.0$\pm$47.5	&	2	\\
HeI	&	0	&	5875.6	&	6002.3$\pm$0.1	&	0.12$\pm$0.02	&	3.6	&	-12.8$\pm$6.0	&	321.0$\pm$14.7	&	2	\\
?	&		&	-	&	6207.1$\pm$1.5	&	0.05$\pm$0.01	&	1.8	&	-	&	532.6$\pm$159.1	&	1	\\
$[$FeVII]	&	99.1	&	6086.9	&	6217.7$\pm$0.1	&	0.34$\pm$0.05	&	11	&	-30.4$\pm$4.5	&	284.2$\pm$8.7	&	2	\\
$[$OI]	&	0	&	6300.3	&	6436.6$\pm$0.1	&	0.54$\pm$0.09	&	14.8	&	9.4$\pm$4.2	&	367.9$\pm$11.5	&	2	\\
$[$SIII]	&	23.3	&	6312.1	&	6447.8$\pm$0.4	&	0.10$\pm$0.03	&	2.8	&	-30.6$\pm$18.5	&	295.8$\pm$36.8	&	2	\\
$[$OI]	&	0	&	6363.8	&	6501.6$\pm$0.3	&	0.15$\pm$0.03	&	4.1	&	16.7$\pm$14.2	&	293.1$\pm$32.6	&	2	\\
$[$FeX]	&	233.6	&	6374.6	&	6511.1$\pm$0.2	&	0.19$\pm$0.03	&	5.2	&	-53.9$\pm$10.3	&	263.2$\pm$19.4	&	2	\\
$[$NII]	&	14.5	&	6548.1	&	6691.8$\pm$0.2	&	0.49$\pm$0.08	&	15	&	100.1$\pm$7.0	&	301.9$\pm$4.9	&	2	\\
H$\alpha$	&	0	&	6562.8	&	6704.2$\pm$0.0	&	3.95$\pm$0.12	&	121.1	&	-16.6$\pm$0.7	&	288.4$\pm$1.8	&	2	\\
$[$NII]	&	14.5	&	6583.4	&	6726.4$\pm$0.0	&	1.47$\pm$0.16	&	45.1	&	33.6$\pm$2.1	&	299.9$\pm$4.9	&	2	\\
?	&		&	-	&	6822.3$\pm$0.7	&	0.04$\pm$0.01	&	1.8	&	-	&	451.6$\pm$61.0	&	1	\\
$[$SII]	&	10.4	&	6716.4	&	6862.2$\pm$0.1	&	0.23$\pm$0.04	&	7.8	&	29.9$\pm$3.8	&	327.9$\pm$9.5	&	2	\\
$[$SII]	&	10.4	&	6730.8	&	6876.6$\pm$0.1	&	0.37$\pm$0.09	&	11.2	&	19.4$\pm$2.7	&	311.0$\pm$6.3	&	2	\\
?	&		&	-	&	7064.8$\pm$1.1	&	0.12$\pm$0.03	&	2	&	-	&	319.2$\pm$86.6	&	1	\\
?	&		&	-	&	7148.2$\pm$3.0	&	0.13$\pm$0.04	&	2	&	-	&	649.3$\pm$182.9	&	1	\\
$[$ArV]	&	59.8	&	7005.9	&	7157.5$\pm$0.5	&	0.13$\pm$0.04	&	2	&	9.5$\pm$23.5	&	251.9$\pm$54.7	&	1	\\
?	&		&	-	&	7197.5$\pm$1.2	&	0.15$\pm$0.03	&	2.1	&	-	&	624.5$\pm$153.6	&	1	\\
HeI	&	0	&	7065.7	&	7217.9$\pm$0.3	&	0.07$\pm$0.02	&	2	&	-18.6$\pm$13.0	&	294.1$\pm$31.7	&	1	\\
$[$ArIII]	&	27.6	&	7135.8	&	7290.1$\pm$0.1	&	0.12$\pm$0.02	&	4	&	5.2$\pm$4.2	&	355.0$\pm$10.8	&	2	\\
?	&		&	-	&	7309.4$\pm$0.8	&	0.04$\pm$0.01	&	1.5	&	-	&	428.4$\pm$82.4	&	1	\\
?	&		&	-	&	7326.5$\pm$0.8	&	0.04$\pm$0.01	&	1.5	&	-	&	348.5$\pm$85.7	&	1	\\
$[$OII]	&	13.6	&	7319.9	&	7478.8$\pm$0.5	&	0.15$\pm$0.03	&	5.1	&	31.2$\pm$20.0	&	521.4$\pm$48.0	&	2	\\
$[$OII]	&	13.6	&	7330.2	&	7490.2$\pm$0.5	&	0.18$\pm$0.06	&	6.1	&	68.7$\pm$20.1	&	340.3$\pm$31.8	&	2	\\
?	&		&	-	&	7537.2$\pm$1.4	&	0.07$\pm$0.02	&	2.5	&	-	&	454.1$\pm$126.6	&	1	\\
?	&		&	-	&	7773.2$\pm$0.5	&	0.03$\pm$0.01	&	1.4	&	-	&	196.4$\pm$29.2	&	1	\\
?	&		&	-	&	7919.9$\pm$1.2	&	0.07$\pm$0.01	&	2.3	&	-	&	677.1$\pm$177.7	&	1	\\
$[$FeXI]	&	262.1	&	7891.8	&	8061.4$\pm$0.2	&	0.13$\pm$0.03	&	4.3	&	-31.9$\pm$6.5	&	237.7$\pm$12.5	&	2	\\

\hline
\end{tabular}								
\label{tab:mrk1388}
\end{table*}

\begin{table*}
\centering
\caption{Line identifications for the SDSS spectrum of III Zw 77. `I$_{P}$ presents the ionization potentials for the securely identified emission species. Flux ratios are relative to H$\beta$ and are not corrected for reddening. Note that the fluxes presented here are from the double-Gaussian fitting model. Line IDs, $\Delta$v and FWHM are determined by fitting a single Gaussian to the emission lines. The total flux of the H$\beta$ emission line is (2.52$\pm$0.09) x 10$^{-14}$ erg s$^{-1}$ cm$^{-2}$ \AA$^{-1}$.The column `Gaussians' indicates the number of Gaussians required to fit the profile of the emission line.}
\begin{tabular}{lcccccccc}
\hline
Line ID	&	I$_P$ (eV)	&	$\lambda$ (\AA )	&	$\lambda$$_{SDSS}$  (\AA ) 	&	F/F$_{H\beta}$	&	EW (\AA )	&	$\Delta$v	&	FWHM	&	Gaussians	\\
\hline																	
$[$OII]	&	13.6	&	3727.4	&	3854.0$\pm$0.1	&	0.63$\pm$0.07	&	8.7	&	-28.2$\pm$7.8	&	413.3$\pm$15.4	&	2	\\
$[$FeVII]	&	99.1	&	3758.9	&	3887.3$\pm$0.1	&	0.49$\pm$0.05	&	6.8	&	17.4$\pm$11.1	&	450.2$\pm$27.5	&	2	\\
$[$FeV]	&	54.8	&	3839.3	&	3968.0$\pm$0.8	&	0.07$\pm$0.01	&	2.1	&	-164.1$\pm$61.2	&	409.0$\pm$109.1	&	1	\\
$[$NeIII]	&	41.0	&	3868.8	&	4000.6$\pm$0.0	&	1.33$\pm$0.24	&	19.2	&	-11.9$\pm$2.9	&	346.2$\pm$6.9	&	2	\\
$[$FeV]	&	54.8	&	3891.3	&	4023.9$\pm$0.2	&	0.41$\pm$0.08	&	5.9	&	-5.6$\pm$16.8	&	486.2$\pm$33.1	&	2	\\
$[$NeIII]	&	41.0	&	3967.5	&	4103.1$\pm$0.2	&	0.16$\pm$0.02	&	2.8	&	21.3$\pm$12.2	&	183.7$\pm$26.7	&	1	\\
?	&		&	-	&	4109.1$\pm$0.5	&	0.49$\pm$0.12	&	8.6	&	-	&	1388.5$\pm$77.0	&		2\\
$[$SII]	&	10.4	&	4068.6	&	4208.2$\pm$0.3	&	0.12$\pm$0.02	&	2.1	&	64.2$\pm$21.7	&	464.2$\pm$53.3	&	1	\\
$[$SII]	&	10.4	&	4076.3	&	4216.2$\pm$0.3	&	0.03$\pm$0.01	&	0.8	&	63.9$\pm$21.6	&	463.2$\pm$53.5	&	1	\\
H$\delta$	&	0	&	4101.7	&	4242.1$\pm$0.1	&	0.25$\pm$0.05	&	4.4	&	35.9$\pm$8.5	&	381.2$\pm$21.0	&	2	\\
H$\delta$$_{B}$	&	0	&	4101.7	&	4233.0$\pm$1.4	&	0.59$\pm$0.03	&	10.4	&	-643.5$\pm$100.1	&	2546.1$\pm$171.9	&	2	\\
$[$OII]	&	13.6	&	4317.2	&	4465.9$\pm$0.5	&	0.15$\pm$0.04	&	3.5	&	99.4$\pm$32.6	&	345.2$\pm$10.5	&	2	\\
H$\gamma$	&	0	&	4340.5	&	4489.0$\pm$0.1	&	0.34$\pm$0.04	&	7.8	&	34.7$\pm$6.7	&	343.0$\pm$10.4	&	2	\\
H$\gamma$$_{B}$	&	0	&	4340.5	&	4479.3$\pm$0.8	&	0.70$\pm$0.05	&	16.1	&	-648.3$\pm$55.4	&	2328.9$\pm$168.4	&	2	\\
$[$OIII]	&	35.1	&	4363.2	&	4512.2$\pm$0.1	&	0.45$\pm$0.05	&	10.4	&	13.9$\pm$4.9	&	340.9$\pm$10.3	&	2	\\
HeI	&	0	&	4471.4	&	4625.7$\pm$0.7	&	0.04$\pm$0.01	&	1.9	&	119.3$\pm$47.4	&	391.5$\pm$79.7	&	1	\\
?	&		&	-	&	4818.0$\pm$0.5	&	0.06$\pm$0.01	&	1.6	&	-	&	279.6$\pm$70.3	&	1	\\
HeII	&	24.6	&	4685.7	&	4845.8$\pm$0.1	&	0.44$\pm$0.07	&	7.0	&	17.0$\pm$8.2	&	469.4$\pm$24.4	&	2	\\
$[$ArIV]	&	40.7	&	4712.6	&	4902.0$\pm$1.2	&	0.03$\pm$0.01	&	1.4	&	8.4$\pm$75.9	&	113.5$\pm$131.9	&	1	\\
H$\beta$	&	0	&	4861.3	&	5027.9$\pm$0.1	&	-	&	15.4	&	50.3$\pm$3.8	&	386.3$\pm$10.4	&	2	\\
H$\beta$$_{B}$	&	0	&	4861.3	&	5016.6$\pm$0.7	&	2.03$\pm$0.09	&	31.3	&	-624.8$\pm$41.3	&	2628.4$\pm$87.8	&	2	\\
$[$FeVII]	&	99.1	&	4893.4	&	5061.5$\pm$0.8	&	0.03$\pm$0.01	&	1.3	&	70.9$\pm$51.8	&	163.9$\pm$69.6	&	1	\\
$[$FeIV]	&	30.7	&	4903.1	&	5070.0$\pm$1.6	&	0.05$\pm$0.01	&	2.7	&	-16.5$\pm$100.4	&	651.9$\pm$249.1	&	1	\\
$[$OIII]	&	35.1	&	4958.9	&	5128.0$\pm$0.1	&	1.94$\pm$0.19	&	29.9	&	1.3$\pm$6.8	&	289.3$\pm$4.8	&	2	\\
$[$OIII]	&	35.1	&	5006.8	&	5177.5$\pm$0.0	&	5.82$\pm$0.29	&	89.6	&	-	&	285.6$\pm$4.7	&	2	\\
$[$FeVII]	&	99.1	&	5159.0	&	5334.9$\pm$0.4	&	0.08$\pm$0.01	&	3.2	&	-0.2$\pm$20.9	&	226.2$\pm$35.0	&	2	\\
$[$FeVII]	&	99.1	&	5276.4	&	5456.9$\pm$0.7	&	0.06$\pm$0.01	&	3.2	&	29.2$\pm$40.0	&	308.4$\pm$88.6	&	2	\\
$[$CaV]	&	67.3	&	5309.1	&	5489.0$\pm$0.9	&	0.22$\pm$0.05	&	8.9	&	-60.7$\pm$52.7	&	975.0$\pm$159.7	&	2	\\
$[$FeVII]	&	99.1	&	5720.7	&	5916.7$\pm$0.1	&	0.23$\pm$0.03	&	5.6	&	46.3$\pm$7.5	&	340.9$\pm$17.4	&	2	\\
HeI	&	0	&	5875.6	&	6077.7$\pm$1.2	&	0.24$\pm$0.04	&	5.8	&	86.4$\pm$60.3	&	697.3$\pm$99.1	&	2	\\
?	&		&	-	&	6281.8$\pm$0.7	&	0.05$\pm$0.01	&	2.9	&	-	&	348.5$\pm$8.8	&	1	\\
$[$FeVII]	&	99.1	&	6086.6	&	6294.9$\pm$0.1	&	0.39$\pm$0.05	&	9.5	&	21.0$\pm$3.9	&	347.6$\pm$8.8	&	2	\\
$[$OI]	&	0	&	6300.3	&	6515.1$\pm$0.1	&	0.13$\pm$0.01	&	3.0	&	-2.2$\pm$6.0	&	236.8$\pm$12.2	&	1	\\
$[$SIII]	&	23.3	&	6312.1	&	6527.0$\pm$0.7	&	0.03$\pm$0.01	&	1.0	&	-15.7$\pm$34.6	&	187.6$\pm$47.0	&	1	\\
$[$OI]	&	0	&	6363.8	&	6581.1$\pm$0.3	&	0.04$\pm$0.01	&	1.1	&	13.6$\pm$14.9	&	188.3$\pm$22.0	&	1	\\
$[$FeX]	&	233.6	&	6374.6	&	6592.8$\pm$0.2	&	0.21$\pm$0.05	&	4.8	&	36.7$\pm$7.1	&	415.8$\pm$18.8	&	2	\\
$[$NII]	&	14.5	&	6548.1	&	6806.8$\pm$0.4	&	0.28$\pm$0.07	&	5.3	&	-50.4$\pm$18.0	&	373.2$\pm$12.4	&	2	\\
H$\alpha$	&	0	&	6562.8	&	6787.9$\pm$0.1	&	4.15$\pm$0.13	&	81.5	&	58.7$\pm$3.4	&	374.4$\pm$12.5	&	2	\\
H$\alpha$$_{B}$	&	0	&	6562.8	&	6773.0$\pm$1.0	&	7.09$\pm$1.84	&	139.2	&	-658.5$\pm$46.7	&	2517.4$\pm$76.5	&	2	\\
$[$NII]	&	14.5	&	6583.4	&	6770.3$\pm$0.4	&	0.89$\pm$0.21	&	15.9	&	-50.5$\pm$18.1	&	375.6$\pm$12.5	&	2	\\
$[$SII]	&	10.4	&	6716.4	&	6945.7$\pm$0.1	&	0.15$\pm$0.03	&	3.0	&	11.7$\pm$3.4	&	189.9$\pm$6.5	&	2	\\
$[$SII]	&	10.4	&	6730.8	&	6960.5$\pm$0.1	&	0.17$\pm$0.03	&	3.1	&	8.8$\pm$3.3	&	207.4$\pm$6.3	&	2	\\
$[$ArV]	&	59.8	&	7005.9	&	7243.0$\pm$1.3	&	0.11$\pm$0.02	&	2.8	&	-73.8$\pm$55.3	&	1090.2$\pm$159.2	&	1	\\
?	&		&	-	&	7282.6$\pm$1.2	&	0.12$\pm$0.03	&	3.0	&	-	&	990.5$\pm$135.5	&	1	\\
HeI	&	0	&	7065.7	&	7307.8$\pm$0.4	&	0.11$\pm$0.02	&	2.8	&	49.5$\pm$17.9	&	477.5$\pm$47.8	&	1	\\
$[$ArIII]	&	27.6	&	7135.8	&	7379.3$\pm$0.2	&	0.08$\pm$0.02	&	2.5	&	8.5$\pm$7.7	&	170.6$\pm$12.3	&	1	\\
$[$OII]	&	13.6	&	7319.9	&	7569.1$\pm$0.5	&	0.06$\pm$0.01	&	2.4	&	-16.7$\pm$22.0	&	298.9$\pm$43.1	&	1	\\
$[$OII]	&	13.6	&	7330.2	&	7581.1$\pm$0.7	&	0.04$\pm$0.01	&	1.8	&	36.7$\pm$30.1	&	234.9$\pm$83.7	&	1	\\
?	&		&	-	&	7628.1$\pm$1.4	&	0.03$\pm$0.01	&	1.7	&	-	&	342.3$\pm$80.1	&	1	\\
$[$FeXI]	&	262.1	&	7891.8	&	8161.5$\pm$0.3	&	0.13$\pm$0.04	&	7.4	&	21.7$\pm$12.1	&	320.2$\pm$28.9	&	2	\\
?	&		&	-	&	8738.4$\pm$1.0	&	0.05$\pm$0.01	&	2.4	&	-	&	304.2$\pm$76.9	&	1	\\

\hline
\end{tabular}								
\label{tab:iiizw}
\end{table*}

\begin{table*}
\centering
\caption{Line identifications for the SDSS spectrum of J1241+44. `I$_{P}$ presents the ionization potentials for the securely identified emission species. Flux ratios are relative to H$\beta$ and are not corrected for reddening. Note that the fluxes presented here are from the double-Gaussian fitting model. Line IDs, $\Delta$v and FWHM are determined by fitting a single Gaussian to the emission lines. The total flux of the H$\beta$ emission line is (4.96$\pm$0.30) x 10$^{-16}$ erg s$^{-1}$ cm$^{-2}$ \AA$^{-1}$.The column `Gaussians' indicates the number of Gaussians required to fit the profile of the emission line.}
\begin{tabular}{lcccccccc}
\hline
Line ID	&	I$_P$ (eV)	&	$\lambda$ (\AA )	&	$\lambda$$_{SDSS}$  (\AA ) 	&	F/F$_{H\beta}$	&	EW (\AA )	&	$\Delta$v	&	FWHM	&	Gaussians	\\
\hline																	
$[$OII]	&	13.6	&	3727.4	&	3885.9$\pm$0.5	&	0.77$\pm$0.18	&	7.7	&	104.2$\pm$44.0	&	433.1$\pm$92.7	&	2	\\
$[$FeVII]	&	99.1	&	3758.9	&	3917.7$\pm$0.2	&	1.09$\pm$0.15	&	13.9	&	17.2$\pm$3.7	&	282.4$\pm$40.2	&	2	\\
$[$NeIII]	&	41.0	&	3868.8	&	4031.7$\pm$0.1	&	1.57$\pm$0.15	&	13.7	&	-27.8$\pm$8.8	&	256.3$\pm$22.2	&	2	\\
H$\gamma$	&	0	&	4340.5	&	4523.2$\pm$0.3	&	0.47$\pm$0.04	&	3.4	&	-31.8$\pm$18.6	&	205.5$\pm$34.8	&	2	\\
$[$OIII]	&	35.1	&	4363.2	&	4547.1$\pm$0.2	&	1.3$\pm$0.13	&	8.3	&	-17.2$\pm$10.8	&	287.9$\pm$29.6	&	2	\\
HeII	&	24.6	&	4685.7	&	4883.4$\pm$0.1	&	0.87$\pm$0.09	&	4.5	&	-3.1$\pm$9.3	&	227.4$\pm$18.4	&	2	\\
$[$NeIII]	&	41.0	&	4815.9	&	5017.9$\pm$0.3	&	0.13$\pm$0.04	&	2.8	&	-73.3$\pm$21.5	&	89.7$\pm$38.8	&	2	\\
H$\beta$	&	0	&	4861.3	&	5066.2$\pm$0.1	&	-	&	5.1	&	-13.2$\pm$5.3	&	178.5$\pm$12.8	&	2	\\
$[$OIII]	&	35.1	&	4958.9	&	5168.2$\pm$0.1	&	1.39$\pm$0.11	&	7.5	&	2.4$\pm$4.8	&	206.1$\pm$11.7	&	2	\\
$[$OIII]	&	35.1	&	5006.8	&	5218.0$\pm$0.0	&	4.34$\pm$0.28	&	22.6	&	-	&	215.8$\pm$4.7	&	2	\\
$[$FeVII]	&	99.1	&	5159.0	&	5375.0$\pm$0.4	&	0.21$\pm$0.07	&	2.1	&	-97.7$\pm$22.4	&	141.3$\pm$67.3	&	2	\\
$[$FeVI]	&	75.0	&	5176.4	&	5394.8$\pm$0.3	&	0.79$\pm$0.18	&	7.9	&	-3.8$\pm$17.8	&	147.0$\pm$38.9	&	2	\\
$[$CaV]	&	67.3	&	5309.1	&	5531.4$\pm$1.1	&	0.27$\pm$0.07	&	2.1	&	-94.0$\pm$59.8	&	564.4$\pm$136.9	&	2	\\
?	&		&	-	&	5766.6$\pm$0.3	&	0.29$\pm$0.06	&	2.2	&	-	&	168.7$\pm$36.4	&	2	\\
?	&		&	-	&	5856.0$\pm$0.4	&	0.27$\pm$0.07	&	2.1	&	-	&	157.2$\pm$47.3	&		2\\
$[$FeVII]	&	99.1	&	5720.7	&	5963.0$\pm$0.2	&	0.64$\pm$0.09	&	4.1	&	44.6$\pm$10.1	&	184.4$\pm$27.7	&	2	\\
HeI	&	0	&	5875.6	&	6124.2$\pm$0.5	&	0.17$\pm$0.04	&	1.9	&	30.6$\pm$12.1	&	159.0$\pm$16.5	&	2	\\
$[$FeVII]	&	99.1	&	6086.9	&	6343.6$\pm$0.1	&	1.11$\pm$0.08	&	5.2	&	-10.0$\pm$4.5	&	209.2$\pm$11.0	&	2	\\
?	&		&	-	&	6443.5$\pm$0.5	&	0.11$\pm$0.04	&	1.6	&	-	&	139.1$\pm$52.0	&		2\\
$[$OI]	&	0	&	6300.3	&	6566.0$\pm$0.2	&	0.61$\pm$0.06	&	3.2	&	-5.8$\pm$7.3	&	186.9$\pm$17.5	&	2	\\
$[$OI]	&	0	&	6363.8	&	6632.4$\pm$0.3	&	0.21$\pm$0.04	&	1.3	&	0.9$\pm$13.5	&	147.9$\pm$2.4	&	2	\\
$[$FeX]	&	233.6	&	6374.6	&	6643.7$\pm$0.1	&	1.88$\pm$0.12	&	8.7	&	5.9$\pm$2.5	&	219.9$\pm$6.2	&	2	\\
$[$ArV]	&	59.8	&	6435.1	&	6707.0$\pm$0.5	&	0.13$\pm$0.04	&	1.5	&	16.0$\pm$25.4	&	176.5$\pm$50.3	&	2	\\
$[$NII]	&	14.5	&	6548.1	&	6823.7$\pm$0.7	&	0.17$\pm$0.06	&	1.3	&	-29.9$\pm$30.0	&	170.5$\pm$51.2	&	2	\\
H$\alpha$	&	0	&	6562.8	&	6839.7$\pm$0.0	&	6.36$\pm$0.21	&	30.5	&	-3.1$\pm$1.1	&	213.7$\pm$2.8	&	2	\\
$[$NII]	&	14.5	&	6583.4	&	6861.2$\pm$0.2	&	0.71$\pm$0.08	&	3.9	&	-2.9$\pm$7.8	&	180.7$\pm$18.5	&	2	\\
$[$SII]	&	10.4	&	6716.4	&	6999.0$\pm$0.5	&	0.11$\pm$0.03	&	1	&	-33.9$\pm$20.1	&	78.1$\pm$97.2	&	2	\\
$[$SII]	&	10.4	&	6730.8	&	7014.8$\pm$0.4	&	0.22$\pm$0.04	&	2	&	-3.3$\pm$17.8	&	181.8$\pm$35.5	&	2	\\
$[$FeXI]	&	262.1	&	6985.2	&	7278.5$\pm$0.3	&	0.24$\pm$0.04	&	2.4	&	-61.5$\pm$13.0	&	142.2$\pm$25.1	&	2	\\
$[$ArV]	&	59.8	&	7005.9	&	7299.5$\pm$0.8	&	0.24$\pm$0.06	&	2.4	&	-82.8$\pm$32.6	&	260.6$\pm$65.5	&	2	\\
$[$ArIII]	&	27.6	&	7135.8	&	7436.6$\pm$0.4	&	0.21$\pm$0.07	&	2	&	-13.3$\pm$16.6	&	144.6$\pm$74.0	&	2	\\
$[$OII]	&	13.6	&	7319.9	&	7629.2$\pm$0.5	&	0.32$\pm$0.08	&	2.2	&	14.1$\pm$20.3	&	195.1$\pm$62.2	&	2	\\
$[$OII]	&	13.6	&	7330.2	&	7639.7$\pm$0.6	&	0.21$\pm$0.06	&	1.9	&	6.8$\pm$25.5	&	175.5$\pm$50.5	&	2	\\
$[$FeXI]	&	262.1	&	7891.8	&	8225.0$\pm$0.1	&	1.45$\pm$0.13	&	7.6	&	4.3$\pm$5.6	&	208.3$\pm$15.6	&	2	\\

\hline
\end{tabular}								
\label{tab:1241}
\end{table*}

\begin{table*}
\centering
\caption{Line identifications for the SDSS spectrum of 1641+43. `I$_{P}$ presents the ionization potentials for the securely identified emission species. Flux ratios are relative to H$\beta$ and are not corrected for reddening. Note that the fluxes presented here are from the double-Gaussian fitting model. Line IDs, $\Delta$v and FWHM are determined by fitting a single Gaussian to the emission lines. The total flux of the H$\beta$ emission line is (2.26$\pm$0.10) x 10$^{-15}$ erg s$^{-1}$ cm$^{-2}$ \AA$^{-1}$. The column `Gaussians' indicates the number of Gaussians required to fit the profile of the emission line.}
\begin{tabular}{lccccccccc}
\hline
Line ID	&	I$_P$ (eV)	&	$\lambda$ (\AA )	&	$\lambda$$_{SDSS}$  (\AA ) 	&	F/F$_{H\beta}$	&	EW (\AA )	&	$\Delta$v	&	FWHM	&	Gaussians	\\
\hline																	
$[$NeV]	&	97.1	&	3345.6	&	4084.2$\pm$0.4	&	0.46$\pm$0.18	&	16.0	&	-174.3$\pm$133.5	&	744.5$\pm$99.6	&	2	\\
$[$NeV]	&	97.1	&	3425.9	&	4182.1$\pm$0.2	&	1.73$\pm$0.13	&	47.9	&	-164.2$\pm$62.5	&	858.8$\pm$37.1	&	2	\\
$[$OII]	&	13.6	&	3727.4	&	4552.4$\pm$0.2	&	0.48$\pm$0.02	&	20.6	&	-8.6$\pm$0.2	&	490.3$\pm$41.3	&	2	\\
$[$FeVII]	&	99.1	&	3758.9	&	4588.8$\pm$0.8	&	0.36$\pm$0.09	&	15.2	&	-149.5$\pm$222.3	&	809.3$\pm$129.1	&	2	\\
$[$NeIII]	&	41.0	&	3868.8	&	4725.7$\pm$0.6	&	0.95$\pm$0.04	&	41.9	&	22.1$\pm$1.3	&	827$\pm$61.4	&	2	\\
$[$FeV]	&	54.8	&	3891.3	&	4752.2$\pm$0.8	&	0.20$\pm$0.08	&	8.3	&	-40.0$\pm$3.8	&	741$\pm$68.9	&	2	\\
$[$NeIII]	&	41.0	&	3967.5	&	4846.0$\pm$0.4	&	0.30$\pm$0.04	&	12.9	&	6.9$\pm$0.3	&	369.9$\pm$58.2	&	2	\\
?	&		&	-	&	4854.0$\pm$0.8	&	0.07$\pm$0.01	&	3.1	&	-102.3$\pm$111.9	&	369.2$\pm$58.1	&	1	\\
H$\delta$	&	0	&	4101.7	&	5009.9$\pm$1.0	&	0.18$\pm$0.04	&	6.6	&	7.2$\pm$0.7	&	665.7$\pm$224.4	&	2	\\
H$\gamma$	&	0	&	4340.5	&	5303.6$\pm$0.8	&	0.40$\pm$0.03	&	15.8	&	122.0$\pm$168.0	&	1222.0$\pm$124.4	&	2	\\
$[$OIII]	&	35.1	&	4363.2	&	5329.0$\pm$0.7	&	0.39$\pm$0.04	&	15.6	&	-14.7$\pm$2.2	&	819.2$\pm$77.7	&	2	\\
$[$OII]	&	13.6	&	4414.9	&	5396.9$\pm$2.7	&	0.15$\pm$0.04	&	5.9	&	251.5$\pm$120.0	&	710.0$\pm$59.2	&	2	\\
HeII	&	24.6	&	4685.7	&	5721.2$\pm$0.7	&	0.30$\pm$0.03	&	9.6	&	-102.3$\pm$111.9	&	680.9$\pm$73.6	&	2	\\
H$\beta$	&	0	&	4861.3	&	5937.8$\pm$0.3	&	-	&	35.0	&	8.3$\pm$0.3	&	721.3$\pm$39.8	&	2	\\
$[$OIII]	&	35.1	&	4958.9	&	6057.3$\pm$0.4	&	1.77$\pm$0.08	&	67.3	&	26.9$\pm$4.3	&	626.3$\pm$45.3	&	2	\\
$[$OIII]	&	35.1	&	5006.8	&	6115.4$\pm$0.1	&	5.31$\pm$0.25	&	201.9	&	-	&	692.3$\pm$16.5	&	2	\\
$[$FeVII]	&	99.1	&	5159.0	&	6302.5$\pm$0.2	&	0.08$\pm$0.01	&	3.5	&	64.1$\pm$10.9	&	546.0$\pm$24.8	&	2	\\
$[$FeVI]	&	75.0	&	5176.4	&	6315.9$\pm$1.1	&	0.11$\pm$0.02	&	4.0	&	-312.7$\pm$200.0	&	543.4$\pm$104.7	&	2	\\
$[$FeVII]	&	99.1	&	5720.7	&	6985.9$\pm$0.9	&	0.25$\pm$0.04	&	10.2	&	-59.3$\pm$63.6	&	788.0$\pm$181.7	&	2	\\
$[$FeVII]	&	99.1	&	6086.9	&	7430.2$\pm$1.0	&	0.50$\pm$0.12	&	19.4	&	-174.5$\pm$78.0	&	904.1$\pm$92.9	&	2	\\
$[$OI]	&	0	&	6300.3	&	7698.9$\pm$2.4	&	0.21$\pm$0.05	&	6.6	&	144.5$\pm$50.0	&	648.2$\pm$237.2	&	2	\\
$[$OI]	&	0	&	6363.8	&	7778.2$\pm$3.4	&	0.07$\pm$0.02	&	1.7	&	210.5$\pm$100.0	&	954.5$\pm$316.7	&	1	\\
$[$FeX]	&	233.6	&	6374.6	&	7785.7$\pm$0.7	&	0.21$\pm$0.05	&	7.2	&	-39.9$\pm$8.2	&	814$\pm$85.2	&	2	\\
$[$NII]	&	14.5	&	6548.1	&	7997.1$\pm$0.5	&	0.57$\pm$0.1	&	17.6	&	-29.0$\pm$10.0	&	364.4$\pm$69.9	&	2	\\
H$\alpha$	&	0	&	6562.8	&	8017.7$\pm$1.6	&	5.38$\pm$0.19 	&	166.0	&	60.8$\pm$15.8	&	476.8$\pm$38.2	&	2	\\
$[$NII]	&	14.5	&	6583.4	&	8040.2$\pm$0.5	&	1.71$\pm$0.11	&	52.8	&	-29.4$\pm$10.3	&	362.0$\pm$69.4	&	2	\\
$[$SII]	&	10.4	&	6716.4	&	8203.6$\pm$1.0	&	0.19$\pm$0.05	&	8.0	&	5.8$\pm$0.7	&	479.7$\pm$170.7	&	2	\\
$[$SII]	&	10.4	&	6730.8	&	8221.3$\pm$1.3	&	0.30$\pm$0.04	&	5.0	&	10.6$\pm$3.2	&	350.3$\pm$84.5	&	2	\\

\hline
\end{tabular}								
\label{tab:1641}
\end{table*}

\section{Unknown line table}

\begin{table*}
\centering
\caption{Emission lines without a secure ID for the SDSS spectra. Information for Q1131+16 is presented to highlight any overlap.}
\begin{tabular}{lccccc}
\hline
$\lambda$$_{rest}$ (\AA )	&	Mrk 1388	&	III Zw 77	&	J1241+44	&	J1641+43	&	Q1131+16	\\
\hline
3973.5	&		&	Y	&		&	Y	&		\\
4219.4	&	Y	&		&		&		&		\\
4230.6	&	Y	&		&		&		&		\\
4243.8	&	Y	&		&		&		&		\\
4659.1	&		&	Y	&		&		&		\\
5413.0	&	Y	&		&		&		&	Y	\\
5440.4	&	Y	&		&		&		&		\\
5460.3	&	Y	&		&		&		&		\\
5533.1	&		&		&	Y	&		&	Y	\\
5618.6	&	Y	&		&	Y	&		&	Y	\\
6075.1	&	Y	&	Y	&		&		&		\\
6182.6	&		&		&	Y	&		&		\\
6678.1	&	Y	&		&		&		&		\\
6915.4	&	Y	&		&		&		&		\\
6997.1	&	Y	&		&		&		&		\\
7043.9	&	Y	&	Y	&		&		&		\\
7154.9	&	Y	&		&		&		&	Y	\\
7171.6	&	Y	&		&		&		&		\\
7377.1	&	Y	&	Y	&		&		&		\\
7608.8	&	Y	&		&		&		&		\\
7752.4	&	Y	&		&		&		&		\\
8450.0	&		&	Y	&		&		&		\\
\hline
\end{tabular}								
\label{tab:?}
\end{table*}

\label{lastpage}

\end{document}